\documentclass[a4paper,11pt]{article}

\usepackage[margin=1in]{geometry} 

\usepackage{amsmath}
\usepackage{amsthm}
\usepackage{amssymb}
\usepackage{lineno}

\usepackage{etoolbox}

\usepackage{graphicx}
\usepackage{bm}
\usepackage{lineno}
\usepackage{caption}
\usepackage{subfigure}	

\usepackage{booktabs}

\usepackage[breaklinks]{hyperref}

\usepackage{multirow}

\usepackage{comment}
\usepackage{xargs}                      

\usepackage[utf8]{inputenc}

\hypersetup{
	unicode,
	colorlinks=true,
	linkcolor=blue,
	filecolor=magenta,      
	citecolor=blue,
	urlcolor=cyan,
	pdfauthor={P. Sperotto, S. Pieraccini and M. A. Mendez},
	pdftitle={A Meshless Method to Compute Pressure Fields from Image Velocimetry},
	pdfsubject={A Meshless Method to Compute Pressure Fields from Image Velocimetry},
	pdfkeywords={article, template, simple},
	pdfproducer={LaTeX},
	pdfcreator={pdflatex}
	colorlinks=true,
	citecolor=blue,
	linkcolor=blue,
	filecolor=magenta,      
	urlcolor=blue,
}

\usepackage[sort&compress,authoryear,round]{natbib}
\usepackage{algorithm, algpseudocode} 
\usepackage{mathrsfs} 

\title{A Meshless Method to Compute \\Pressure Fields from Image Velocimetry}
\author{Pietro Sperotto$^{1,2}$ \and Sandra Pieraccini$^2$ \and Miguel A. Mendez$^1$}

\date{
	$^1$von Karman Institute for Fluid Dynamics, Sint-Genesius-Rode, Belgium \\%
	$^2$Dipartimento di Ingegneria Meccanica e Aerospaziale, Politecnico di Torino, Italy
}

\begin{document}
	\maketitle
	
	\begin{abstract}
		
		We propose a meshless method to compute pressure fields from image velocimetry data, regardless of whether this is available on a regular grid as in cross-correlation based velocimetry or on scattered points as in tracking velocimetry. The proposed approach is based on Radial Basis Functions (RBFs) regression and relies on the solution of two constrained least square problems. The first one is the regression of the measurements to create an analytic representation of the velocity field. This regression can be constrained to impose boundary conditions (e.g. no-slip velocity on a wall or inlet conditions) or differential constraints (e.g. the solenoidal condition for an incompressible flow). The second one is the meshless integration of the pressure Poisson equation, achieved by seeking a solution in the form of a RBF expansion and using constraints to impose boundary conditions.
		
		We first illustrate the derivation of the two least square problems and the numerical techniques implemented for their solution. Then, we showcase the method with three numerical test cases of growing complexity. These are a 2D Gaussian Vortex, a 2D flow past a cylinder from CFD and a 3D Stokes flow past a sphere. For each case, we consider randomly sampled vector fields simulating particle tracking measurements and analyze the sensitivity to noise and seeding density.

		\vspace{7mm}
		
	\end{abstract}

	
	
	\clearpage
	\tableofcontents
	\clearpage

	\section{Introduction} \label{sec:1}
	Modern image-based velocimetry can provide velocity fields with sufficient resolution to allow for computing pressure fields. This enables non-intrusive measurements of aerodynamic loading or sound fields, and considerable experimental insights into fluids dynamics.
	
	Many methods have been developed to this end, initially mostly based on Particle Image Velocimetry (PIV) and more recently adapted and enhanced by advances in Lagrangian Particle Tracking (LPT). An extensive literature review and a comparative analysis between various methods can be found in \cite{Charonko2010,van2013piv,Gent2017,Pan2016,de2012instantaneous,McClure2017,Liu2020}. Broadly, pressure integration methods can be classified into Eulerian and Lagrangian, depending on how the material acceleration is computed.
	
	Eulerian methods use the local time derivatives and velocity gradients and can be further subdivided into directional approaches integrating the pressure gradient from the Navier Stokes Equation (e.g., \cite{Jakobsen1997,Kngeter1999PIVWH,Liu2006,Wang2019}) or global approaches integrating the Poisson equation (e.g., \cite{gurka1999computation,Ghaemi2012,Pan2016}). These methods have been implemented in many variants. For example, one could include turbulence modelling via Reynolds Averaged Na\-vier Stokes (RANS) formulations to compute averaged pressure fields (e.g., \cite{gurka1999computation,Oudheusden2007}) or leverage Taylor's frozen turbulence hypothesis to compute instantaneous pressures \citep{de2012instantaneous,laskari2016full,van2019pressure}. Within the class of Eulerian methods, more sophisticated methods include solvers based on pres\-sure-velocity algorithms from CFD \citep{Gunaydinoglu2019,Francisco} or immersed boundary techniques \citep{Pirnia2020}.
	
	Lagrangian or pseudo-Lagrangian methods compute the material acceleration along the trajectories of fluid particles. Following \cite{de2012instantaneous}, pseudo-LAgrangian (pLA) approaches use pseudo-track\-ing, in the sense that the particle trajectories are reconstructed from (Eulerian) velocity fields (e.g., \cite{Schneiders2016a,Novara2013}). On the other hand, fully Lagrangian methods are based on the direct determination of particle trajectories. These have been recently enabled by advances in tracking techniques \citep{Schanz2016}, nowadays capable of providing accurate trajectories with high seeding densities. Lagrangian approaches can be further distinguished in techniques that interpolate the particle acceleration onto a Cartesian grid \citep{Geseman,Huhn2016,Huhn2018,Schneiders2016} and techniques that integrate the pressure on scattered data such as the Voronoi integration proposed by \cite{Neeteson2015}.
	
	Once the material acceleration is computed, most of the aforementioned approaches are based on ``classic'' numerical techniques to integrate the relevant equations. By ``classic'', we here consider those methods based on Finite Differences, Finite Volumes or Finite Elements that require building a computational mesh. This is a difficult task, especially in the presence of curved boundaries and/or laser reflections, and requires advanced interpolation methods \citep{Agarwal2021} to map the scattered data onto the computational grid. 
	
	Recent advances in data assimilation and machine learning are currently opening new perspectives for meshless methods to solve Partial Differential Equations (PDEs). Among the most notable examples, we mention the use of Artificial Neural Networks (ANN) to discover and solve PDEs \citep{Lagaris1998,Sirignano2018,Raissi2019,Li2020}, recently popularized as \emph{physics-informed} neural networks (see \cite{Raissi2019}).

	Promoting this paradigm shift is the fact that many parametric models from the machine learning literature can be easily differentiated with respect to their inputs. Therefore, their prediction can be easily constrained to respect PDEs and related initial and boundary conditions, as well as other differential conditions (e.g. solenoid or potential fields). Thus, the problem of solving a PDE can be converted into a constrained optimization problem: given a parametric representation of the form $\boldsymbol{y}=f(\boldsymbol{u},\boldsymbol{w})$, linking some input field $\boldsymbol{u}$ to some output field $\boldsymbol{y}$, one seeks to identify the set of parameters $\boldsymbol{w}$ such that $\boldsymbol{y}$ solves a PDE (i.e. minimizes some properly defined residuals).
	The framework applies equally well to Artificial Neural Networks or to Radial Basis Functions (RBFs). Compared to the ANN, the RBF leads to least square that are considerably easier to handle because of their linearity with respect to the model parameters. It is thus not surprising that the meshless integration of PDEs via RBFs has a long history in computational engineering. The idea was introduced by \cite{Kansa1990,Kansa1990a} and has grown into a mature approach with an extensive literature (see \cite{Fornberg2015,cmes,Chen2002,Chen2003,Sarler}). RBF-based meshless integration extend classic pseudo-spectral methods \citep{Fornberg1996}, in which the parametrization is usually based on Fourier or Chebyshev expansions, and can be seen as a special class of collocation methods.

	In the literature of image velocimetry, RBFs have been used for their robust interpolation \citep{Casa2013}, to compute derivatives \citep{Karri2009}, as regression tools for super-resolution \citep{Ratz} and to support the physics-informed interpolation from scattered to uniform grids \citep{Schneiders2016}, also enabling super-resolution. Nevertheless, to the author's knowledge, their implementation for the mesh-less integration of PDEs has not yet been fully exploited. This work proposes a meshless method based on RBFs to compute pressure fields from scattered (and noisy) velocity fields. The input velocity fields can be 2D or 3D and can result from PTV or PIV measurements. We focus on integrating Poisson equation in an Eulerian formalism, but we note that the proposed algorithm is essentially a tool to solve PDEs and can be generalized to a Lagrangian formalism. Similarly, the integration could be used to compute instantaneous pressure fields if time-resolved data is available (or Taylor's hypothesis invoked) or include RANS modelling for turbulence. These extensions are currently under investigation and will be presented in a dedicated contribution.
	
	\textcolor{black}{The proposed method differs from the ``track-based'' approaches proposed by \cite{Gesemann2015} and \cite{Bobrov2021} in that our approach is purely Eulerian, i.e. it requires velocity information in various points but no trajectories. It it thus} conceptually similar to the ``second-generation'' flowfit by \cite{Geseman}, in that it also formulates the pressure integration as a sparse least square problem. However, our method differ in the basis selection and collocation, in the cost function formulation and in definition of penalties and constraints.
	
	The second generation flow fit \cite{Geseman} uses uniformly collocated B-splines and constructs a single cost function including the accuracy of the velocity regression and the pressure integration. The result is a nonlinear regression problem. Moreover, physical constraints (e.g. boundary conditions or solenoidal condition) are included as penalties (regularizations) and not as \emph{hard} constraints. Our approach uses scattered truncated Gaussian RBFs, collocated by a clustering algorithm, and splits the velocity regression and the pressure integration into two different problems. This produces two \emph{linear} least square problems. Moreover, physical constraints are included both as penalties and as hard constraints, by using Lagrange multiplies.

	
	
	We illustrate the method on three synthetic test cases of growing complexity, for which the background truth is available. This let us test its robustness against measurement noise and seeding concentration (sparsity) of the measured velocity field. The mathematical background of the RBF integration is described in Section \ref{sec:2}, while Section \ref{sec:3} briefly introduces the selected test cases. Section \ref{sec:4} collects the results; conclusions and perspectives are given in Section \ref{sec:5}.

	\section{Mathematical Framework}\label{sec:2}
	
	We begin by introducing Radial Basis Functions (RBFs) as tools for approximating a function $f :  \mathbb{R}^d \mapsto \mathbb{R}$, focusing on the cases $d=2$ and $d=3$. The function approximation is built from data and can be constrained via standard tools from constrained optimization. 
	
	The general formulation is described in Section \ref{sec:2p1}. We introduce in Section \ref{sec:2p2} the problem of deriving an approximation of velocity fields using constraints (or penalties) to impose (or to promote) boundary conditions and physical priors (e.g. divergence-free). The same framework is then used in Section \ref{sec:2p3} for the meshless integration of the Poisson equation. Section \ref{sec:2p4} completes the presentation of the method illustrating the clustering algorithm used to collocate the RBFs. Finally, Section \ref{sec:2p5} presents our current approach to solve the large systems of equations produced in \ref{sec:2p2} and \ref{sec:2p3}.

	\subsection{Constrained Regression via RBFs}\label{sec:2p1}
	
	Among the many possible RBFs (see \cite{Fornberg2015} for more background), we here consider Gaussians of the form
	
	\begin{equation}
		\label{eq1}
		\varphi_{k}(\boldsymbol{x}| \boldsymbol{x}_k^*, c_k)=\exp \left(-c_k^{2}\left\|\boldsymbol{x}-\boldsymbol{x}_{\boldsymbol{k}}^{*}\right\|^{2}\right)
	\end{equation} where $c_k>0$ is the {\em shape parameter} and $\boldsymbol{x}^*_k\in\mathbb{R}^{d}$ is the {\em collocation point}. In this section, we consider both $c_k$ and $ \boldsymbol{x}_k^*$ as given: their identification is discussed in Sec. \ref{sec:2p4}. The basis element $\varphi_k$ is thus solely function of $\boldsymbol{x}$ (the symbol $|$ separates variables from parameters) and is a \emph{radial} function because it only depends on the distance from the collocation points. Note that we restrict the treatment to {\em isotropic} bases, as they depend on one shape parameter only.
	
	The function approximation is a linear combination of $n_c$ RBFs:
	
	\begin{equation}
		\label{eq2}
		f(\boldsymbol{x})\approx \tilde{f}(\boldsymbol{x})=\sum_{k=1}^{n_c} w_{k}\, \varphi_{k}(\boldsymbol{x}| \boldsymbol{x}_k^*, c_k)\,,
	\end{equation} with the {\em weights} $w_k$ to be identified from data. 
	
	In the machine learning literature, eq. \eqref{eq2} can be seen as a linear artificial neural network \citep{Schwenker2001,Broomhead1988MultivariableFI} with a single hidden layer, having the RBFs in \eqref{eq1} as activation functions and a single output with linear activation. This parallelism opens the path towards an arsenal of stochastic optimization techniques and bridges with the recent advances in physics informed neural networks \citep{Raissi2019,Li2020}. However, the linearity of the model \eqref{eq2} with respect to the weights makes the regression problem considerably simpler than in ANN-based formulation.
	
	Following the machine learning literature, the identification of the weights is hereinafter referred to as \emph{training}, and the data used at the scope is referred to as \emph{training set}. In this work, this is a set of $n_p$ pairs
	$\{(\boldsymbol{x}_i, f_i)\}_{i=1,\ldots,n_p}$
	with $\boldsymbol{x}_i=(x_i,y_i,z_i)$ and $f_i= f(\boldsymbol{x}_i)$, produced, for example, by 3D PTV measurements. The coordinates of the sampling points can be arranged into a matrix $\boldsymbol{X}\in\mathbb{R}^{n_p\times 3}$ and the treatment that follows holds regardless of whether these points are randomly scattered as in PTV or over a grid as in PIV.
	
	The stability of the learning is greatly enhanced if a set of $n_\gamma \ll n_c$ polynomial functions $\gamma_k(\boldsymbol{x})$ is added to the basis so that \eqref{eq2} becomes 
	
	\begin{equation}\label{RBFstandard}
		\tilde{f}(\boldsymbol{x})=\sum_{k=1}^{n_\gamma} w_{k} \gamma_{k}(\boldsymbol{x})+\sum_{k=1}^{n_c} w_{n_\gamma+k} \varphi_{k}(\boldsymbol{x}| \boldsymbol{x}_k^*, c_k)\,.
	\end{equation} Nevertheless, in the interest of conciseness in the notation, we treat these additional terms as elements of the same basis of $n_b=n_\gamma+n_c$ functions. Accordingly, we write the function approximation \eqref{RBFstandard} on the training set as 
	
	\begin{equation}
		\label{approx}
		\boldsymbol{f}\simeq \tilde{\boldsymbol{f}}=\boldsymbol{\Phi} (\boldsymbol{X})\,\boldsymbol{w}, 
	\end{equation} where $\boldsymbol{f}=(f_1,f_2\dots f_{n_p})^T$ is the vector collecting all the training \emph{targets}, $\boldsymbol{w}=(w_1,w_2\dots w_{n_b})^T$ is the vector containing the set of weights (regardless of whether these are associated to RBFs or polynomial terms) and we let $\boldsymbol{\Phi} (\boldsymbol{X})\in\mathbb{R}^{n_b\times n_p}$ denote the matrix obtained by evaluating the $n_b$ basis functions at the $n_p$ training points:

	\begin{equation*}
		\boldsymbol{\Phi}(\boldsymbol{X})=\left(\begin{array}{cccccc}
			\gamma_{1}\left(\boldsymbol{x}_{1}\right)&\cdots &\gamma_{n_{\gamma}}\left(\boldsymbol{x}_{1}\right)&\varphi_{1}\left(\boldsymbol{x}_{1}\right) & \cdots & \varphi_{n_c}\left(\boldsymbol{x}_{1}\right) \\
			\vdots & \cdots & \vdots & \vdots & \vdots & \vdots\\
			\gamma_{1}\left(\boldsymbol{x}_{n_p}\right) & \cdots & \gamma_{n_\gamma}\left(\boldsymbol{x}_{n_p}\right)&\varphi_{1}\left(\boldsymbol{x}_{n_p}\right) & \cdots & \varphi_{n_c}\left(\boldsymbol{x}_{n_p}\right)
		\end{array}\right).
	\end{equation*}
	
	Training the RBF model \eqref{RBFstandard} corresponds to solving the linear system \eqref{approx}, in a least square sense, for the weights $\boldsymbol{w}\in \mathbb{R}^{n_b}$. This is a classic problem involving the minimization of a quadratic cost function:
	
	\begin{equation}
		\label{cost2}
		J(\boldsymbol{w})=||\boldsymbol{f}-\boldsymbol{\Phi} (\boldsymbol{X})\,\boldsymbol{w}||^2_2\,,
	\end{equation} where $||\cdot ||_2$ denotes the $l_2$ norm. Since such a problem is notoriously ill-posed \citep{Bishop2006}, it is common practice adding a regularization, namely a penalty on the magnitude of the weights. We consider a Tikhonov \citep{Kress1998} regularization of the form $\alpha ||\boldsymbol{w}||^2_2$, with penalty $\alpha$ computed as described in Sec. \ref{sec:2p5}. However, since this regularization is only used to ensure numerical stability, the introduction of this term is deferred to the next sections and we here focus on the general problem formulation. Moreover, we leave the treatment open to a second penalty to promote certain conditions (e.g. divergence-free) for the resulting function. We include such a condition in the form of a linear operator $\boldsymbol{C}_s\boldsymbol{w}=0$, with $\boldsymbol{C}_s\in\mathbb{R}^{n_s\times n_b}$, acting on $n_s$ points. Thus, an additional regularization term $\alpha_s ||\boldsymbol{C}_s\boldsymbol{w}||^2_2$ is added.
	
	Finally, we further extend the treatment to include constraints to the approximation. In particular, we consider two sets of linear equality constraints written as $\boldsymbol{C}_1\boldsymbol{w}=\boldsymbol{c}_1$ and $\boldsymbol{C}_2\boldsymbol{w}=\boldsymbol{c}_2$, with $\boldsymbol{C}_1\in\mathbb{R}^{n_{\lambda_1}\times n_b}$ and $\boldsymbol{C}_2\in\mathbb{R}^{n_{\lambda_2}\times n_b}$ linear operators acting on $n_{\lambda_1}$ and $n_{\lambda_2}$ points respectively, and $\boldsymbol{c}_1\in\mathbb{R}^{n_{\lambda_1}}$ and $\boldsymbol{c}_2\in\mathbb{R}^{n_{\lambda_2}}$. As detailed in the next sections, we use these constraints to impose Dirichlet, Neumann or mixed boundary conditions in the pressure integration, or to impose physical constraints to the regression of the velocity field.
	
	In other words, the problem is formulated as
	
	\begin{eqnarray}\label{const_problem}
		&\min & J(\boldsymbol{w})\\
		&\text{s.t.} & \boldsymbol{C}_1\boldsymbol{w}=\boldsymbol{c}_1, \, \boldsymbol{C}_2\textbf{w}=\boldsymbol{c}_2
	\end{eqnarray}
	
	The solution of problem \eqref{const_problem} can be pursued by solving the Karush-Kuhn-Tucker optimality conditions \citep{JorgeNocedal2006,Chong2013}. Let $J^*(\boldsymbol{w},\bm{\lambda})$ denote the Lagrangian
	
	\begin{equation}
		\label{Lagrangian}
		J^*(\boldsymbol{w},\bm{\lambda})=||\boldsymbol{f}-\boldsymbol{\Phi} (\boldsymbol{X})\,\boldsymbol{w}||^2_2+\bm{\lambda}^T_1\bigl(\boldsymbol{C}_1\boldsymbol{w}-\boldsymbol{c}_1\bigr)+    \bm{\lambda}^T_2\bigl(\boldsymbol{C}_2\boldsymbol{w}-\boldsymbol{c}_2\bigr)+\alpha_s ||\boldsymbol{C}_s \boldsymbol{w}||^2_2\,\,,
	\end{equation} where $\bm{\lambda}_1 \in \mathbb{R}^{n_{\lambda 1}}$ and $\bm{\lambda}_2 \in \mathbb{R}^{n_{\lambda_2}}$ are Lagrange multipliers associated to the linear equality constraints and $\bm{\lambda}=(\bm{\lambda}_1,\bm{\lambda}_2)^T \in\mathbb{R}^{n_{\lambda}}$ with $n_{\lambda}=n_{\lambda_{1}}+n_{\lambda_{2}}$. 
	The KKT optimality conditions for this problem are obtained canceling the gradients with respect both to weights $(\nabla_{\boldsymbol{w}} J^{*})$, and to the multipliers $(\nabla_{\bm{\lambda}} J^{*})$. This leads to the following system of equations:
	
	\begin{eqnarray*}         2\bigl(\boldsymbol{\Phi}^T\boldsymbol{\Phi}+\alpha_s \boldsymbol{C}^T_s\boldsymbol{C}_s\bigr) \boldsymbol{w} +\boldsymbol{C}^T_1\bm{\lambda}_1+\boldsymbol{C}^T_2\bm{\lambda}_2
		&=&2\boldsymbol{\Phi}^T \boldsymbol{f}_i \\
		\boldsymbol{C}_1\boldsymbol{w}
		&=&\boldsymbol{c}_1 \\
		\boldsymbol{C}_2\boldsymbol{w}
		&=&\boldsymbol{c}_2 \,,
	\end{eqnarray*}
	having used the short-hand notation $\boldsymbol{\Phi}$ for $\boldsymbol{\Phi}(\boldsymbol{X})$. This system can be conveniently written as 
	
	\begin{equation}
		\label{Blocks}
		\left(  \begin{array}{cc}
			\boldsymbol{A}  & \boldsymbol{B} \\
			\boldsymbol{B^T}  & \boldsymbol{0}
		\end{array}\right) \left(\begin{array}{c}
			\boldsymbol{w}  \\
			\bm{\lambda} 
		\end{array}\right)=\left(\begin{array}{c}
			\boldsymbol{b}_1  \\
			\boldsymbol{b}_2
		\end{array}\right)\,,
	\end{equation} with $\boldsymbol{A}=2\boldsymbol{\Phi}^T\boldsymbol{\Phi}+2\alpha_s \boldsymbol{C}^T_s\boldsymbol{C}_s \in\mathbb{R}^{n_b\times n_b}$, $\boldsymbol{B}=\left (\boldsymbol{C}^T_1,\boldsymbol{C}^T_2\right) \in\mathbb{R}^{n_b\times n_{\lambda}}$, $\boldsymbol{b_1}=2\boldsymbol{\Phi}^T \boldsymbol{f}_i\in \mathbb{R}^{n_b}$ and $\boldsymbol{b}_2=(\boldsymbol{c}_1,\boldsymbol{c}_2)^T\in\mathbb{R}^{n_\lambda }$.
	
	The solution of \eqref{Blocks} leads to a minimum of $J(\boldsymbol{w})$ as long as its Hessian $\nabla^2_{\boldsymbol{w}} J(\boldsymbol{w})=\boldsymbol{A}$ is positive semi-definite. Both the problems of velocity fields approximation and meshless pressure computation leads to constrained least square problems of the form in \eqref{Blocks}. It is worth pointing out that once the training is complete, the approximation in \eqref{approx} is analytical: it can be evaluated in \emph{any} new set of points and its derivatives are analytically available. This enables the meshless integration of PDEs.
	
	\subsection{Approximating Velocity Fields}\label{sec:2p2}
	We now set the problem of approximating velocity fields in the framework introduced in the previous section. We present the derivation for the case of a 2D field and refer the reader to Appendix \ref{AppendixA} for the problem set in 3D. The dataset now consists of two scalar fields for the velocity components, i.e. $U=(u_i,v_i)$, sampled over a set of $n_p$ points. The collocation points for the Gaussian RBFs are now on the plane $\boldsymbol{x}^*_k=(x^*_k,y^*_k)$. Eq. \eqref{eq2} and its matrix form in \eqref{approx} can be conveniently extended to vector fields if the data is reshaped into column vectors. Namely, let $\boldsymbol{u},\boldsymbol{v}\in\mathbb{R}^{n_p}$ be the vectors collecting all the sampled values of $u_i$ and $v_i$ in the grid points $\boldsymbol{x}_i$ stored in $\boldsymbol{X}\in\mathbb{R}^{n_p \times 2}$. Let $\boldsymbol{U}=(\boldsymbol{u};\boldsymbol{v})\in\mathbb{R}^{2 n_p}$ be a column vector with the semicolon ; denoting vertical concatenation. Similarly, let $\boldsymbol{w}_U=(\boldsymbol{w_u};\boldsymbol{w_v})\in\mathbb{R}^{2 n_b}$ be the colum vector concatenating the weights $\boldsymbol{w_u},\boldsymbol{w_v}\in\mathbb{R}^{n_b}$ for the RBFs expansion of the velocity components $u$ and $v$. Then, the approximation of the velocity field in the available points is:
	\begin{equation}
		\label{approx_U}
		{\boldsymbol{U}}=\left(  \begin{array}{c}
			\boldsymbol{u}   \\
			\boldsymbol{v} 
		\end{array}\right)\approx \tilde{\boldsymbol{U}}=\left(  \begin{array}{c}
			\tilde{\boldsymbol{u}}   \\
			\tilde{\boldsymbol{v}} 
		\end{array}\right) = \left(  \begin{array}{cc}
			\boldsymbol{\Phi}(\boldsymbol{X}) & \boldsymbol{0}  \\
			\boldsymbol{0} & \boldsymbol{\Phi}(\boldsymbol{X})
		\end{array}\right)\left(  \begin{array}{c}
			\boldsymbol{w_u}   \\
			\boldsymbol{w_v}
		\end{array}\right)=\boldsymbol{\Phi}_U(\boldsymbol{X})\boldsymbol{w}_U\,,
	\end{equation} where the same matrix of basis function $\boldsymbol{\Phi}(\boldsymbol{X})$ is used for both components, assuming that these are available at the \emph{same} points in $\boldsymbol{X}$. To ease the connection with the previous section, we introduce the basis matrix $\boldsymbol{\Phi}_U(\boldsymbol{X})\in\mathbb{R}^{2n_p\times 2n_b}$. Note that the regression could be carried out for both fields independently, but this would prevent using constraints and penalties that links them. The general cost function is maintained as a single objective and quadratic function:
	
	\begin{equation}
		J(\boldsymbol{w}_U)=||\boldsymbol{U}-\boldsymbol{\Phi}_U(\boldsymbol{X})\boldsymbol{w}_U||^2_2\,.
	\end{equation} 
	
	Concerning the penalties and constraints for this problem, besides the Tikhonov regularization, we seek to impose boundary conditions and physical constraints. Focusing on incompressible flows, the main physical constraint is that of a divergence-free flow. Other conditions can be considered as long as the corresponding operator is linear.
	
	The divergence-free condition is set  both as a constraint and as a penalty in different domain points. This choice offers a trade-off between the accuracy of the approximation and the amount of data required: large amounts of constraints increase the problem's size and require large sets of RBFs; penalties, albeit less restrictive, do not increase the problem size. To define the divergence operator, note that the derivatives of the approximated field are readily available from the derivatives of the basis functions. For the RBF, for example, one has the following partial derivatives along $x$ ($\partial_x$) and along $y$ ($\partial_y$):

	\begin{subequations}
		\label{eq10}
		\begin{gather}
			\partial_x \varphi_{k}(\boldsymbol{x}| \boldsymbol{x}_k^*, c_k)=-2 c^2_k (x-x^*_k)\varphi_{k}(\boldsymbol{x}| \boldsymbol{x}_k^*, c_k)\\
			\partial_y \varphi_{k}(\boldsymbol{x}| \boldsymbol{x}_k^*, c_k)=-2 c^2_k (y-y^*_k)\varphi_{k}(\boldsymbol{x}| \boldsymbol{x}_k^*, c_k)\,
		\end{gather}
	\end{subequations} 
	
	Therefore, the derivatives of the approximation are:

	\begin{subequations}
		\label{eq11}
		\begin{gather}
			\partial_x u (\boldsymbol{x}_i)\approx\sum_{k=0}^{n_c} w_{u_k}\, \partial_x\,\varphi_{k}(\boldsymbol{x}| \boldsymbol{x}_k^*, c_k)\rightarrow \boldsymbol{\Phi}_x(\boldsymbol{X}) \boldsymbol{w}_u\\
			\partial_y v (\boldsymbol{x}_i)\approx\sum_{k=0}^{n_c} w_{v_k}\, \partial_y\,\varphi_{k}(\boldsymbol{x}| \boldsymbol{x}_k^*, c_k)\rightarrow \boldsymbol{\Phi}_y(\boldsymbol{X}) \boldsymbol{w}_v\,,
		\end{gather}
	\end{subequations} where the matrices $\boldsymbol{\Phi}_x(\boldsymbol{X}),\boldsymbol{\Phi}_y(\boldsymbol{X})\in\mathbb{R}^{n_p\times n_b}$ collect basis functions' derivatives along their columns:

	\begin{subequations}
		\label{eq12}
		\begin{gather}
			\boldsymbol{\Phi}_x(\boldsymbol{X})=[\partial_x \gamma_1,\dots \partial_x \gamma_{n_\gamma},\partial_x \phi_{1}\dots \partial_x \phi_{n_c}]\\
			\boldsymbol{\Phi}_y(\boldsymbol{X})=[\partial_y \gamma_1,\dots \partial_y \gamma_{n_\gamma},\partial_y \phi_{1}\dots \partial_y \phi_{n_c}]\,.
		\end{gather}
	\end{subequations}
	
	The divergence of the velocity field $\nabla \cdot \boldsymbol{U}(\boldsymbol{X}_{\nabla})$ in a set of points $\boldsymbol{X}_\nabla\in\mathbb{R}^{n_\nabla\times 2}$ is approximated by  $\boldsymbol{D}_{\nabla}(\boldsymbol{X}_\nabla)\in\mathbb{R}^{n_\nabla \times 2 n_b}$. The divergence free condition applied to the RBF is:
	
	\begin{equation}
		\label{divergence}
		\nabla \cdot \left(  \begin{array}{c}
			\boldsymbol{u}(\boldsymbol{X}_\nabla)   \\
			\boldsymbol{v}(\boldsymbol{X}_\nabla) 
		\end{array}\right)\approx\left(  \begin{array}{cc}
			\boldsymbol{\Phi}_x(\boldsymbol{X}_\nabla) & \boldsymbol{\Phi}_y(\boldsymbol{X}_\nabla)
		\end{array}\right)\left(  \begin{array}{c}
			\boldsymbol{w_u}   \\
			\boldsymbol{w_v}
		\end{array}\right)=\boldsymbol{D}_{\nabla}(\boldsymbol{X}_\nabla)\, \boldsymbol{w}_U=\boldsymbol{0}\,,
	\end{equation} where $\boldsymbol{\Phi}_x(\boldsymbol{X}_\nabla),\boldsymbol{\Phi}_y(\boldsymbol{X}_\nabla)\in\mathbb{R}^{n_{\nabla}\times n_b}$ are the matrices of basis function derivatives at the points $\boldsymbol{X}_\nabla\in\mathbb{R}^{n_\nabla\times 2}$, and $\boldsymbol{0}\in\mathbb{R}^{2 n_{\nabla}}$ is the zero vector of appropriate size. 
	
	Equation \eqref{divergence} is introduced in \eqref{Lagrangian} as both a penalty ($\boldsymbol{C}_s$) and a constraint ($\boldsymbol{C}_1$), although generally at different points. We use $\boldsymbol{X}_\nabla\in\mathbb{R}^{n_\nabla \times 2 }$ to denote matrix collecting the coordinates of the points in which the divergence-free is a constraint.
	
	
	Finally, other constraints are added to impose boundary conditions. Let $\boldsymbol{X}_D\in\mathbb{R}^{n_D\times 2}$ collect the set of points over which $n_D$ Dirichlet conditions must be enforced, and let $\boldsymbol{c}_D\in\mathbb{R}^{2n_D}$ collect all the corresponding values for both $u$ and $v$ components, vertically concatenated. The associated linear operator $\boldsymbol{D}(\boldsymbol{X}_D)\in\mathbb{R}^{n_D\times 2 n_b}$ is:
	

	\begin{equation}
		\label{Dirvelocity}
		\left(  \begin{array}{cc}
			\boldsymbol{\Phi}(\boldsymbol{X}_D) & \boldsymbol{0}  \\
			\boldsymbol{0} & \boldsymbol{\Phi}(\boldsymbol{X}_D)
		\end{array}\right)\left(  \begin{array}{c}
			\boldsymbol{w_u}   \\
			\boldsymbol{w_v}
		\end{array}\right)=\boldsymbol{D}(\boldsymbol{X}_D)\, \boldsymbol{w}_U=\boldsymbol{c}_D\,.
	\end{equation} Similarly, let $\boldsymbol{X}_N\in\mathbb{R}^{n_N\times 2}$ collect the set of points over which Neumann conditions must be enforced, and let $\boldsymbol{c}_N=(\boldsymbol{c}_{Nu};\boldsymbol{c}_{Nv})\in\mathbb{R}^{2n_N}$ collect all the corresponding values.

		The associated linear operator $\boldsymbol{N}_{n}(\boldsymbol{X}_N)\in\mathbb{R}^{2 n_N\times 2 n_b}$ is:


		\begin{equation}
			\label{Neuvelocity}
			\left(  \begin{array}{cc}
				\boldsymbol{\Phi}_{\boldsymbol{n}}(\boldsymbol{X}_N) & \boldsymbol{0}  \\
				\boldsymbol{0} & \boldsymbol{\Phi}_{n}(\boldsymbol{X}_N)
			\end{array}\right)\left(  \begin{array}{c}
				\boldsymbol{w_u}   \\
				\boldsymbol{w_v}
			\end{array}\right)=\boldsymbol{N}_{\boldsymbol{n}}(\boldsymbol{X}_N)\, \boldsymbol{w}_U=\boldsymbol{c}_N\,,
		\end{equation} where $\boldsymbol{\Phi}_{\boldsymbol{n}}(\boldsymbol{X}_N)\in\mathbb{R}^{n_N\times n_b}$ is the matrix collecting the projection of the gradient along the normal to the surface for which the boundary condition must be set. Denoting as $\boldsymbol{n}=(n_x,n_y)$ the normal vector at a given location and $\boldsymbol{n}_x(\boldsymbol{X}_N),\boldsymbol{n}_y(\boldsymbol{X}_N) \in\mathbb{R}^{n_N}$ the collections of normal vectors at the points where the condition is to be imposed, we have 
		
		\begin{equation}
			\label{Normal_Derivative}
			\boldsymbol{\Phi}_{n}(\boldsymbol{X}_N)=\boldsymbol{\Phi}_{x}(\boldsymbol{X}_N)\odot \boldsymbol{n}_x(\boldsymbol{X}_N)+  \boldsymbol{\Phi}_{y}(\boldsymbol{X}_N)\odot \boldsymbol{n}_y(\boldsymbol{X}_N)
		\end{equation} where $\odot$ denote the entry by entry (Shur) product acting along the columns.
		
		Finally, assembling the problem in the form of \eqref{Blocks}, the full problem for the velocity approximation becomes:

		\begin{equation}
			\label{Blocks_V}
			\left(  \begin{array}{cc}
				\boldsymbol{A}_U  & \boldsymbol{B}_U \\
				\boldsymbol{B}_U^T  & \boldsymbol{0}
			\end{array}\right) \left(\begin{array}{c}
				\boldsymbol{w}_U  \\
				\bm{\lambda}_U 
			\end{array}\right)=\left(\begin{array}{c}
				\boldsymbol{b}_{U1}  \\
				\boldsymbol{b}_{U2}
			\end{array}\right)\,,
		\end{equation} where $\bm{\lambda}_U=(\lambda_\nabla,\lambda_D,\lambda_N)^T\in\mathbb{R}^{n_\lambda}$ is the vector of the Lagrange multipliers for all constraints, i.e. $n_\lambda=n_\nabla+2n_D+2n_N$, and:
		\begin{subequations}
			\label{Prob_U}
			\begin{gather}
				\boldsymbol{A}_U= 2\boldsymbol{\Phi}_U^T\boldsymbol{\Phi}_U+2\alpha_\nabla \boldsymbol{D}_{\nabla}^T\boldsymbol{D}_{\nabla} \in\mathbb{R}^{2 n_b\times 2 n_b}\label{a_u_v}\\
				\boldsymbol{B}_U=\left (\boldsymbol{D}^T_\nabla,\boldsymbol{D}^T,\boldsymbol{N}^T_{\boldsymbol{n}}\right) \in\mathbb{R}^{2n_b\times n_{\lambda}}\label{b_u_v}\\
				\boldsymbol{b}_{U1}=2\boldsymbol{\Phi}_U^T \boldsymbol{U}\in \mathbb{R}^{2n_b}\label{c_u_v}\\
				\boldsymbol{b}_{U2}=(\boldsymbol{0},\boldsymbol{c}_D\,,\boldsymbol{c}_N)^T\in\mathbb{R}^{n_\lambda },
			\end{gather}
		\end{subequations} and having used the short hand notation  $\boldsymbol{\Phi}$ for $\boldsymbol{\Phi}(\boldsymbol{X})$, $\boldsymbol{N}_{\boldsymbol{n}}$ for $\boldsymbol{N}_{\boldsymbol{n}}(\boldsymbol{X}_N)$, etc. It is worth highlighting that one has $\boldsymbol{D}_{\nabla}(\boldsymbol{X})$ in \eqref{a_u_v} but 
		$\boldsymbol{D}_{\nabla}(\boldsymbol{X}_{\nabla})$ in \eqref{b_u_v}. Moreover, Neumann-like conditions can be redundant or conflicting with the divergence-free constraints if the latter are applied on boundaries. One should not have common points between the sets represented by $\boldsymbol{X}_N$ and $\boldsymbol{X}_\nabla$. However, it is possible to have both Dirichlet and divergence-free conditions on the same points. In this case, the common $n_{\lambda}$ points will be included in both $\boldsymbol{X}_D$ and $\boldsymbol{X}_{\nabla}$ and thus in the counting of both conditions (i.e to the counting $n_{\nabla}$ and $n_D$).
		
		\subsection{Computing Pressure Fields}\label{sec:2p3}
		
		We now consider the pressure computation by integrating the Poisson equation for incompressible and stationary flows. This equation is derived by taking the divergence of the momentum equation and reads
		
		\begin{equation}
			\label{Poisson}
			\Delta p=-\rho \nabla \cdot \left(\boldsymbol{U}\cdot \nabla \boldsymbol U\right)\,,
		\end{equation} where $\Delta p$ is the Laplacian of the pressure field $p$, $\rho$ is the fluid density, $\boldsymbol{U}$ is the velocity field and $\nabla \boldsymbol U$ is the velocity gradient tensor. The divergence-free assumption makes the equation independent from the viscous and unsteady terms; these nevertheless play a role at the boundary condition.
		
		For unsteady flows, the material derivative of the velocity field is required to compute instantaneous pressure fields (entering through the Neumann boundary conditions) while Reynolds stresses could be included to compute average pressure fields from turbulent flows \citep{van2013piv,gurka1999computation,van2019pressure}. None of these generalization poses additional difficulties to the proposed meshless integration and hence their implementation is postponed to future work. 
		
		We here illustrate the RBF integration of \eqref{Poisson} and its formulation in the framework of Sec. \ref{sec:2p1} in the case of a 2D flow. The generalization to the 3D case is reported in Appendix \ref{AppendixB}. In 2D, with the velocity field denoted as $\boldsymbol{U}=(u,v)$, the right hand side of \eqref{Poisson} reads:
		
		\begin{equation}
			\label{eq21}
			\nabla \cdot \left(\boldsymbol{U}\cdot \nabla \boldsymbol U\right)=(\partial_x u)^2+2\partial_x v \,\partial_y u+(\partial_y v)^2\,.
		\end{equation} 
		
		Once the RBF approximation of the velocity field is computed as described in Sec.\ref{sec:2}, the evaluation of \eqref{eq21} on a set of $\boldsymbol{X}\in\mathbb{R}^{n_p\times 2}$ points can be approximated as 
		
		\begin{equation}
			\label{source}
			\nabla \cdot \left(\boldsymbol{U}(\boldsymbol{X})\cdot \nabla \boldsymbol U(\boldsymbol{X})\right)\approx \left( \boldsymbol{\Phi}_x\boldsymbol{w}_u\right)^2+
			\boldsymbol{\Phi}_y\boldsymbol{w}_u\odot\boldsymbol{\Phi}_x\boldsymbol{w}_v+\left( \boldsymbol{\Phi}_y\boldsymbol{w}_v\right)^2\,,
		\end{equation} where the squares are to be computed entry by entry. We compute the source term \eqref{source} on the available $n_p$ points and store the results into a column vector $\boldsymbol{s}\in\mathbb{R}^{n_p}$.
		
		Writing the pressure field at every location $\boldsymbol{X}$ as in \eqref{approx}:
		
		\begin{equation}
			\boldsymbol{p}\approx \tilde{\boldsymbol{p}}=\boldsymbol{\Phi}(\boldsymbol{X})\boldsymbol{w}_p\,    
		\end{equation} with $\boldsymbol{w}_p\in\mathbb{R}^{n_b}$ the weights for the pressure, the approximation of the Laplace operator is:
		\begin{equation}
			\label{Laplace_RBF}
			\Delta \boldsymbol{p}\approx \boldsymbol{L}(\boldsymbol{X}) \boldsymbol{w}_p\,,
		\end{equation} where $\boldsymbol{L}(\boldsymbol{X})=[\Delta \gamma_1(\boldsymbol{X}),\dots \Delta \gamma_{n_\gamma}(\boldsymbol{X}),\Delta \varphi_{1}(\boldsymbol{X})\dots \Delta \varphi_{n_c}(\boldsymbol{X})] \in\mathbb{R}^{n_p\times n_b}$ is the matrix collecting the Laplacian of every element of the basis functions along its columns. For the Gaussian RBFs, at every point $\mathbf{x}_i\in\mathbf{X}$ this reads
		
		\begin{equation}
			\label{Laplac_PHI}
			\Delta\varphi_k(\boldsymbol{x}_i)=4c^2_k\left[(\boldsymbol{x}_i-\boldsymbol{x}_k)^2-1\right]\varphi_k(\boldsymbol{X})\,,
		\end{equation} and one recovers $\Delta\varphi_k(\boldsymbol{X})\in \mathbb{R}^{n_p}$ when \eqref{Laplac_PHI} is applied to all the set of points. 
		
		The RBF approximation of \eqref{Poisson} is thus $\boldsymbol{L}(\boldsymbol{X}) \boldsymbol{w}_p=\boldsymbol{s}$, and the cost function to minimize is:
		
		\begin{equation}
			\label{cost_P}
			J(\boldsymbol{w}_p)=||\boldsymbol{L}(\boldsymbol{X}) \boldsymbol{w}_p-\boldsymbol{s}||^2_2\,.
		\end{equation}
		
		Among the constraints for the pressure integration, we consider Dirichlet and Neumann conditions. Using the same notation as in the previous section, these read:
		
		\begin{subequations}
			\label{Conds}
			\begin{gather}
				\label{Cond_A}
				\boldsymbol{\Phi}(\boldsymbol{X}_D)\boldsymbol{w}_p=\boldsymbol{c}_D\\
				\label{Cond_B}
				\boldsymbol{\Phi}_{\boldsymbol{n}}(\boldsymbol{X}_N)\boldsymbol{w}_p=\boldsymbol{c}_N\,.
			\end{gather}
		\end{subequations}
		
		In the pressure integration problem for a steady flow, the pressure gradient required to impose Neumann conditions ($\boldsymbol{c}_N$ in \eqref{Cond_B}) is computed from Navier-Stokes equation \citep{gurka1999computation}:
		
		\begin{equation}
			\label{Projection_P}
			\nabla p\cdot \boldsymbol{n}=\partial_{n}p= \left(-\boldsymbol{U}\cdot \nabla \boldsymbol{U}+\mu\Delta\boldsymbol{U}+\rho \boldsymbol{g}\right)\cdot\boldsymbol{n}
		\end{equation} where $\boldsymbol{n}$ is the vector normal to the surface at hand, $\mu$ is the dynamic viscosity and $\boldsymbol{g}$ is the gravity acceleration. This equation can be easily computed from the vector field approximation and then set in form of \eqref{Conds}. The reader is referred to \cite{Gresho1987} for a discussion on the relevant boundary conditions for the pressure computation in an incompressibe flow.
		
		It is worth noticing that the points used to integrate the pressure field in \eqref{Laplace_RBF} and the points selected to impose the boundary conditions in \eqref{Conds} need not be the same as the ones chosen for the velocity field approximation. Nevertheless, for the first implementation proposed in this work, the same points and the same bases are used.
		
		To conclude, the pressure integration problem in the template from Sec. \ref{sec:2p1} reads:
		
		\begin{equation}
			\label{Blocks_P}
			\left(  \begin{array}{cc}
				\boldsymbol{A}_P  & \boldsymbol{B}_P \\
				\boldsymbol{B}_P^T  & \boldsymbol{0}
			\end{array}\right) \left(\begin{array}{c}
				\boldsymbol{w}_p  \\
				\bm{\lambda}_p 
			\end{array}\right)=\left(\begin{array}{c}
				\boldsymbol{b}_{P1}  \\
				\boldsymbol{b}_{P2}
			\end{array}\right)\,,
		\end{equation} where $\bm{\lambda}_p=(\lambda_D,\lambda_N)^T\in\mathbb{R}^{n_\lambda}$ is the vector of the Lagrange multipliers for all constraints, i.e. $n_\lambda=n_D+n_N$ and:
		
		\begin{subequations}
			\label{Prob_P}
			\begin{gather}
				\boldsymbol{A}_P= 2\boldsymbol{L}^T\boldsymbol{L} \in\mathbb{R}^{n_b\times n_b}\label{a_u}\\
				\boldsymbol{B}_P=\left (\boldsymbol{D}^T,\boldsymbol{N}^T_{\boldsymbol{n}}\right) \in\mathbb{R}^{n_b\times n_{\lambda}}\label{b_u}\\
				\boldsymbol{b}_{P1}=2\boldsymbol{L}^T \boldsymbol{s}\in \mathbb{R}^{n_b}\label{c}\\
				\boldsymbol{b}_{P2}=(\boldsymbol{0},\boldsymbol{c}_D\,,\boldsymbol{c}_N)^T\mathbb{R}^{n_\lambda }.
			\end{gather}
		\end{subequations} 
		
		The notation from the previous sections is used.
		
		\subsection{Clustering to Collocate}\label{sec:2p4}
		
		The optimal choice of the collocation points $\boldsymbol{x}^*_k$ and shape parameters $c_k$ has been investigated by several authors (see \cite{Hardy1971,Franke1982,Kansa1992,Sarra2017,Rippa1999}) but no universally accepted method is available.
		
		It is rather intuitive that RBFs should well cover the domain of interest and that large overlapping between basis elements hurts the conditioning of the matrix $\boldsymbol{A}$. A common practice is thus that of randomly placing the RBFs in the domain (see \cite{Kansa1992,Sarra2017}) and selecting the shape factor proportionally to the average distance to the $k$ nearest points (see \cite{Hardy1971,Franke1982}). A more complex formulation, proposed by \cite{Rippa1999} and extended by \cite{Karri2009}, consists in minimizing a cost function $J(\boldsymbol{x}^*_k,c_k)$ that measures the accuracy of the RBF approximation. Although this leads to a nonlinear optimization problem, the availability of analytic gradients $\partial_{\boldsymbol{x}_k} J$ and $\partial_{c_k} J$ allows for the effective implementation of powerful gradient-based optimizers. Nevertheless, this optimization requires considerable computational resources and time for large problems (e.g., in 3D tracking velocimetry).
		
		In this work, we propose a much simpler and computationally cheaper approach. The main idea is to ensure that all RBFs are ``supported" by approximately the same number of data points. In other words, defining as $\Omega_k$ the area within which a RBF is $\varphi_k(\boldsymbol{x}_i|x^*_k,c_k)>\varepsilon_t$, with $\varepsilon_t$ a user-defined threshold, we compute $x^*_k$ and $c_k$ such that an average of $n_K$ particles fall inside $\Omega_k$. This problem can be solved using clustering techniques \citep{Bishop2006,Nielsen2016}, which we implement using the fast mini-batch version of the K-means algorithm by \cite{Sculley2010}. 
		
		Briefly, the K-means clustering aims at partitioning a set of $n_p$ vectors $\boldsymbol{x}_i$ (here the coordinates at which the velocity vectors are available) into $K$ clusters $\mathcal{C}_k$ with $k=1,\dots, K$. Each cluster contains $|\mathcal{C}_k|$ particles and is such that $\sum_{k}|\mathcal{C}_k|=n_p$ and $\mathcal{C}_i\cap \mathcal{C}_j=0 \,\forall i\neq j$. Each cluster identify vectors with certain degree of similarity, defined in terms of some distance metrics. In this work, we consider the Euclidean distance. Let $\boldsymbol{\mu}_k$ denote the centroid of cluster $\mathcal{C}_k$, the clustering problem aims at minimizing the intra-cluster variance 
		
		\begin{equation}
			\label{J_Cluster}
			J(\boldsymbol{\mu}_1,\dots \boldsymbol{\mu}_N)=\sum^{K}_{k=1}\,\,\sum_{\boldsymbol{x}_i\in\mathcal{C}_k} ||\boldsymbol{x}_i-\boldsymbol{\mu}_k||^2\quad \mbox{with}\quad \boldsymbol{\mu}_k=\frac{1}{|\mathcal{C}_k|}\sum_{j\in\mathcal{C}_k} \boldsymbol{x}_j\,.
		\end{equation}
		
		The classic iterative algorithm by \cite{Lloyd1982} solves this minimization starting with a random set of cluster centroids. At each iteration, the cluster assignment is followed by a new computation of centroids until convergence. Because the function \eqref{J_Cluster} is usually not convex, the algorithm is repeated a number of times, and various initialization techniques have been proposed to escape local minima (see \cite{Solis-Oba2006}). The mini-batch version of the K-means algorithm implemented in this work uses stochastic gradient descent and allows for considerable saving in the computational cost \citep{Sculley2010}.

\begin{figure}[ht]
			\centering
			\includegraphics[width=0.3\columnwidth]{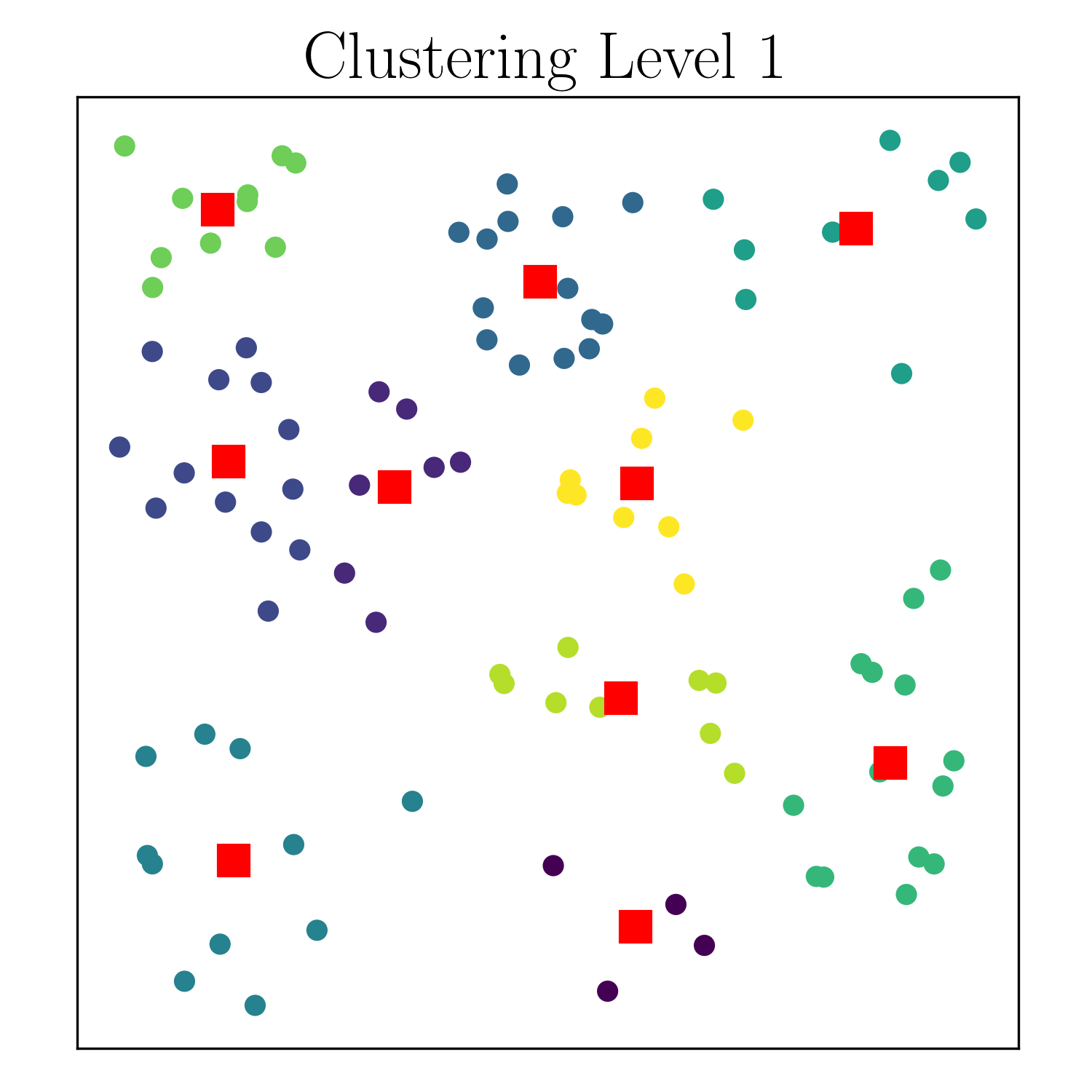}
			\includegraphics[width=0.3\columnwidth]{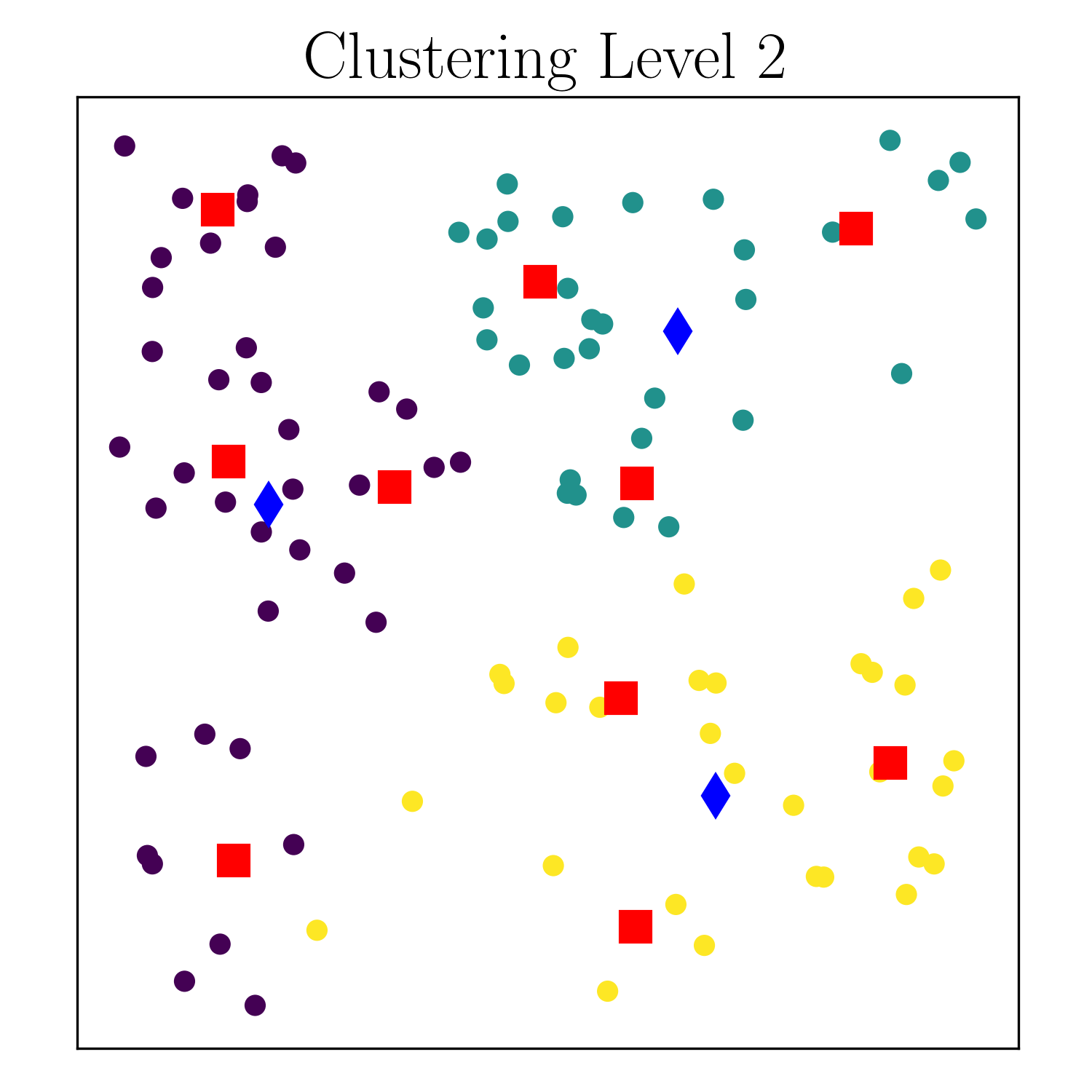}
			\caption{Illustration of the proposed hierarchical clustering to collocate RBFs and to compute their shape parameter. The ilustration takes $n_p=100$ particles with $\boldsymbol{n}_K=[10,30]$. The red square markers are the centroids produced in the first level; the blue diamond markers are the ones at the second level. In both figures, the particles are coloured based on the cluster they belong to. }
			\label{fig:clusteringsketch}
		\end{figure}

		The proposed collocation technique is essentially an agglomerative clustering approach \citep{Nielsen2016} using the K-means algorithm iteratively and at multiple levels. The procedure is illustrated in Figure \ref{fig:clusteringsketch} for the RBF collocation in $n_p=100$ points using two levels. At the first level (figure on the left), we set $n_k=10$, resulting in $K=10$ clusters. The centroids are shown with square markers. At the second level (figure on the right), the clustering is repeated on the centroids of the previous level. If one expect approximately $n_k=3$ centroids per cluster, then $K=\mbox{floor}(100/(3\cdot 10))=3$ clusters are used (with $\mbox{floor}$ the rounding down operator). The centroids belonging to the second cluster are shown with blue diamond markers.
		
		In general, denoting as $n_l$ the number of levels and with $K_j$ the number of clusters at the $j$-th level, with $j=1,\dots,n_{l}$, one has 
		
		\begin{equation}
			K_j=\frac{n_p}{\prod^j_{n=1} n_{K}^{(n)}}\,,
		\end{equation} where $n_{K}^{(n)}$ is the expected number of points per cluster in the $n$-th level. In a compact notation, we use the vector $\boldsymbol{n}_K=[n_{K}^{(1)},\dots, n_{K}^{(n_l)}]\in \mathbb{R}^{n_l}$.

		All the centroids are taken as collocation points regardless of their level, i.e. $\boldsymbol{x}^*_k=\boldsymbol{\mu}_k$. The shape factors are computed at each level in such a way that $\varphi_k(\boldsymbol{\mu}^{(j)}_i|\boldsymbol{\mu}^{(j)}_k,c_k)=\varepsilon_t$, where $\boldsymbol{\mu}^{(j)}_k$ is the collocation point of interest at level $j$ and $\boldsymbol{\mu}^{(j)}_i$ is the nearest collocation point at the same level. Therefore, one has $c^{(j)}_k=\sqrt{-\ln(\varepsilon_t)}/\min\left(|\boldsymbol{\mu}^{(j)}_i-\boldsymbol{\mu}^{(j)}_k|\right)$ with $k\neq j$. 
		

		\begin{figure}[ht]
			\centering
			\includegraphics[width=0.7\columnwidth]{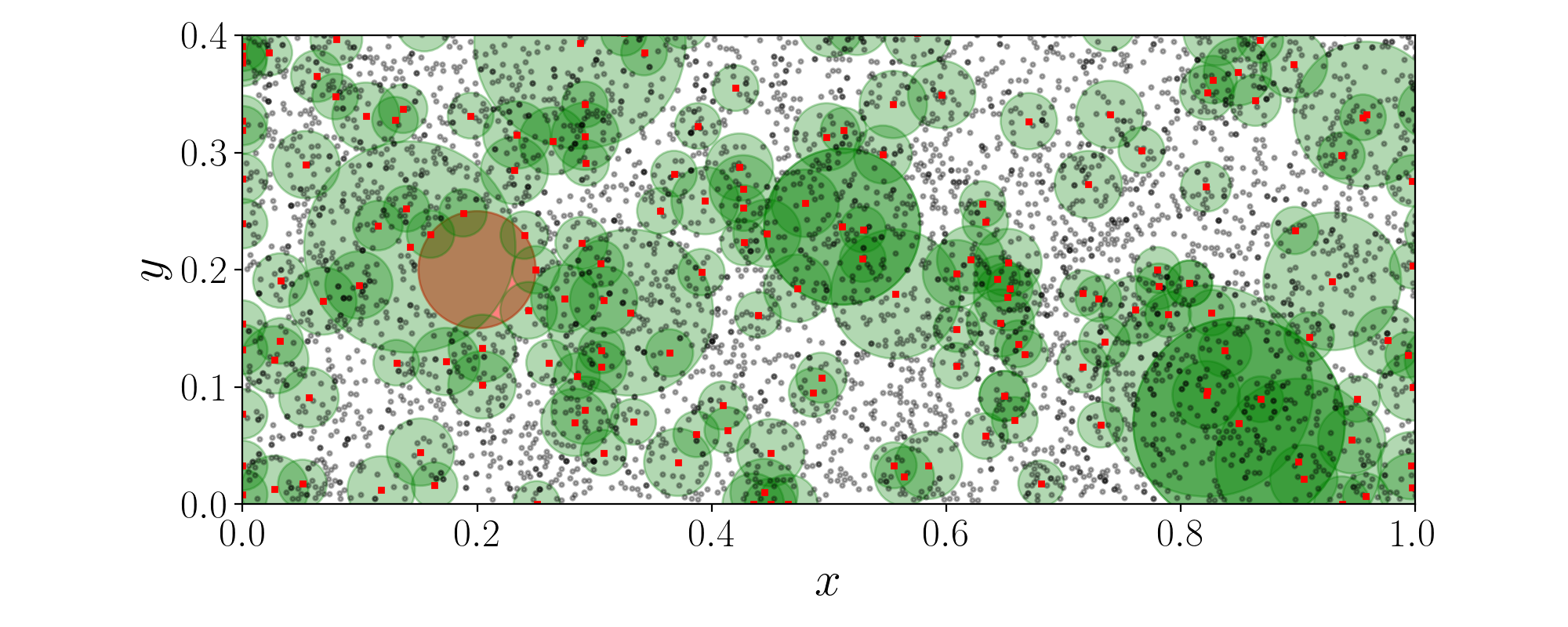}\\
			\includegraphics[width=0.6\columnwidth]{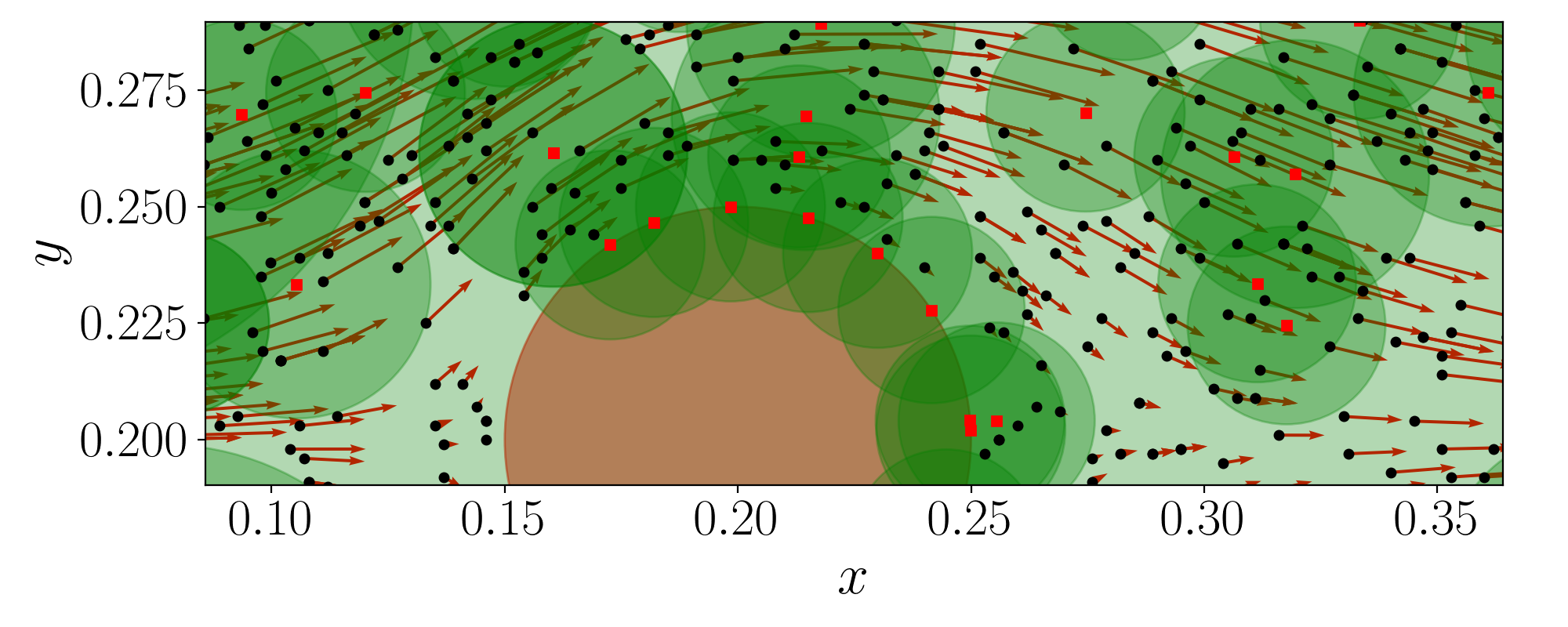}
			\caption{Test case 2. Results from the clustering technique for collocation and shape factor selection; the cylinder is identified by red color. Top: $300$ randomly selected collocation points (red square markers) together with their region $\Omega_k$ such that $\varphi_k>0.85$. The black markers identify the seeding particles. Bottom: a zoomed view with quiver plot of the available data for the velocity field.}
			\label{Fig_Clust_1}
		\end{figure}
		
		Figure \ref{Fig_Clust_1} illustrates the result of the clustering for the second test case analyzed in this work (see Sec. \ref{sec:3p3}). The figure on the top shows the result of the clustering considering $n_p=18755$ points and $\boldsymbol{n}_K=[4,8]$, i.e. having two sets with
		$K_1=n_p/8\approx 2417$ and $K_2=n_p/(8\cdot 4)\approx 604$ clusters. The top figures show $300$ randomly selected RBFs, with the green circles denoting the regions $\Omega_k$ such that $\varphi_k>0.85$. The bottom figure shows a closed view near the cylinder, together with the region $\Omega_k$ of about 30 randomly chosen radial basis functions. The large variety of shape factors is evident.
		
		
		Some conservative conditions are added to avoid overly large $c_k$ (resulting in too small RBFs), as these make the problem ill-posed \citep{Cheng_2003}. Firstly, if a cluster contains only one point, its shape factor is set to the smallest of the same level, i.e. $min(c^{(j)}_k)$. The same limiting factor is used if the points within a RBF are less than $n_{K}^{(j)}$. Secondly, an upper limit is set to the value of $c_k$. This is provided by the user, and all shape factors above this value are set equal to this value. Finally, because constraints remove degrees of freedom to the regression problem, a basis function is given to each constraint (i.e. every constraint point is also a collocation point). The same criteria limiting the shape factors are used for these basis elements, but computing the $c_k$ from the nearest neighbour regardless of the clustering level. Figure \ref{Fig_Clust_1} illustrates several basis functions located on the cylinder walls.

		\subsection{Numerical Methods}\label{sec:2p5}
		
		The previous sections have shown that both the velocity regression and the pressure integration require solving a linear system of the form in \eqref{Blocks}. This kind of system arises in quadratic problems with linear equality constraints, and it is known as the Karush-Kuhn-Tucker (KKT) system. The reader is referred to \cite{JorgeNocedal2006} for a vast literature on both direct and iterative solution methods.
		
		It is possible to show that the KKT system admits a unique solution if  $\boldsymbol{B}\in\mathbb{R}^{n_b\times n_{\lambda}}$ is full rank and the reduced Hessian $\boldsymbol{Z}^T\boldsymbol{A}\boldsymbol{Z}$, with $\boldsymbol{Z}\in\mathbb{R}^{n_b\times(n_b-n_\lambda)}$ a basis for the nullspace of $\boldsymbol{B}^T$ (i.e. $\boldsymbol{B}^T\boldsymbol{Z}=\boldsymbol{0}$), is positive definite. This condition is nevertheless hardly met in practice, notably because it is difficult to avoid some redundancy in the constraints. A natural solution is thus the use of SVD or a rank revealing QR factorization of $\boldsymbol{B}$ to remove redundant constraints, but this is generally expensive in 3D problems where $\boldsymbol{B}$ is large. 
		
		Therefore, the proposed approach relies on a direct heuristic method that has no guarantees of uniqueness but has so far provided the best compromise between solution accuracy, robustness, and memory requirements. Assuming that the regularization parameter ensures the positive definiteness of $\boldsymbol{A}$, from the first equation in \eqref{Blocks} one gets 
		
		\begin{equation}
			\label{w_equation}
			\boldsymbol{w}=\boldsymbol{A}^{-1}\bigl(\boldsymbol{b_1}-\boldsymbol{B}\boldsymbol{\lambda}\bigr)\,.
		\end{equation}
		
		Inserting this in the second equation, rearranging terms and defining $\boldsymbol{M}=\boldsymbol{B}^T\boldsymbol{A}^{-1}\boldsymbol{B}$, $\boldsymbol{b}^*_2=\boldsymbol{B}^T\boldsymbol{A}^{-1}\boldsymbol{b}_1-\boldsymbol{b}_2$ and $\boldsymbol{b}^*_1=\boldsymbol{b}_1-\boldsymbol{B}\boldsymbol{\lambda}$, the linear system in \eqref{Blocks} is decoupled in the two lower-dimensional systems
		
		\begin{subequations}
			\label{Blocks2}
			\begin{gather}
				\boldsymbol{M}\bm{\lambda} = \boldsymbol{b}^*_2\\
				\boldsymbol{A} \boldsymbol{w}  = \boldsymbol{b}^*_1
			\end{gather}
		\end{subequations} 
		to be solved sequentially. The solution method now hinges on the symmetry and the positive definitiveness of the matrices $\boldsymbol{A}\in\mathbb{R}^{n_b\times n_b}$ and $\boldsymbol{M}\in\mathbb{R}^{n_\lambda\times n_\lambda}$, ensured by an appropriate regularization.
		
		Let $\alpha_A$ and $\alpha_M$ denote the regularization parameters for $\boldsymbol{A}$ and $\boldsymbol{M}$, such that the regularization of these matrices reads $\boldsymbol{A}\leftarrow\boldsymbol{A}+\alpha_A\boldsymbol{I}$ and $\boldsymbol{M}\leftarrow\boldsymbol{M}+\alpha_M\boldsymbol{I}$, with $\boldsymbol{I}$ the identity matrix of appropriate size. Ideally, one would set these parameters as the smallest positive values guaranteeing positive definiteness of the corresponding matrix. However, since estimating this parameters would be a too expensive task for large datasets, we rely on a cheap estimation of the condition number using the infinity norm. For the matrix $\boldsymbol{A}$, for instance, letting $\kappa(\boldsymbol{A},2)$ denote the spectral condition number, $\lambda_m(\boldsymbol{A})$ the smallest eigenvalue of $\boldsymbol{A}$,
		$\lambda_M(\boldsymbol{A})$ is the largest eigenvalue of $\boldsymbol{A}$,
		and assuming $\alpha_A\approx \lambda_m(\boldsymbol{A})$, one has (see, for example \cite{Rannacher2018})
		
		\begin{equation}
			\label{cond}
			\kappa(\boldsymbol{A},2)=\frac{\lambda_M(\boldsymbol{A})}{\lambda_m(\boldsymbol{A})}\approx\frac{\lambda_M(\boldsymbol{A})}{\alpha_A}\leq \frac{\sqrt{n_b}||\boldsymbol{A}||_{\infty}}{\alpha_A}\rightarrow \alpha_A \lesssim  \frac{\sqrt{n_b}||\boldsymbol{A}||_{\infty}}{\kappa(\boldsymbol{A},2)}=\mbox{tol}\sqrt{n_b}||\boldsymbol{A}||_{\infty}\,\,
		\end{equation} where $\mbox{tol}$ is a user defined estimate of the tolerated $1/\kappa(\boldsymbol{A},2)$ (usually $\mbox{tol}=10^{-12}$ when working in float64 and $\mbox{tol}=10^{-5}$ when working in float32). The same procedure is used for estimating $\alpha_M$ to regularize the matrix $\boldsymbol{M}$ in \eqref{Blocks2}.
		
		The successful regularization enables solving both systems in \eqref{Blocks2} using the Cholesky decomposition. Let $\boldsymbol{A}=\boldsymbol{L}_A\boldsymbol{L}_A^T$ be the Cholesky decomposition of $\boldsymbol{A}$.
		Then, letting $\boldsymbol{R}$ denote the product $\boldsymbol{R}=\boldsymbol{L}_A^{-1}\boldsymbol{B}$ we have
		
		\begin{equation}
			\label{Chol}
			\boldsymbol{M}=\boldsymbol{B}^T \boldsymbol{A}^{-1} \boldsymbol{B}=\boldsymbol{B}^T (\boldsymbol{L}_A^{-1})^T\boldsymbol{L}_A^{-1} \boldsymbol{B} =\boldsymbol{R}^T \boldsymbol{R}\,,
		\end{equation}
		We remark that the inversion of the lower triangular matrix $\boldsymbol{L}_A$ is only formal as $\boldsymbol{R}$ is computed by solving triangular systems, each one requiring $\simeq \frac12 n^2_b$ operations.
		
		While the Cholesky decomposition in \eqref{Chol} is the most expensive operation (with an operation count of $1/6 n^3_b$) of the numerical approach, the computed factors serve three purposes. Indeed, besides allowing for the rapid computation of $\boldsymbol{M}=\boldsymbol{B}^T\boldsymbol{A}^{-1}\boldsymbol{B}$ following equation \eqref{Chol}, it can also be exploited in the computation of the right-hand-side $\boldsymbol{b}^*_2$:
		\begin{equation}\label{b2s}
			\boldsymbol{b}^*_2=\boldsymbol{B}^T\boldsymbol{A}^{-1}\boldsymbol{b}_1-\boldsymbol{b}_2=\boldsymbol{B}^T (\boldsymbol{L}_A^{-1})^T\boldsymbol{L}_A^{-1} \boldsymbol{b}_1-\boldsymbol{b}_2=\boldsymbol{R}^T \boldsymbol{L}_A^{-1} \boldsymbol{b}_1-\boldsymbol{b}_2
		\end{equation}
		thus only requiring one triangular solve and a matrix-vector multiplication.
		Furthermore, once the solution of the system $\boldsymbol{M}\boldsymbol{\lambda}=\boldsymbol{b}^*_2$ is found, the factors allow to compute the weights in \eqref{w_equation} cheaply, by solving again two triangular systems. The complete algorithm is summarized in the listing \ref{PIETRO_ALG}.
		\begin{algorithm}
			\begin{algorithmic}[1]
				\State Assembly $\boldsymbol{A}$, $\boldsymbol{B}$, $\boldsymbol{b_1}$ and $\boldsymbol{b_2}$ as in \eqref{Blocks}.
				\State Compute $||\boldsymbol{A}||_{\infty}$ and $\alpha_A$ from \eqref{cond}. Then regularize $\boldsymbol{A}\leftarrow \boldsymbol{A}+\alpha \boldsymbol{I} $
				\State Compute the Cholesky factorization $\boldsymbol{A}=\boldsymbol{L}_A\boldsymbol{L}_A^T$
				\State Compute  $\boldsymbol{R}$ solving a triangular system for each column of $\boldsymbol{B}$
				\State Compute $\boldsymbol{M}=\boldsymbol{R}^T\boldsymbol{R}$,  $||\boldsymbol{M}||_{\infty}$ and $\alpha_M$ as in \eqref{cond}. Then regularize $\boldsymbol{M}\leftarrow \boldsymbol{M}+\alpha_M \boldsymbol{I} $
				\State Compute the Cholesky decomposition $\boldsymbol{M}=\boldsymbol{L}_M\boldsymbol{L}_M^T$
				\State Compute the r.h.s $\boldsymbol{b}_2^*$ as in \eqref{b2s}
				\State Solve two triangular systems for $\boldsymbol{\lambda}=(\boldsymbol{L}_M^{-1})^T\boldsymbol{L}_M^{-1}\boldsymbol{b}_2^*$
				\State Solve two triangular systems for $\boldsymbol{w}=(\boldsymbol{L}_A^{-1})^T\boldsymbol{L}_A^{-1}\bigl(\boldsymbol{b_1}-\boldsymbol{B}\boldsymbol{\lambda}\bigr)$
				\caption{Direct Method for the systems in Eq. \eqref{Blocks2}}\label{PIETRO_ALG}\end{algorithmic}
		\end{algorithm}

		\section{Selected Test Cases}\label{sec:3}
		

		\subsection{Test case 1: a Gaussian Vortex in 2D}\label{sec:3p1}
		
		As a first test case we consider a Gaussian vortex field. This is a classic benchmark problem (see \cite{de2012instantaneous,Azijli2016,McClure2017}), characterized by smooth gradients in both velocity and pressure fields. The conditions are those in \cite{de2012instantaneous}, with a velocity field in polar coordinate ${U}=(u_r,u_\theta)$:
		
		\begin{equation}
			\label{vortex_EQ1}
			u_r=0 \quad \mbox{and}\quad u_\theta=\frac{\Gamma}{2\pi r}\Biggl( 1-e^{-r^2/c_\theta}\Biggr)\,,
		\end{equation} where $\Gamma$ is the circulation, $c_\theta=r^2_c/\gamma$, with $r_c$ the radial distance from the vortex center where the largest velocity is reached, and $\gamma=1.25643$. This is a Lamb-Oseen vortex with $4\nu t=c_{\theta}$.
		
\begin{figure}[ht]
			\centering
			\begin{subfigure}[\label{fig:High_D}]{
					\includegraphics[width=.3\textwidth ]{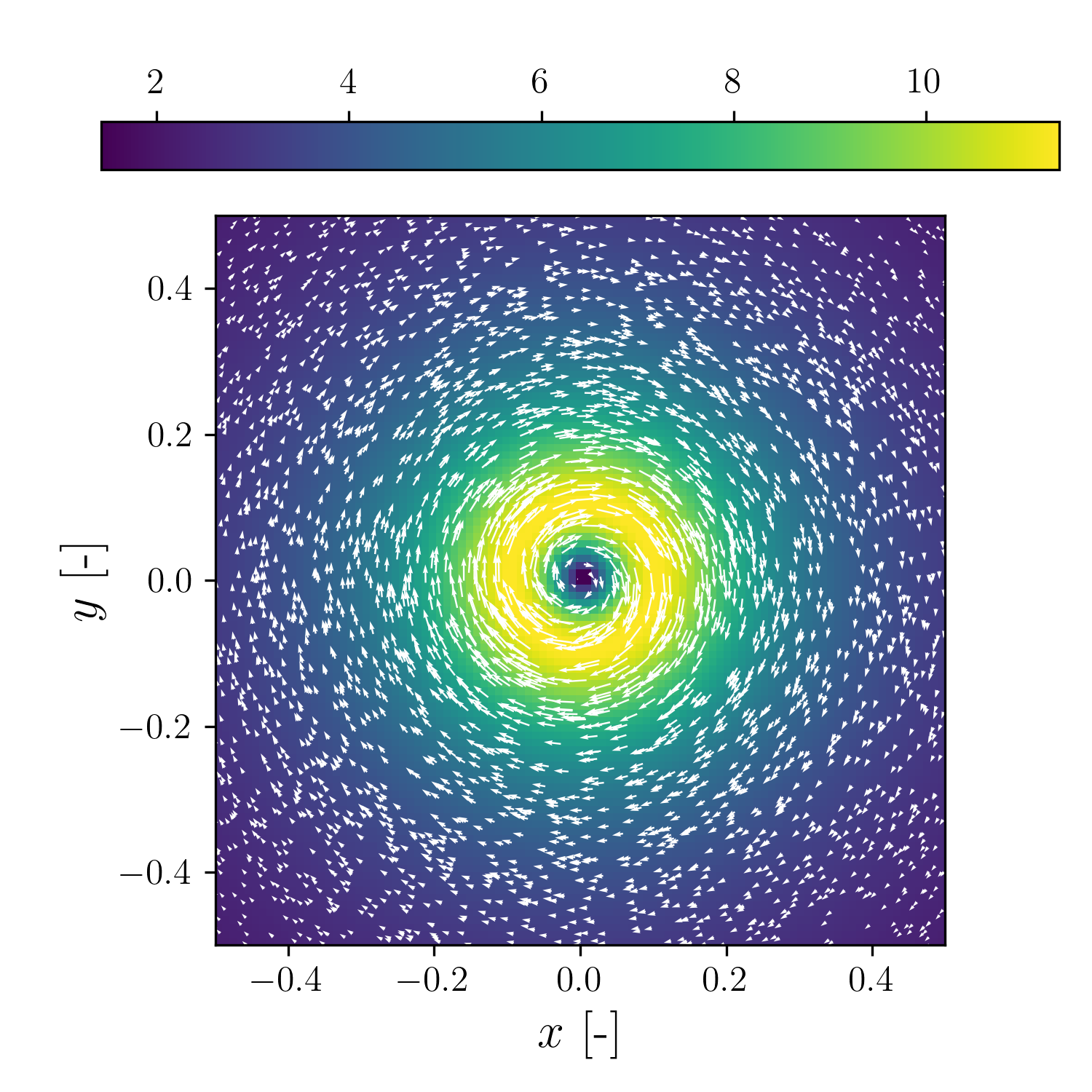}
				}
			\end{subfigure}
			\begin{subfigure}[\label{fig:Low_D}]{
					\includegraphics[width=.3\textwidth ]{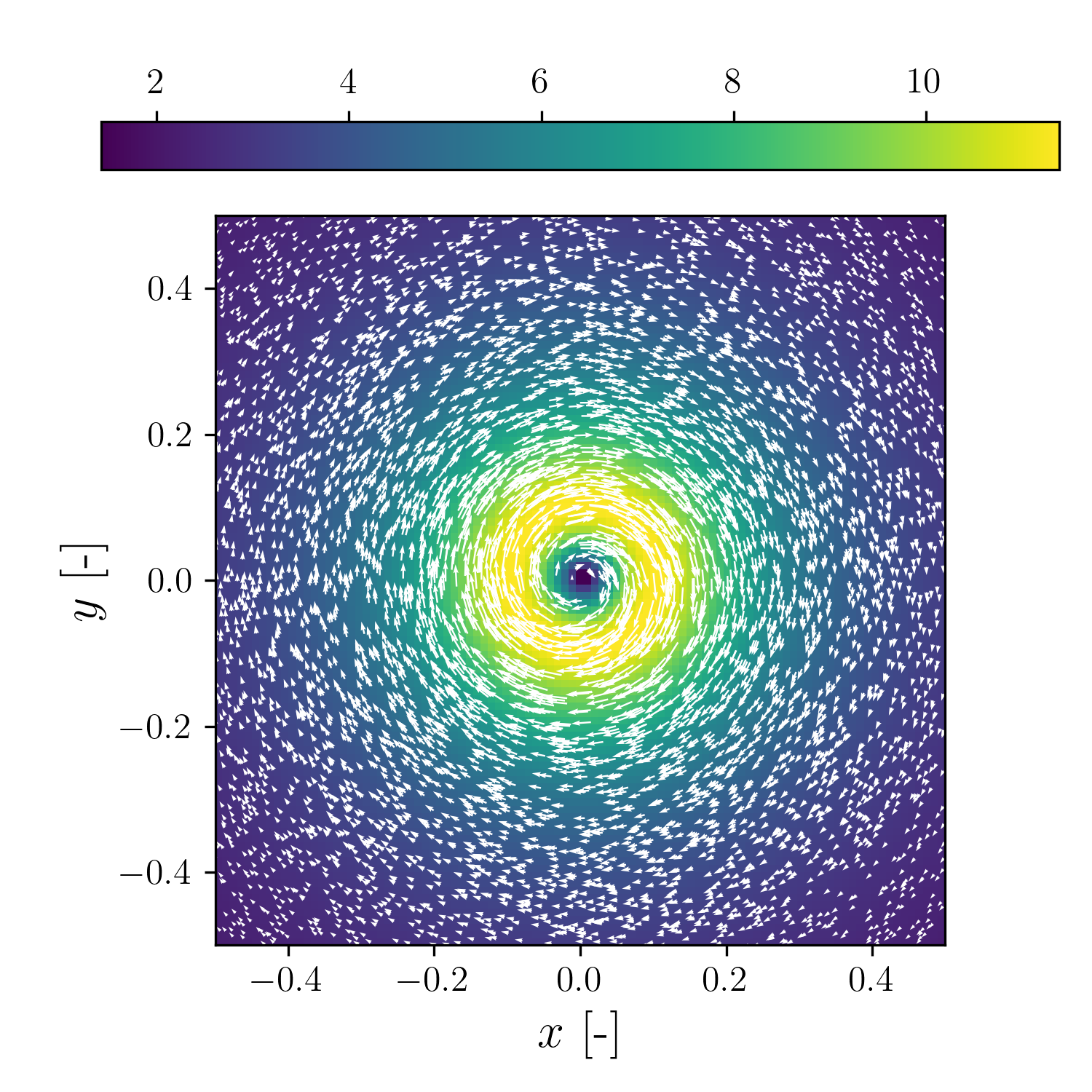}
				}
			\end{subfigure}
			\begin{subfigure}[\label{fig:raw_100_noise}]{
					\includegraphics[width=.31\textwidth ]{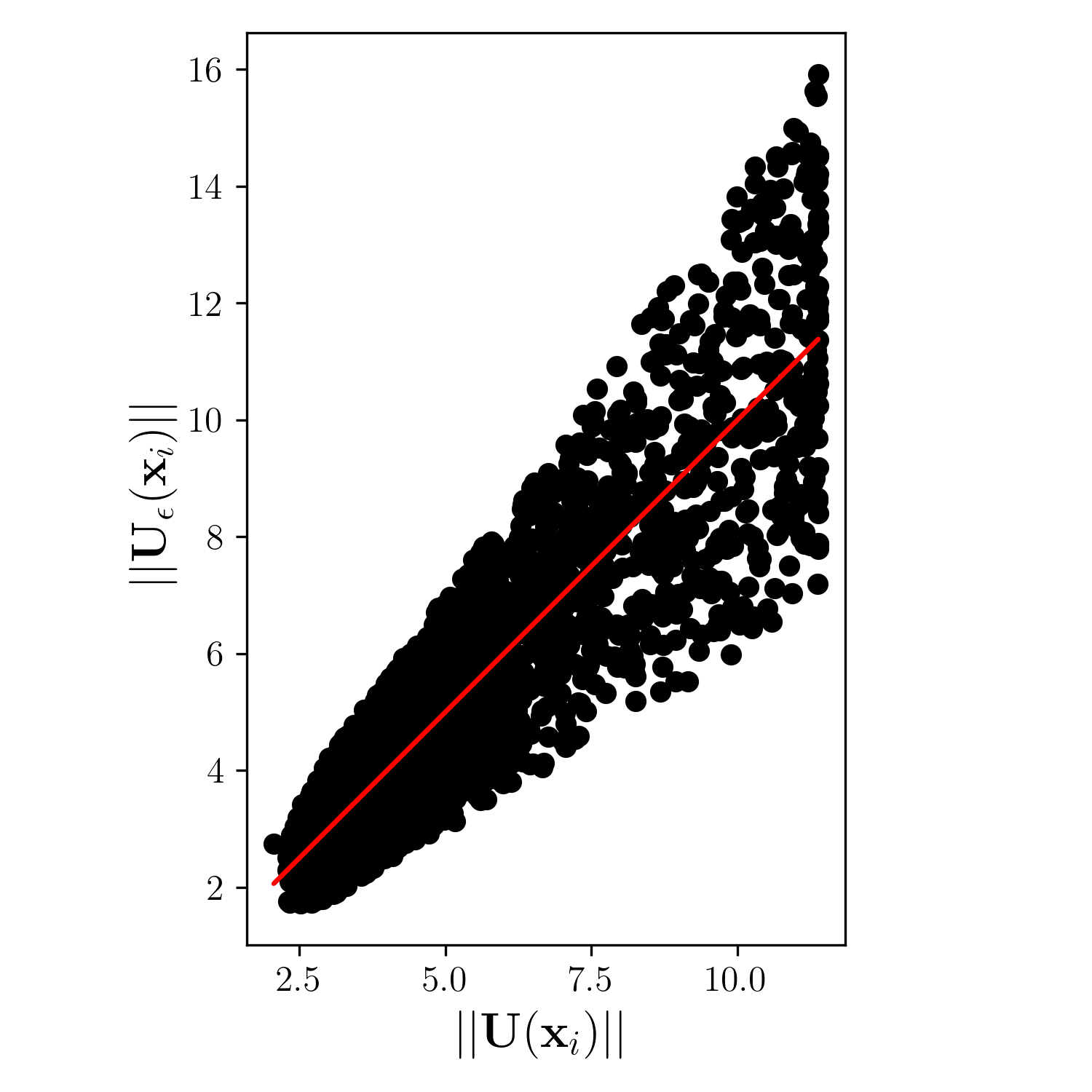}
				}
			\end{subfigure}
			\caption{Test case 1. Velocity fields for the lowest (a) and the highest (b) seeding densities used to test the pressure integration. The number of particles is $n_p=3145$ and $n_p=5242$, respectively; velocity is randomly sampled from \eqref{vortex_EQ1}. Figure (c):  for the highest density case, velocity distributions of the noisy field versus the original field, with the highest noise level tested ($p=0.4$).} 
			\label{fig:test_case_1}
		\end{figure}

		The domain of interest is $\boldsymbol{x}\in[-0.5,0.5]\times [-0.5,0.5]$ and is taken as dimensionless. Assuming that this is mapped onto a sensor with $1024\times 1024$ pixels allows for analyzing the impact of the seeding concentration in terms of the familiar source density $N_p$ (in particles per pixel, ppp). In this test case, we consider a seeding concentration in the range $N_p=[0.003,0.005]$, corresponding to a number of particles in the range $n_p=[3145, 5242]$. Particles are randomly distributed within the domain, sampled with a uniform distribution, and their velocity is taken according to \eqref{vortex_EQ1}.
		Figures \ref{fig:High_D} and \ref{fig:Low_D} show the cases with the smallest and the largest seeding density. The contour map in the background is the velocity magnitude (dimensionless).

		In addition to the impact of the seeding density, we consider the impact of random noise. This is introduced as $\boldsymbol{U}_{\varepsilon}(\boldsymbol{x}_i)=\boldsymbol{U}(\boldsymbol{x}_i)(1+q\,\boldsymbol{w}(\boldsymbol{x}_i))$, where $\boldsymbol{U}(\boldsymbol{x}_i)$ is the ideal velocity field from \eqref{vortex_EQ1}, $q$ is the noise level and $\boldsymbol{w}(\boldsymbol{x}_i)\in [-1,1] \times [-1,1]$ is a bi-dimensional field of uniformly distributed noise. For this specific test case, characterized by a smooth velocity field and no solid surfaces, the algorithm proved capable of handling errors as large as $q=0.4$. Figure \ref{fig:raw_100_noise} plots the distribution of noisy fields versus the expected field, showing the large impact of the noise for a case with $N_p=0.005$ and $q=0.4$. This noise level is clearly unrealistic, but it allows to showcase the robustness of the approach for problems with smooth velocity and pressure fields. The results are presented in section \ref{sec:4p1}.

		The Navier-Stokes equations in polar coordinate give $\partial_r p=\rho/r u^2_\theta$ and $\partial_\theta p=0$. The integration of the pressure gradient from $-\infty$ to $r$ gives 
		
		\begin{equation}
			\label{Pressure_vortex}
			p(r)=-\frac{1}{2}\rho u^2_\theta-\frac{\rho \Gamma^2}{4 \pi ^2 c_\theta}\Bigl[E_1\biggl(\frac{r^2}{c_\theta}\biggr)-E_1\biggl(\frac{2r^2}{c_\theta}\biggr)\Bigr]\,,
		\end{equation} where $E_1(x)$ is the exponential integral
		
		\begin{equation}
			E_1(x)=\int^{\infty}_{x}\frac{e^{-t}}{t} dt\,.
		\end{equation}
		
\begin{figure}[ht]
			\centering
			\includegraphics[width=0.45\columnwidth,trim={0.4cm 0.6cm 0.4cm 0},clip]{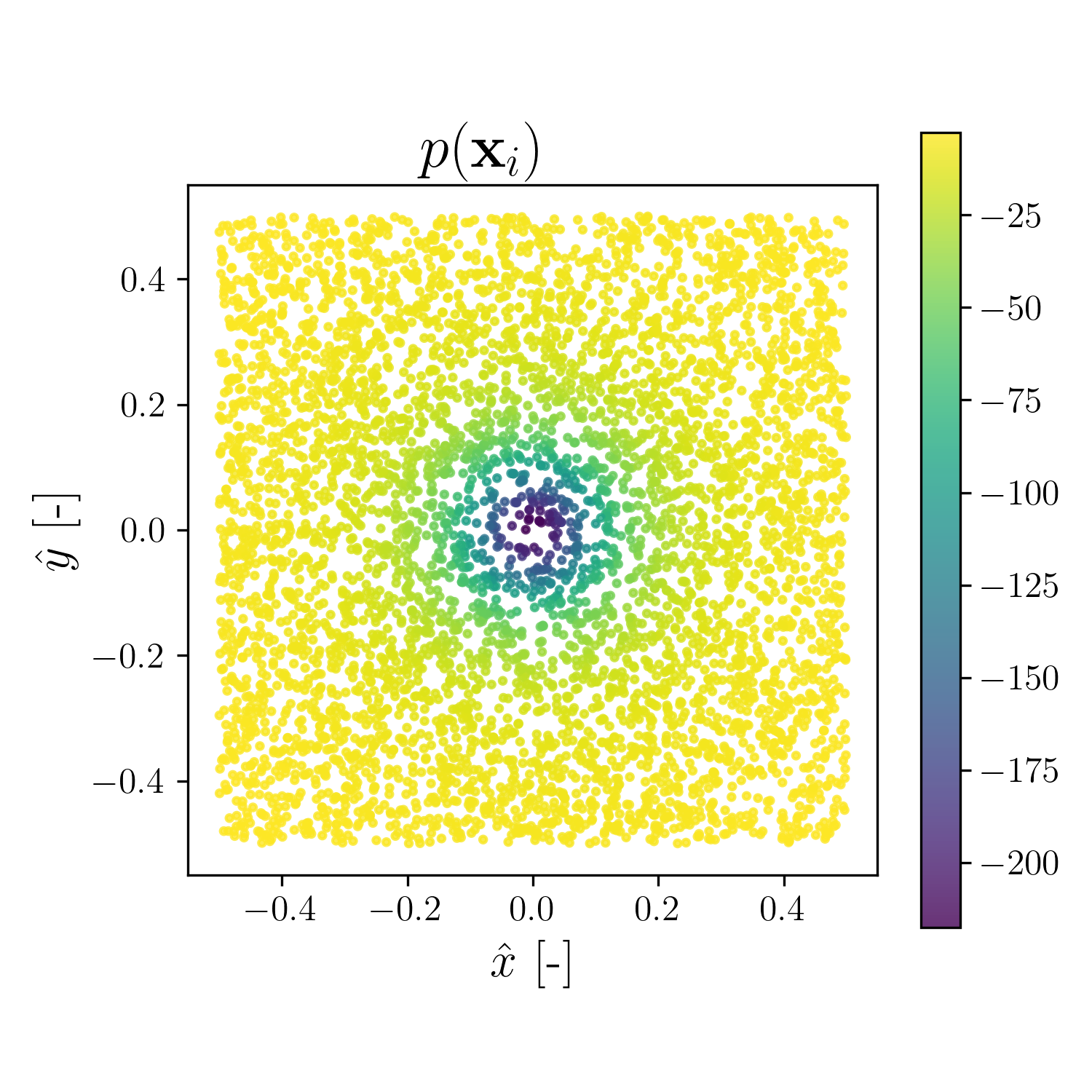}
			\vspace{-4mm}
			\caption{Test case 1. Pressure field from \eqref{Pressure_vortex}, made visible by coloring the markers of a scatter plot. The seeding density in the plot corresponds to $N_p=0.005$, i.e. $n_p=5242$ particles.}
			\label{Pressure_Vortex}
		\end{figure}

		The pressure field is shown in Figure \ref{Pressure_Vortex} over the $n_p=5242$ scattered points which corresponds to $N_p=0.005$, i.e. the highest seeding density considered for this example. The contour of the pressure field is made visible by coloring the markers, located at the particle positions, with the pressure values in equation \eqref{Pressure_Vortex}. Most of the particles are in a region of nearly uniform pressure and most of the pressure variation is located at approximately $r<0.2$.

		\subsection{Test case 2: a 2D Flow Past a Cylinder from CFD}\label{sec:3p2}
		
\begin{figure}[ht]
			\centering
			\includegraphics[width=0.8\columnwidth]{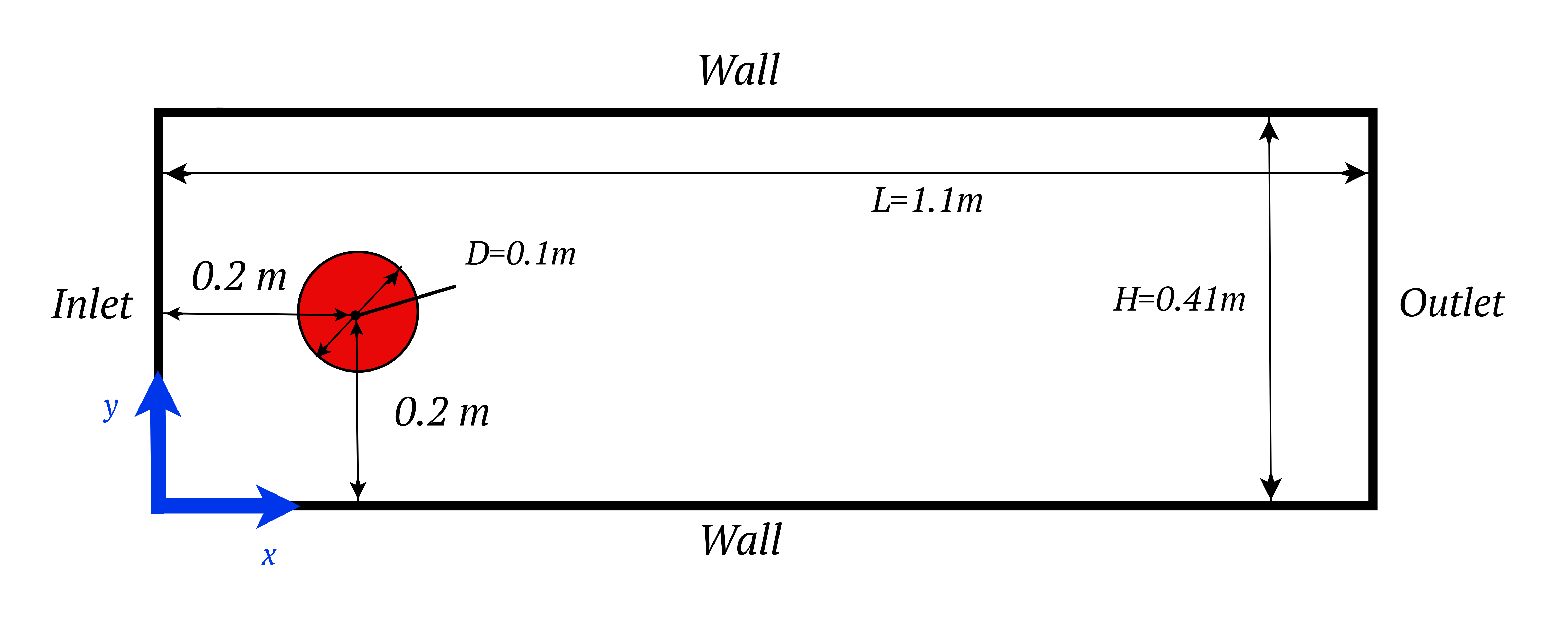}
			\vspace{-4mm}
			\caption{Test case 2. Inlet Schematic, recalling the relevant dimensions and boundary conditions. The dataset is taken from \cite{Rao2020}, who share it at \url{https://github.com/Raocp/PINN-laminar-flow}.}
			\label{Cylinder_set_UP}
		\end{figure}

\begin{figure}[ht]
			\centering
			\includegraphics[width=1\textwidth]{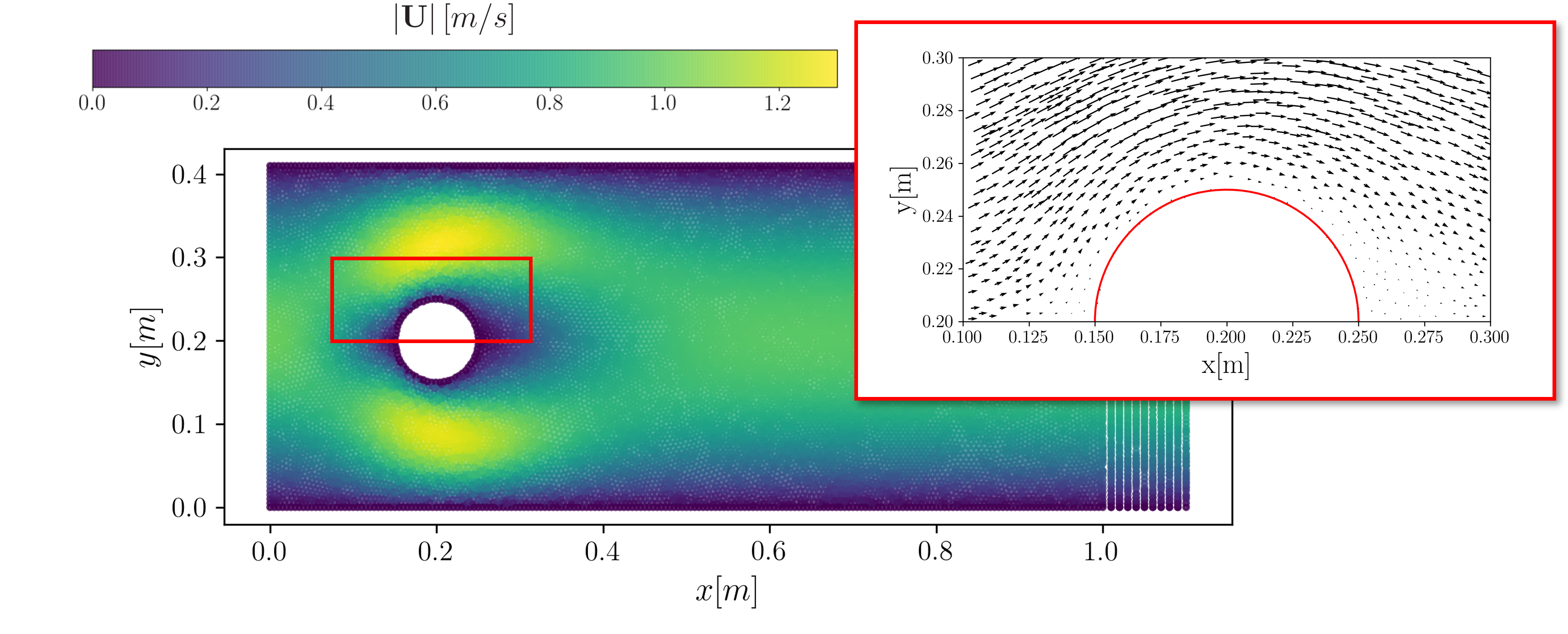}
			\vspace{-6mm}
			\caption{Test case 2. Overview of the flow field with $n_p=18755$. The velocity magnitude is plotted in all the sampling location in the form of a scatter plot; the high particle density makes the plot appear continuous. The zoom shows a quiver plot around the cylinder.}
			\label{Snapshot_Cylinder}
		\end{figure}

\begin{figure}[ht]
			\centering
			\includegraphics[width=0.95\textwidth]{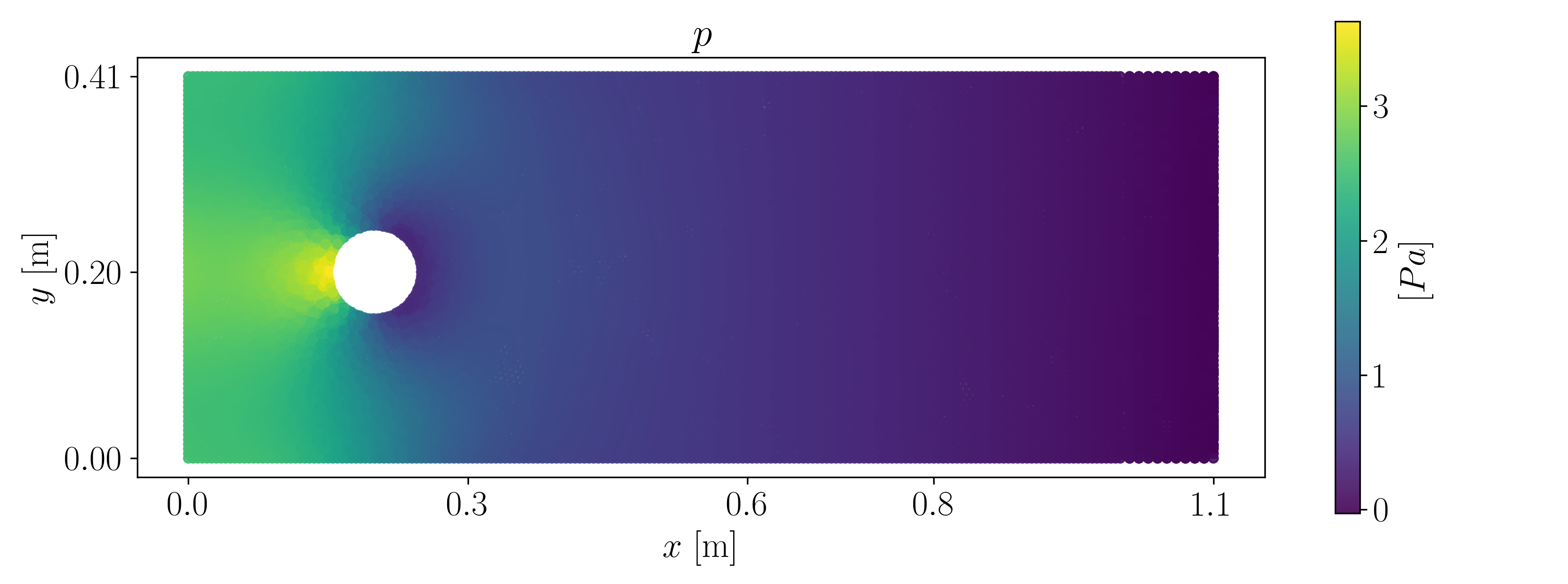}
			\vspace{-4mm}
			\caption{Test case 2. Overview of the pressure field with $n_p=18755$.}
			\label{Pressure_Cylinder}
		\end{figure}

		The second selected test case is a 2D problem featuring curved walls and demanding for boundary conditions in both the velocity regression and the pressure integration. This is the laminar and incompressible flow past a cylinder in a 2D channel. The flow configuration and the relevant dimensions and boundaries are shown in Figure \ref{Cylinder_set_UP}. \textcolor{black}{Figure \ref{Snapshot_Cylinder} shows a snapshot for the velocity field, with a zoomed view around the cylinder wall, while Figure \ref{Pressure_Cylinder} shows the associated pressure field.}
		
		This is a classic benchmark test case (see \cite{Schaefer1996,John2002}) presented in various tutorials (e.g., \cite{LangtangenLogg2017}). Nevertheless, in this work we use the dataset released by \cite{Rao2020}. This dataset was used to train a mixed-variable scheme for physics informed neural networks (PINNS), a valid alternative to the meshless approach proposed in this work and to which we compare our regression results. 
		
		We here recall the main parameters for this dataset and refer the reader to the original publication for more details. The velocity profile on the inlet is set as 
		
		\begin{equation}
			\label{Inlet}
			u(0,y)=4 \frac{U_M}{H^2} \Bigl (H-y\Bigr)y 
		\end{equation} with $U_M=1$m/s and $H=0.41$m.
		The fluid density is taken as $\rho=1$kg/m$^3$ and the dynamic viscosity is taken as $\mu=2\cdot 10^{-2}$kg/m$^3$. The velocity vector is set to zero at all walls and the pressure is set to $p=0$ at the outlet. 
		
		
		\subsection{Test case 3: the 3D Stokes Flow past a Sphere}\label{sec:3p3}
		
		The third selected test case is the classic 3D axisymmetric Stokes flow with uniform free stream velocity $U_\infty$ past a sphere of radius $R$. In spherical coordinates ($\boldsymbol{r}=(r,\theta,\varphi)$), the relevant components of the velocity field ($\boldsymbol{v}=(v_r,v_\theta,0)$) reads (see for example \cite{Bird})

		\begin{subequations}
			\label{Stokes_V}
			\begin{gather}
				v_r=U_\infty\Biggl[1-\frac{3}{2}\biggl(\frac{R}{r}\biggr)+\frac{1}{2}\biggl(\frac{R}{r}\biggr)^3\Biggr]\cos(\theta)\\
				v_\theta=U_\infty\Biggl[1-\frac{3}{4}\biggl(\frac{R}{r}\biggr)-\frac{1}{4}\biggl(\frac{R}{r}\biggr)^3\Biggr]\sin(\theta)\,\,,
			\end{gather}
		\end{subequations} where the free stream flows along the $z$ direction, $\theta$ is the polar coordinate in the ($x,y$) plane, and $\varphi$ is the polar coordinate in the ($y,z$) plane, hence the symmetry of the flow is such that all derivatives along ${\varphi}$ vanish. The pressure field, taking $p_\infty=0$ as reference in the far field, is 
		
		\begin{equation}
			\label{Stokes_P}
			p(r,\theta)=-\frac{3}{2}\Biggl( \frac{\mu U_{\infty}}{R}\Biggr)\Biggl(\frac{R}{r}\Biggr)^2\cos(\theta)\,.
		\end{equation}
		
		Both the velocity and the pressure fields are here scaled to their dimensionless counterpart $\hat{p}=pD/(\mu U_\infty)$ and $\hat{\mathbf{v}}=(v_r/U_\infty,v_\theta/U_\infty)$ and the radial coordinate becomes $\hat{r}=r/D$. Therefore, the dimensionless stagnation pressure becomes $\hat{p}(0.5,0)=3$. 
		
\begin{figure}[ht]
			\centering
			\begin{subfigure}[\label{fig3D_Stokes_a}]{
					\includegraphics[width=0.45\textwidth,trim={2cm 0 1.6cm 0},clip]{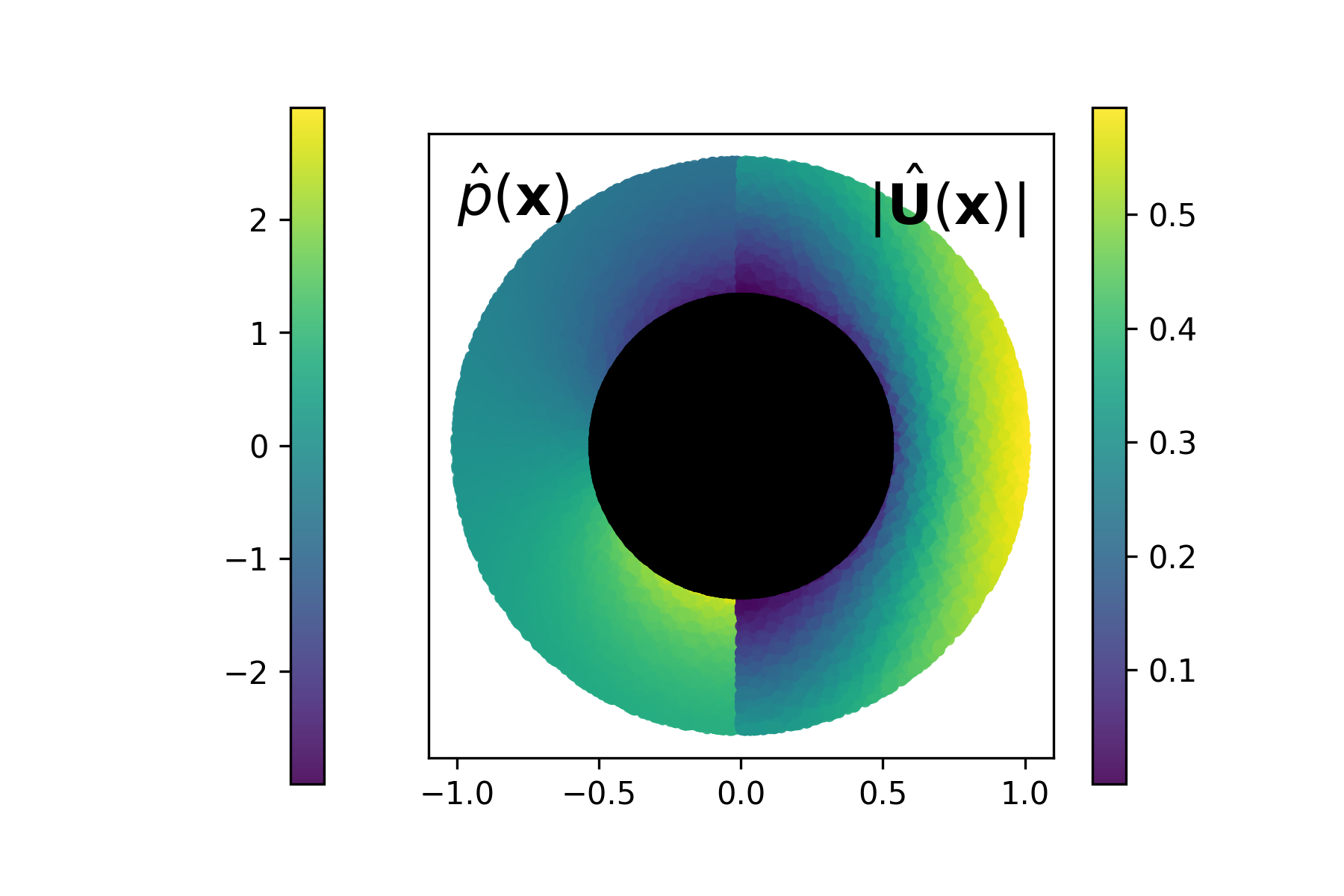}
				}
			\end{subfigure}
			\begin{subfigure}[\label{fig3D_Stokes_b}]{
					\includegraphics[width=0.48\textwidth,trim={3cm 0 3cm 0},clip]{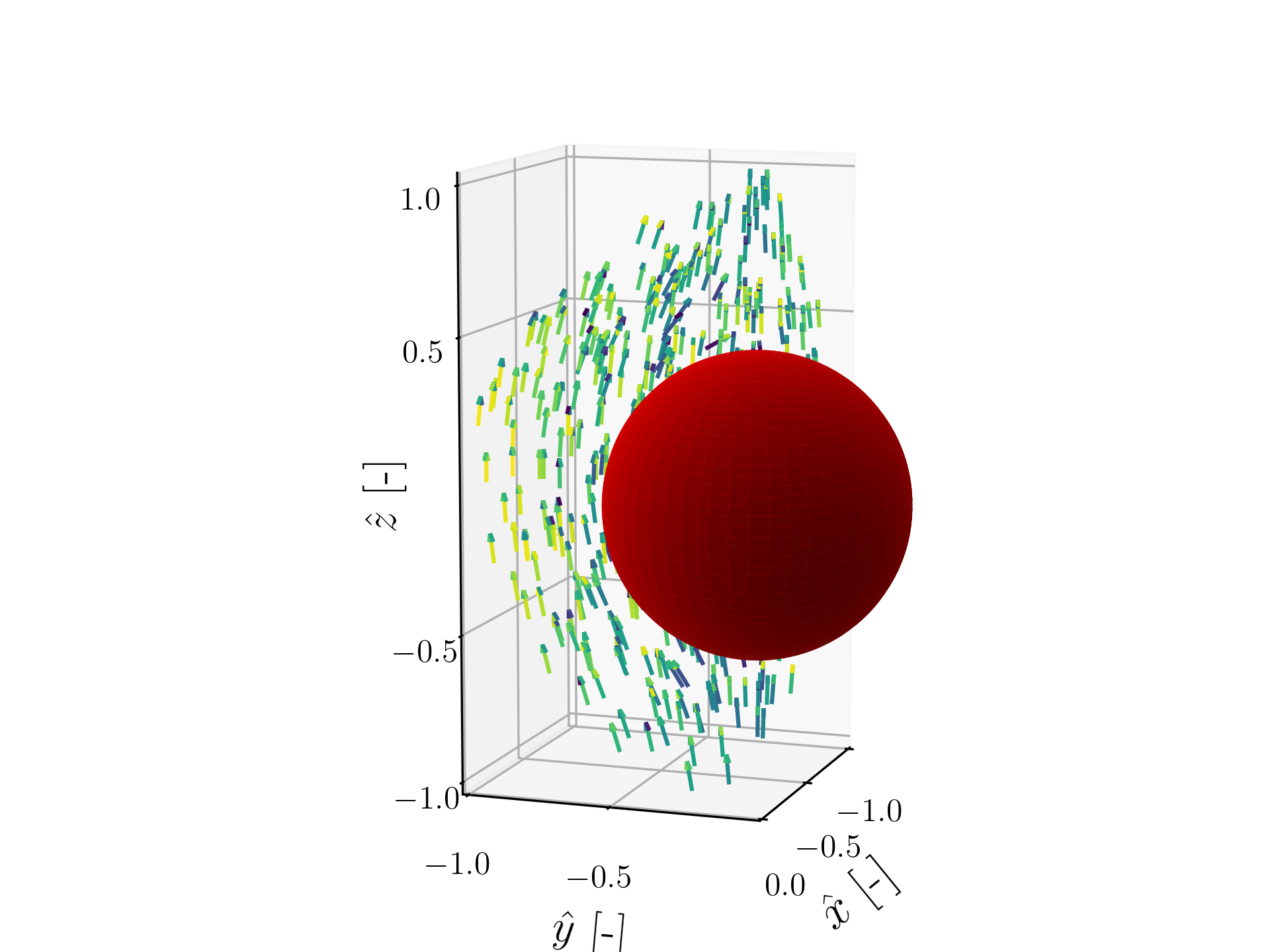}
				}
			\end{subfigure}
			\caption{Test case 3. (a) Contour-plot of the dimensionless pressure (left) and dimensionless velocity magnitude (right) on the ($y,z$) plane. (b) View of the flow seeded with $n_p=450$ particles in $\varphi\in[0,\pi/2]$ and $\theta \in[-\pi,\pi]$.} 
			\label{fig:3D_Stokes}
		\end{figure}

		Figure \ref{fig3D_Stokes_a} shows the contour-plot of the dimensionless pressure field $\hat{p}$ (on the half side) and the dimensionless velocity magnitude $\hat{\mathbf{v}}$ (on the right side) on the ($y,z$) plane. Both fields are shown in a domain with $\hat{r}\in[0.5,1]$. Within the same domain, Figure \ref{fig3D_Stokes_b} shows a quiver plot of the velocity field using $n_p=450$ particles within $\varphi\in[0,\pi/2]$ and $\theta \in[-\pi,\pi]$.

		This test case was analyzed with a number of particles ranging from $n_p=10000$ to $n_p=20000$. This is thus a relatively small 3D problem if compared to the experiments performed by \cite{Huhn2018}, but in line with other authors working on the regression of 3D Lagrangian tracking on regular grids (see for examples \cite{Agarwal2021}). Recalling that a 3D problem leads to matrices of the order of $\sim 3 n_b\times 3  n_b$, it is easy to see that the memory demands grow considerably even if the proposed clustering algorithm is such that $n_b\approx n_p/2$.
		
		A particle count of the order of $n_p\approx 20 000$ was considered appropriate to the scope of this work, namely providing a first proof of concept to the constrained RBF for regression and pressure computation on a 3D problem of realistic size. Accordingly, we present a test case with $n_p=18300$, taking the volume spherical shell within $\hat{r}=[0.5,1]$ as the integration volume. The choice of such a small volume makes the pressure integration particularly challenging because the boundary conditions at the open boundary ($\hat{r}=1$) requires an accurate evaluation of all the gradients of the velocity field. A larger domain or a much less `viscous' flow would allow the velocity field to become almost potential sufficiently far from the wall and would considerably simply the setting of the boundary conditions. The results for the velocity regression and the pressure integration for this test case are presented in section \ref{sec:4p3}.
		
		\section{Results}\label{sec:4}
		
		The three analyzed test cases represent problems of vastly different scales in terms of computational cost and complexity of the associated least-square problems. The first test case needs only a few constraints. The second has curved walls requiring more constraints and the third, being 3D, leads to a fairly large least square problem. Table \ref{Table_I} recalls the main parameters controlling the size of the problem in terms of number of particles, number of constraints and number of employed RBFs. Because these test cases were analyzed with various conditions in terms of seeding density and clustering approach, these numbers refer to one condition per case, namely the one considered the most representative. The implementation details and the results for each case are discussed in the following subsections.

		\begin{table}[ht]
			\centering
			\resizebox{0.8\textwidth}{!}{%
				\begin{tabular}{@{}c|c|cl|cc|@{}}
					\cmidrule(l){2-6}
					& N. of particles ($n_p$) & \multicolumn{2}{c|}{Size of the basis ($n_b$)} & \multicolumn{2}{c|}{N of Constraints   ($n_\lambda$)} \\ \cmidrule(l){2-6} 
					& V/P & \multicolumn{2}{c|}{V/P} & \multicolumn{1}{c|}{V} & P \\ \midrule
					\multicolumn{1}{|c|}{Case 1: Vortex Flow} & 5242 & \multicolumn{2}{c|}{1158} & \multicolumn{1}{c|}{196} & 196 \\ \midrule
					\multicolumn{1}{|c|}{Case 2: 2D Cylinder Wake} & 18755 & \multicolumn{2}{c|}{4203} & \multicolumn{1}{c|}{1348} & 748 \\ \midrule
					\multicolumn{1}{|c|}{Case 3: 3D Stokes Flow} & 18297 & \multicolumn{2}{c|}{10381} & \multicolumn{1}{c|}{13323} & 6990 \\ \bottomrule
				\end{tabular}%
			}
			\caption{Size of the test cases analyzed in this work in terms of number of particles $n_p$, number of RBFs $n_b$ and constraints $n_{\lambda}$. The letter V denotes the velocity regression problem, the letter P denotes the pressure integration problem.}
			\label{Table_I}
		\end{table}

		\begin{table}[ht]
			\centering
			\resizebox{\textwidth}{!}{%
				\begin{tabular}{@{}c|c|cc|cc|cc|cc|@{}}
					\cmidrule(l){2-10}
					& Time for Clustering & \multicolumn{2}{c|}{Time to Prepare   $\mathbf{A}$} & \multicolumn{2}{c|}{\begin{tabular}[c]{@{}c@{}}Timing to prepare $\mathbf{M}$\\       (lines 4 and 5)\end{tabular}} & \multicolumn{2}{c|}{\begin{tabular}[c]{@{}c@{}}Timing to compute $\boldsymbol{\lambda}$\\      (line 7)\end{tabular}} & \multicolumn{2}{c|}{\begin{tabular}[c]{@{}c@{}}Timing to compute $\boldsymbol{w}$ \\      (Line 8)\end{tabular}} \\ \cmidrule(l){2-10} 
					& V/P & \multicolumn{1}{c|}{V} & P & \multicolumn{1}{c|}{V} & P & \multicolumn{1}{c|}{V} & P & \multicolumn{1}{c|}{V} & P \\ \midrule
					\multicolumn{1}{|c|}{Case 1} & 2.2 & \multicolumn{1}{c|}{0.86} & 0.66 & \multicolumn{1}{c|}{0.24} & 0.04 & \multicolumn{1}{c|}{0.002} & 0.001 & \multicolumn{1}{c|}{0.01} & 0.002 \\ \midrule
					\multicolumn{1}{|c|}{Case 2} & 4.3 & \multicolumn{1}{c|}{2.61} & 1.94 & \multicolumn{1}{c|}{2.19} & 0.27 & \multicolumn{1}{c|}{0.16} & 0.028 & \multicolumn{1}{c|}{0.035} & 0.012 \\ \midrule
					\multicolumn{1}{|c|}{Case 3} & 12.9 & \multicolumn{1}{c|}{320} & 90.3 & \multicolumn{1}{c|}{633} & 42.8 & \multicolumn{1}{c|}{16.7} & 3.6 & \multicolumn{1}{c|}{10.9} & 0.078 \\ \bottomrule
				\end{tabular}%
			}
			\caption{Timing (in seconds) for some of the key steps in the algorithm (cf. Algorithm 1) for the three test cases considered in this work. The lines involved in the timing are recalled in the relevant columns. As for Table \ref{Table_I}, V denotes the velocity regression and P denotes the pressure integration.}
			\label{Table_II}
		\end{table}
		
		All computations were performed on a modest desktop computer with 32 GB of installed RAM, running with an Intel(R) Core(TM)i7-3770 CPU (3.GHz). The implementations were coded in Python 3.7. It is important to note that the current implementation is not yet fully optimized, and significant performance gains are possible using low-level programming languages or code parallelization. We nevertheless share the timing results in table \ref{Table_II} for the critical operations from Algorithm 1. These numbers are given only to have an order of magnitude of the computational costs involved in the velocity regression and pressure computation. While the main computational concern of the proposed method in its current form is the memory demand for large scale problems, it is interesting to note that the computing time for a 2D problem like the one in case 2 is of the order of $9$ seconds for the velocity regression and $2.2$ seconds for the pressure computation while for the most expensive 3D test case considered in this work (with $n_p=18297$ particles with $n_b=10381$ RBFs in 3D) the timing goes to approximately $15$ minutes for the velocity regression and $2$ minutes for the pressure integration. The same 3D case (including both velocity regression and pressure computation) runs in about $4$ minutes on a single AMD EPYC 7742 64-core processors and 528 GB of RAM.
		
		\subsection{Test Case 1: Gaussian Vortex}\label{sec:4p1}
		
		A total of $n_\lambda=196$ constraints is chosen for both the regression of the velocity fields and the pressure integration. These constraints ($50$ on each side) are equally spaced along the boundaries of the square domain. On these points, the divergence-free conditions is imposed as a constraints of the velocity regression. Moreover the implementation find automatically the repeated point to ease the input procedure. We thus have $n_{\Delta}=196$ , and $n_{D}=n_N=0$ in \eqref{Prob_U}. A divergence-free penalty is set in the whole domain with $\alpha_{\nabla}=1$.
		
		The same boundary points are used to impose Neumann boundary conditions for the pressure integration (cf equation \eqref{Projection_P}). Moreover, a Dirichlet boundary condition is set in one point, on the top left corner. This simulates a local pressure probe and enables a unique solution to the Poisson equation. We thus have $n_N=200$ and $n_D=1$ in \eqref{Prob_P}. The cluster-level vector is $\boldsymbol{n}_M=[6,10]$ with $\varepsilon_t=0.88$ and the maximum capping taken as $c_k=0.7$ (see Sec. \ref{sec:2p4}).
		
		To explore the impact of seeding concentration and noise, a grid with $N_p=[0.003,0.005]$ and $q=[0,0.3]$ (see \ref{sec:3p1} for definitions) with $20\times 20$ elements was analyzed. Figure \ref{Grid_P} shows the $l_2$ error for the velocity ($E_U$) and the pressure ($E_P$) computations, defined as 
		
		\begin{equation}
			\label{errors}
			E_U=\frac{||\boldsymbol{u}(\boldsymbol{X})-\bm{\Phi}(\boldsymbol{X}) \boldsymbol{w}_u||_2+||\boldsymbol{v}(\boldsymbol{X})-\bm{\Phi}(\boldsymbol{X}) \boldsymbol{w}_v||_2}{||\boldsymbol{u}(\boldsymbol{X})||_2+||\boldsymbol{v}(\boldsymbol{X})||_2}\quad \mbox{and} \quad E_P=\frac{||\boldsymbol{p}(\boldsymbol{X})-\bm{\Phi}(\boldsymbol{X}) \boldsymbol{w}_p||_2}{||\boldsymbol{p}(\boldsymbol{X})||_2}
		\end{equation} where $||\cdot||_2$ is the $l_2$ norm of a vector, $(\boldsymbol{u}(\boldsymbol{X}),\boldsymbol{v}(\boldsymbol{X}))$ is the theoretical velocity field and $\boldsymbol{p}(\boldsymbol{X})$ is the theoretical pressure field on the scattered points $\boldsymbol{X}$, arranged as column vectors.

		It is worth noticing that the error is computed with respect to the theoretical (noiseless) data, thus also evaluating the RBF expansion's filtering capabilities.
		
		\begin{figure}[h!]
			\centering
			\begin{subfigure}[\label{Aa}]{
					\includegraphics[width=0.46\textwidth]{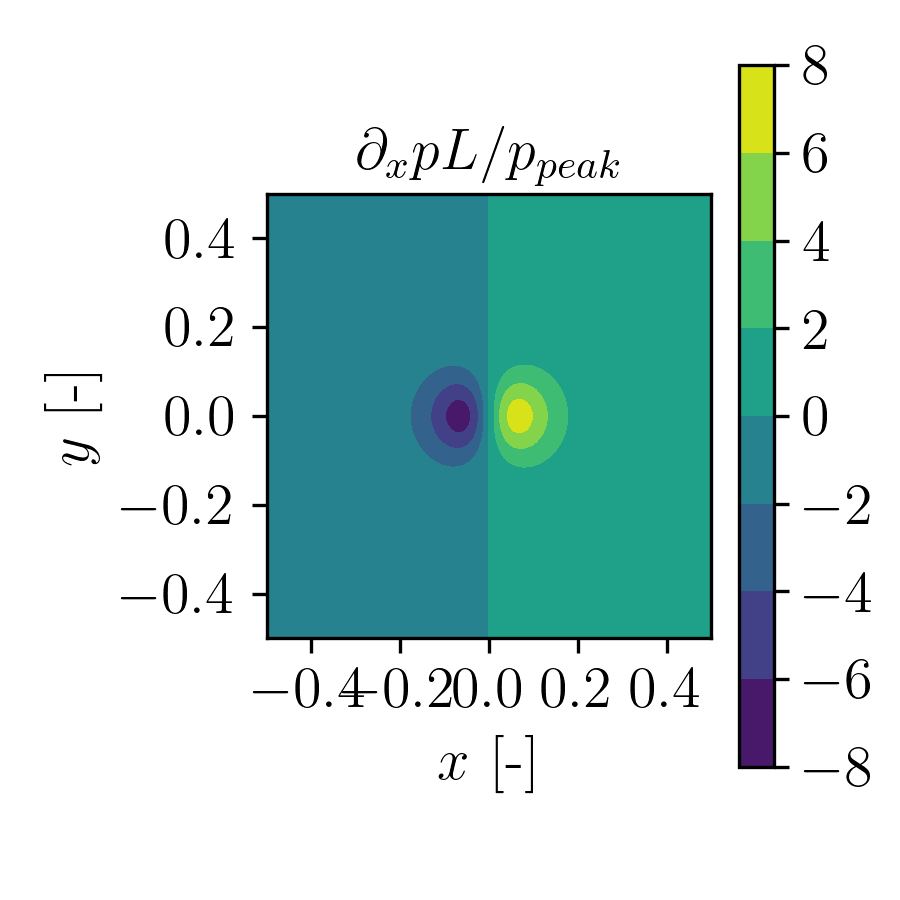}}
			\end{subfigure}
			\begin{subfigure}[\label{Ab}]{
					\includegraphics[width=0.46\textwidth]{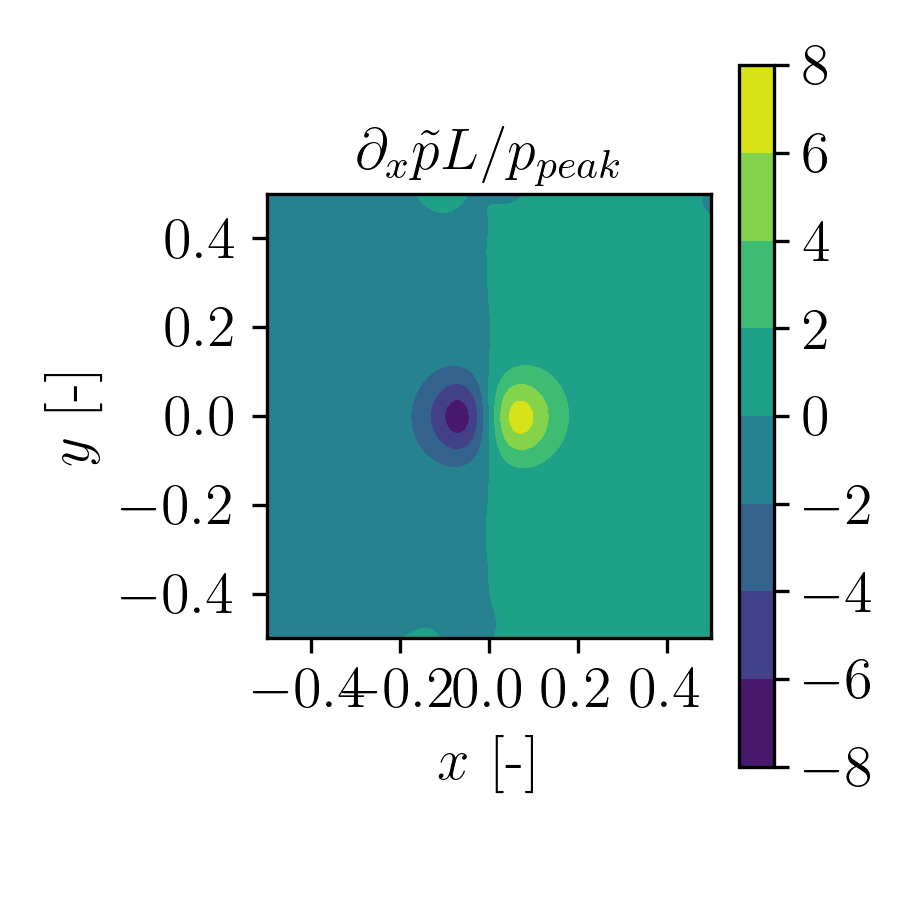}}
			\end{subfigure}
			\begin{subfigure}[\label{Ac}]{
					\includegraphics[width=0.65\textwidth]{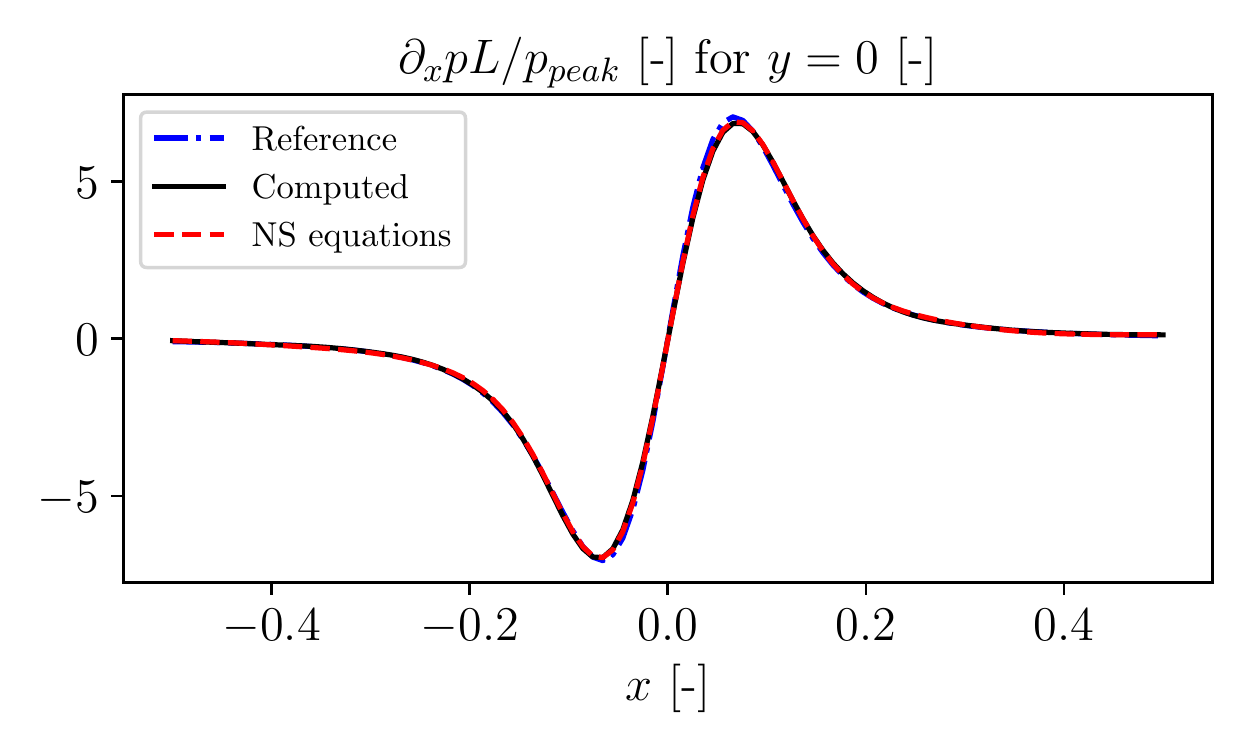}}
			\end{subfigure}	
			\caption{Comparison of the pressure gradient reconstruction with the background truth. Figure \ref{Aa} shows the normalized pressure gradient $\partial_x p$, scaled with respect to $p_{peak}/L$, while Figure \ref{Ab} shows the results obtained by differentiating the RBF solution. Figure \ref{Ac} show the pressure gradient profile along $y=0$.} 
			\label{Aggiunta_vortex}
		\end{figure}

		\begin{figure}[ht]
			\centering
			\begin{subfigure}[\label{fig:Err_V_T1}]{
					\includegraphics[height=6.7cm,trim={2cm 0 1cm 0},clip]{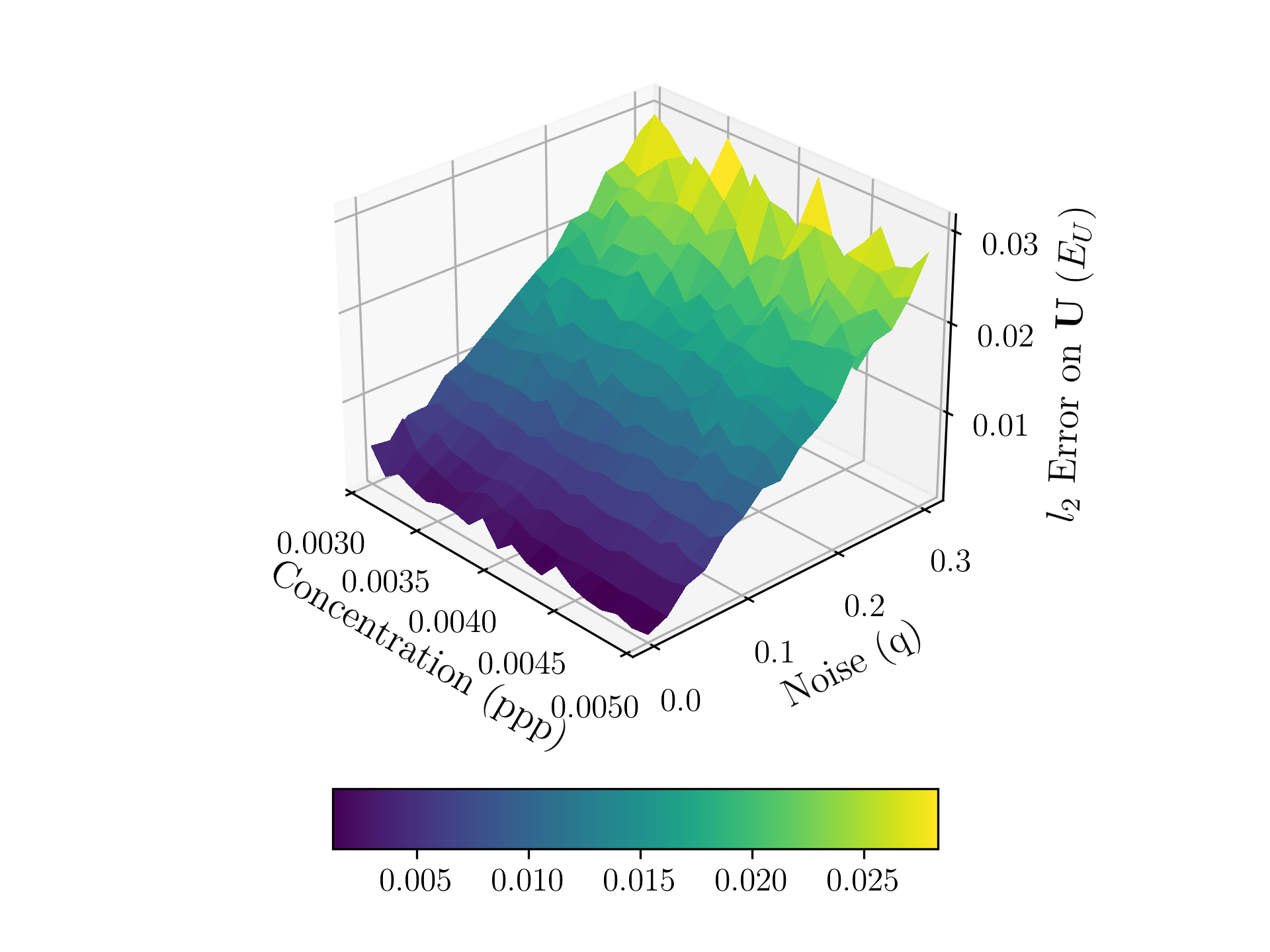}
				}
			\end{subfigure}
			\begin{subfigure}[\label{fig:Err_P_T1}]{
					\includegraphics[height=6.7cm,trim={2cm 0 1cm 0},clip]{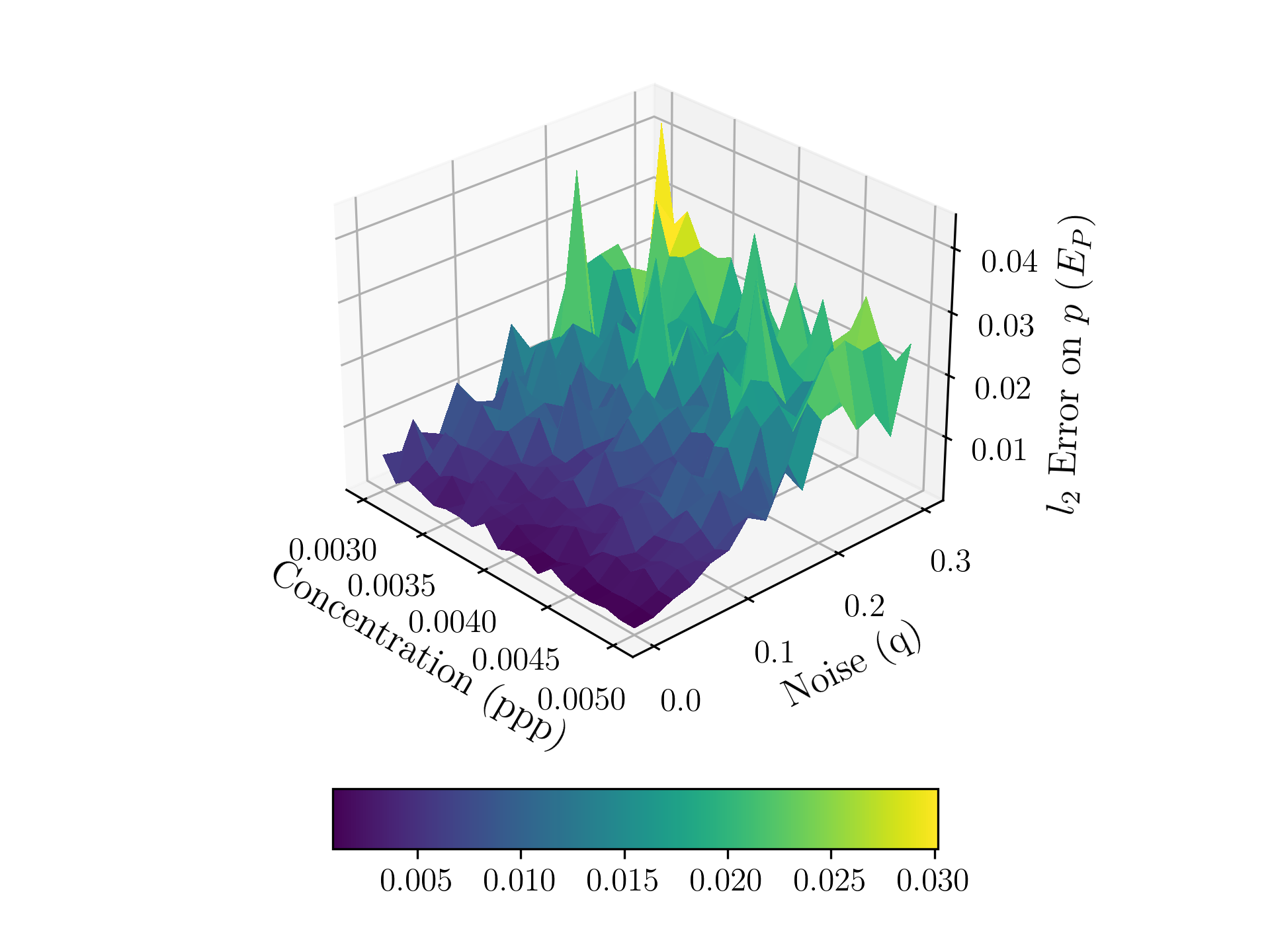}
				}
			\end{subfigure}
			\caption{Test case 1. $l_2$ errors in the velocity field reconstruction (left) and pressure integration (right) as defined in \ref{errors} as a function of the particle concentration $N_p$ (in particles per pixels) and noise level $q$ (see definitions in Sec. \ref{sec:3p1}) .} 
			\label{Grid_P}
		\end{figure}

		\begin{figure}[h!]
			\centering
			\begin{subfigure}[\label{fig:Error_V_VORTEX}]{
					\includegraphics[width=0.4\textwidth,trim={0.4cm 0.6cm 0.4cm 0},clip]{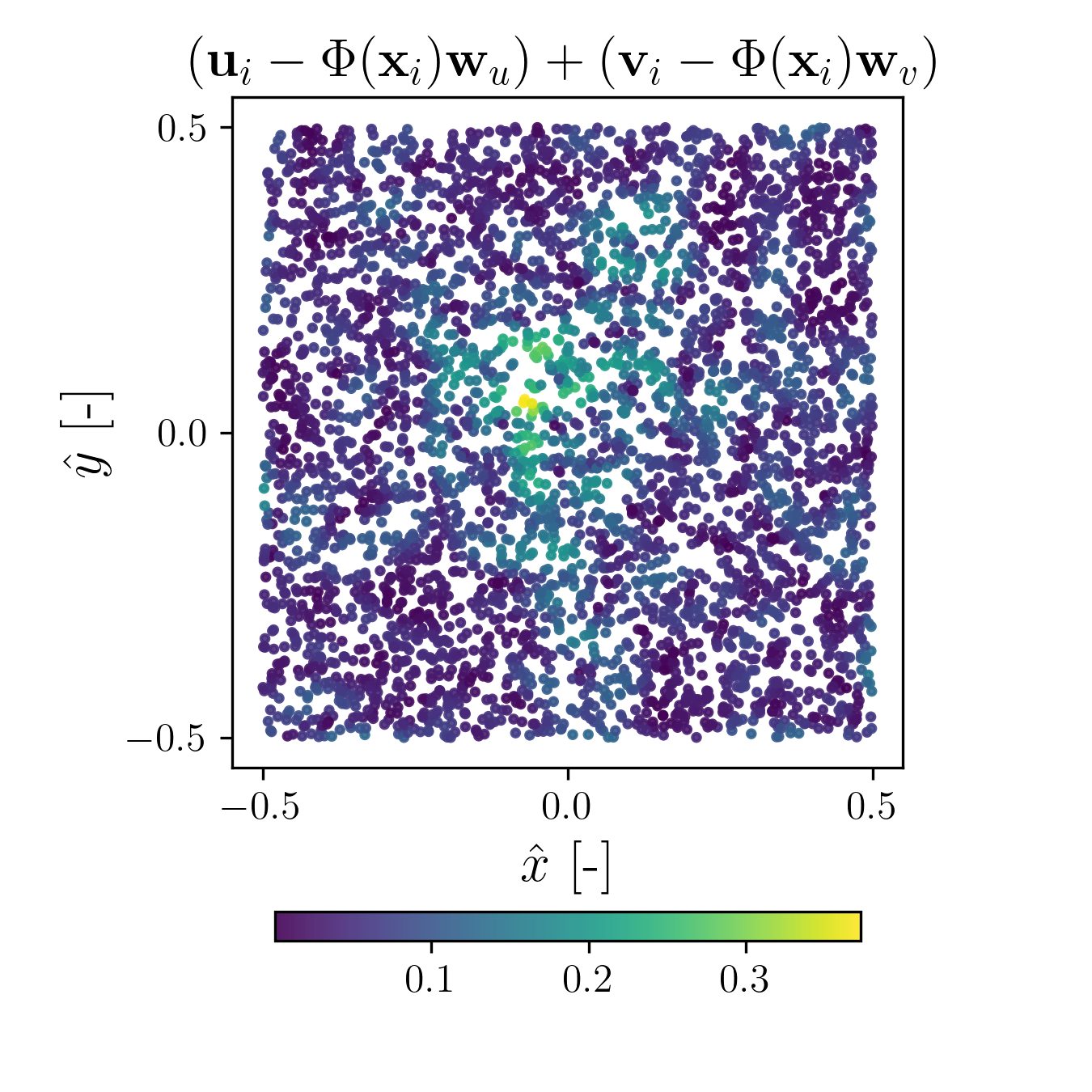}}
			\end{subfigure}
			\begin{subfigure}[\label{fig:Error_P_VORTEX}]{
					\includegraphics[width=0.4\textwidth,trim={0.4cm 0.6cm 0.4cm 0},clip]{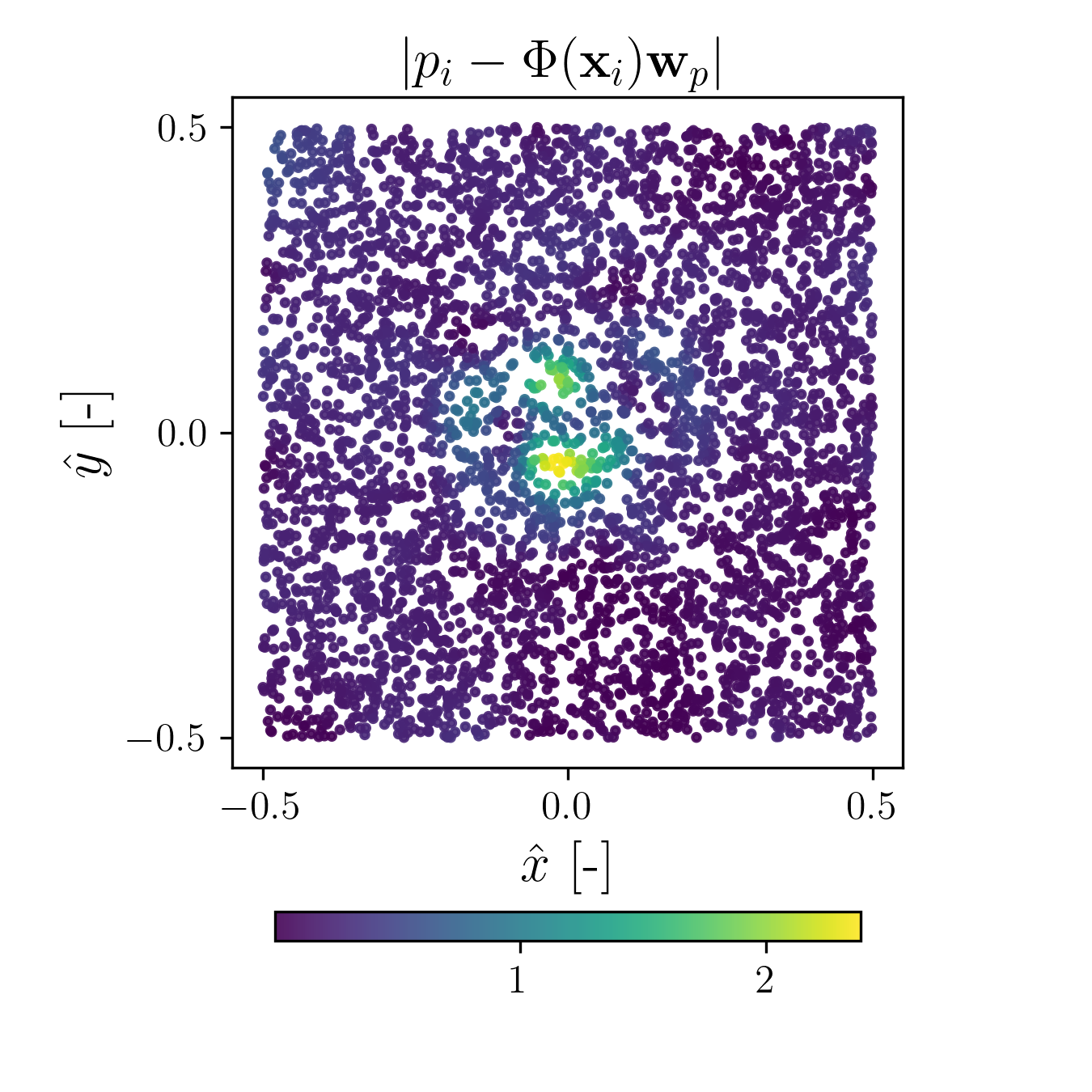}
				}
			\end{subfigure}
			\caption{Test case 1. Local absolute error distributions for the velocity regression (a) and the pressure computation (b) with $N_{pp}=0.004$ and $q=0.15$ (cf. Section \ref{sec:3p1}). The local error is everywhere below $1\%$ for the pressure reconstruction. } 
			\label{Err_Vortex}
		\end{figure}
		\color{black}

		Overall, the (global) $l_2$ error is below $2\%$ for both velocity and pressure reconstruction even at noise levels as high as $q=30\%$ and seeding densities as low as $N_p=0.003$ (corresponding to $n_p=3150$ particles). Moreover, as long as a sufficient number of particle is available, the accuracy of the velocity reconstruction is found to be independent from the seeding concentration and approximately linearly dependent on the noise level. 
		
		\textcolor{black}{Figure \ref{Aggiunta_vortex} illustrates the robustness of the method in the gradient reconstruction. Figure \ref{Aa} shows the pressure gradient along the $x$ direction by differentiating \eqref{Pressure_vortex}; Figure \ref{Ab} shows the same quantity computed analytically from the RBF reconstruction. This test case considers $N_{pp}=0.004$ with a noise level of $q=25\%$. The agreement is remarkable, as also shown in Figure \ref{Ac}, which shows the profile of the pressure gradient from the analytic formula (blue dash-dotted line), from the RBF computation (continuous black line) and the projection of the NS equation evaluated via the RBF expansion (red dashed lines). The curves are nearly indistinguishable. Compared to the results presented in the literature for this test case (see e.g. \cite{McClure2017b}, who showed results with comparable noise levels), the results appears much smoother and more accurate. This appears remarkable if one consider that this test case is usually solved with at least one edge of the domain implementing Dirichlet boundary conditions (see \cite{de2012pressure}, while \cite{McClure2017b} uses Dirichlet boundary conditions on the four sides). }

		To conclude, considering an intermediate test case with $N_{pp}=0.004$ and $q=0.15$, Figure \ref{fig:Error_V_VORTEX} shows the distribution of absolute error for the velocity field $(\boldsymbol{u}_i-\bm{\Phi}(\mathbf{x}_i)\boldsymbol{w}_u)^2+(\boldsymbol{v}_i-\bm{\Phi}(\boldsymbol{x}_i)\boldsymbol{w}_v)^2$ while Figure \ref{fig:Error_P_VORTEX} shows the distribution of absolute error for the pressure field $\boldsymbol{p}_i-\bm{\Phi}(\boldsymbol{x}_i) \boldsymbol{w}_p$ (on the right). The error is mostly present in the region of high shear, but it is overall negligible for both the velocity regression and the pressure computation.
		
		\subsection{Test Case 2: 2D Flow Past a Cylinder}\label{sec:4p2}
		
		This test case involves all the set of boundary conditions implemented in this work. In each of the solid walls, $150$ (equally spaced) constraints are taken for both the regression of the velocity and the pressure integration. In the velocity regression, the divergence-free condition is imposed in all these points (leading to $n_{\nabla}=748$ once common points are removed), while Dirichlet conditions are imposed only at walls walls and at the inlet. Therefore, we have $n_{D}=600$ points with both conditions and a total of $n_\lambda=n_{\nabla}+2 n_{D}=1948$ constraints (cf. Table \ref{Table_I}).
		
\begin{figure}[ht]
			\centering
			\includegraphics[width=0.45\textwidth ]{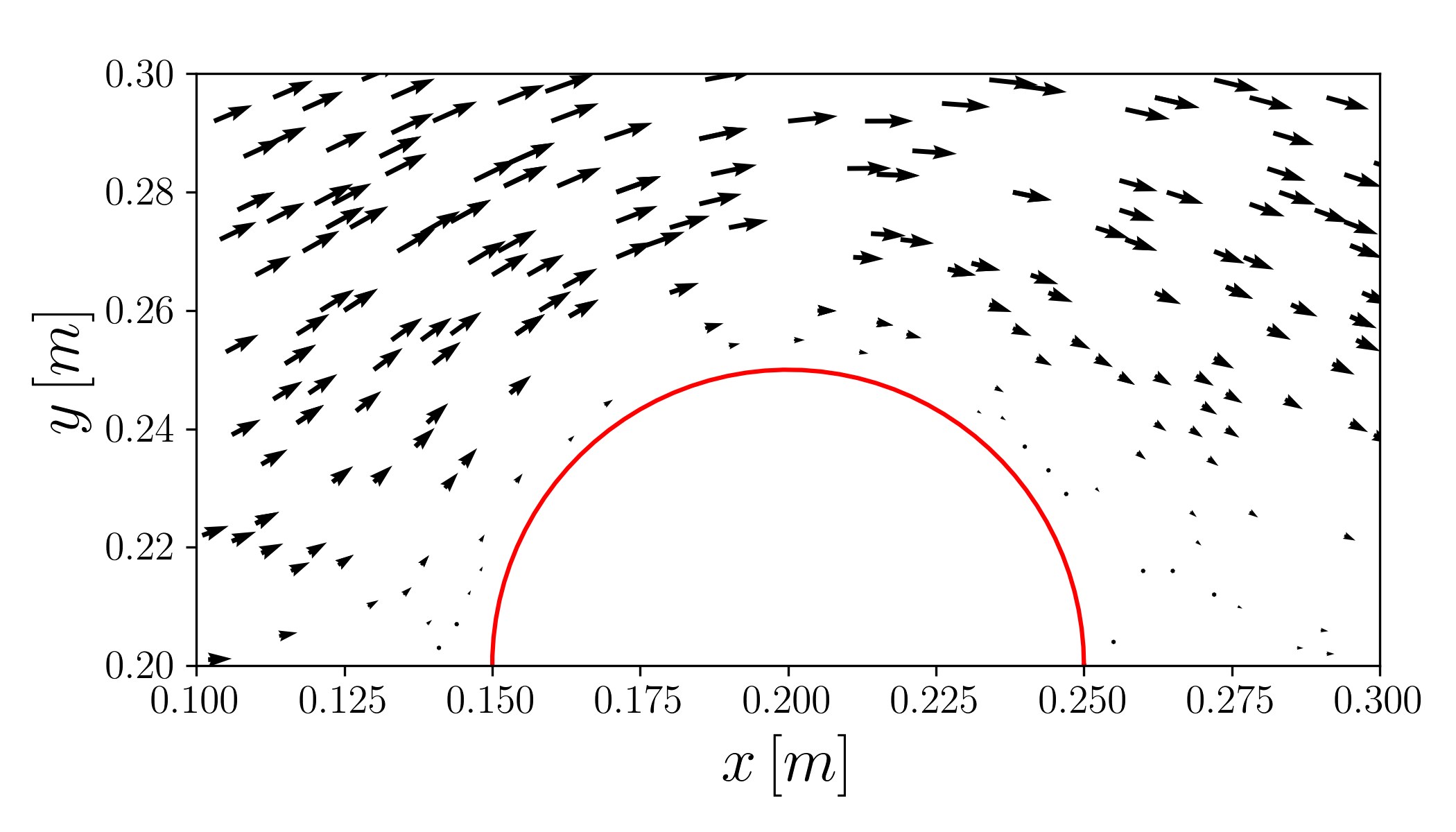}\hspace{2mm}
			\includegraphics[width=.45\textwidth ]{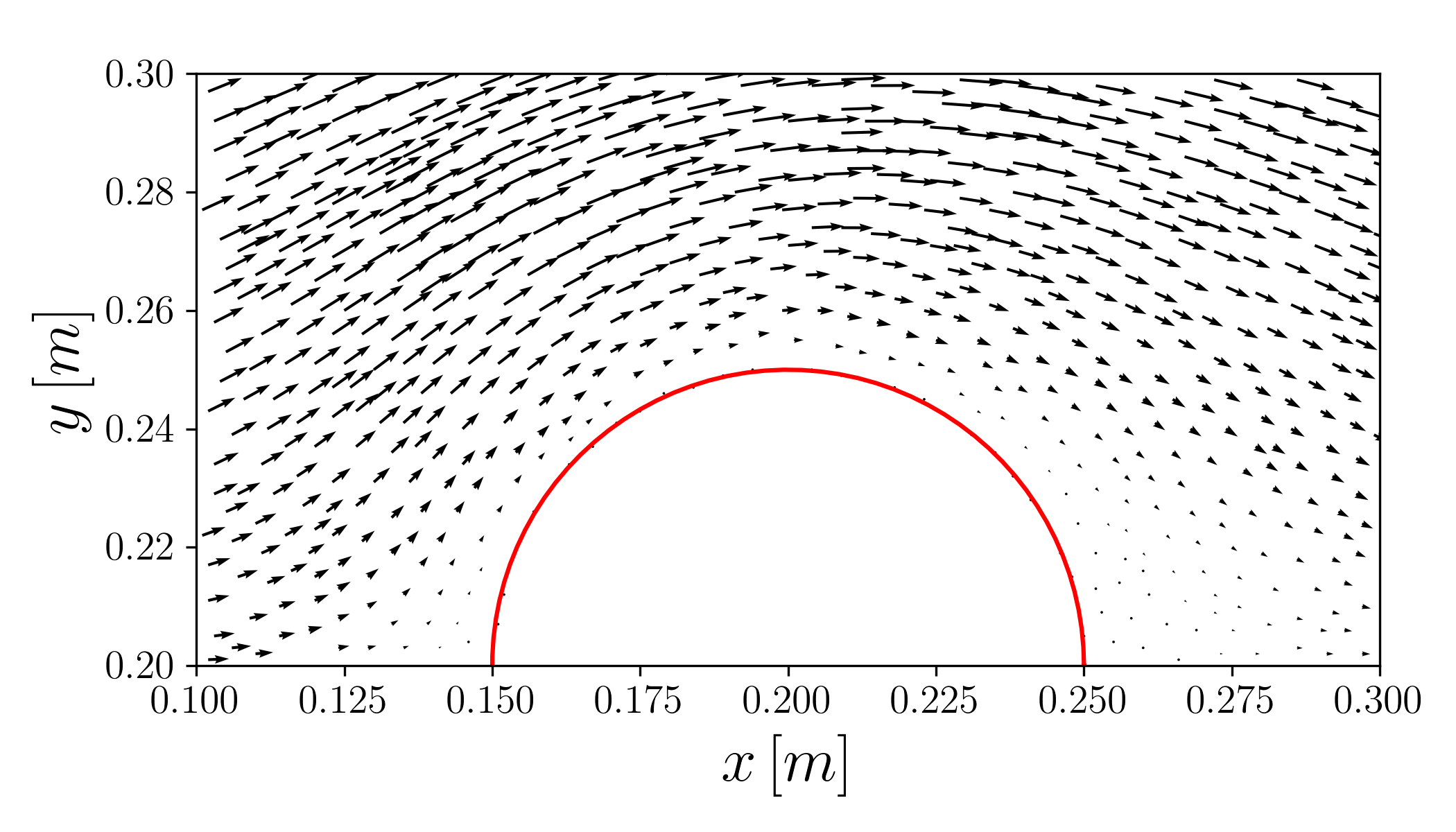}
			\caption{Test case 2. Zoomed view of the scattered velocity data used at the lowest (left, with $n_p=6000$) and the highest (right, with $n_p=18755$) seeding densities.} 
			\label{close_up_cyl}
		\end{figure}

		For the pressure integration, Neumann conditions from the momentum equation (cf. equation \eqref{Projection_P}) in all of these points. Moreover, a Dirichlet condition is set on the top corner, at the inlet. This test case was analyzed with a number of particles ranging from $n_p=6000$ to $n_p=18755$ points and two noise levels, namely $q=0$ (no noise) and $q=0.1$. Figure \ref{close_up_cyl} shows a close-up near the cylinder for the cases with the lowest and the highest concentration; it is worth pointing out that the lowest seeding only leaves a few dozens of particles in the region $r\in[R,1.2R]$, which is the one where most of the velocity gradients are located. It is thus somewhat surprising that the velocity field is reasonably well reconstructed even in these conditions.

\begin{figure}[ht]
			\centering
			\begin{subfigure}[\label{Vq0}]{
					\includegraphics[width=0.45\textwidth]{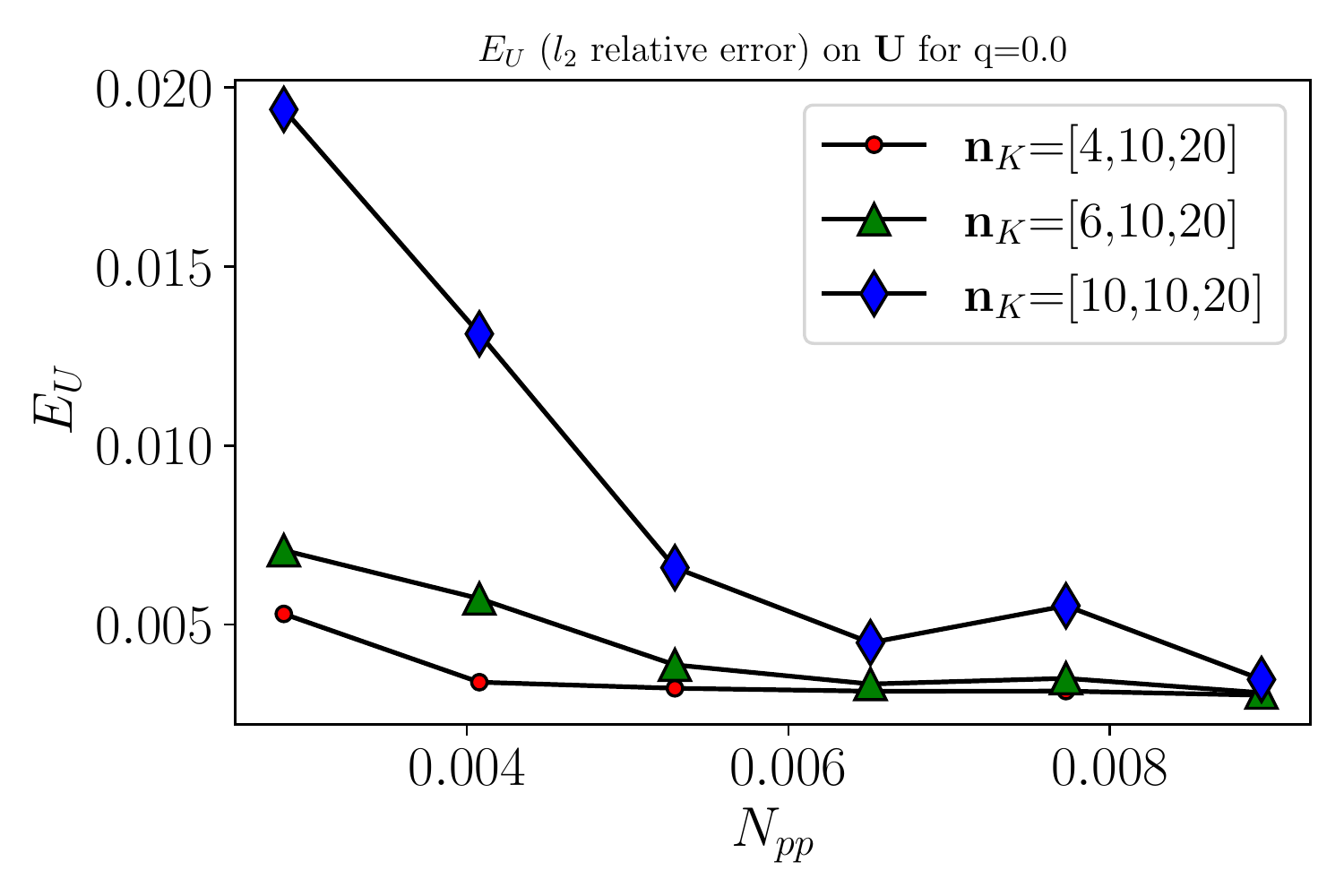}
				}
			\end{subfigure}
			\begin{subfigure}[\label{Pq0}]{
					\includegraphics[width=0.45\textwidth]{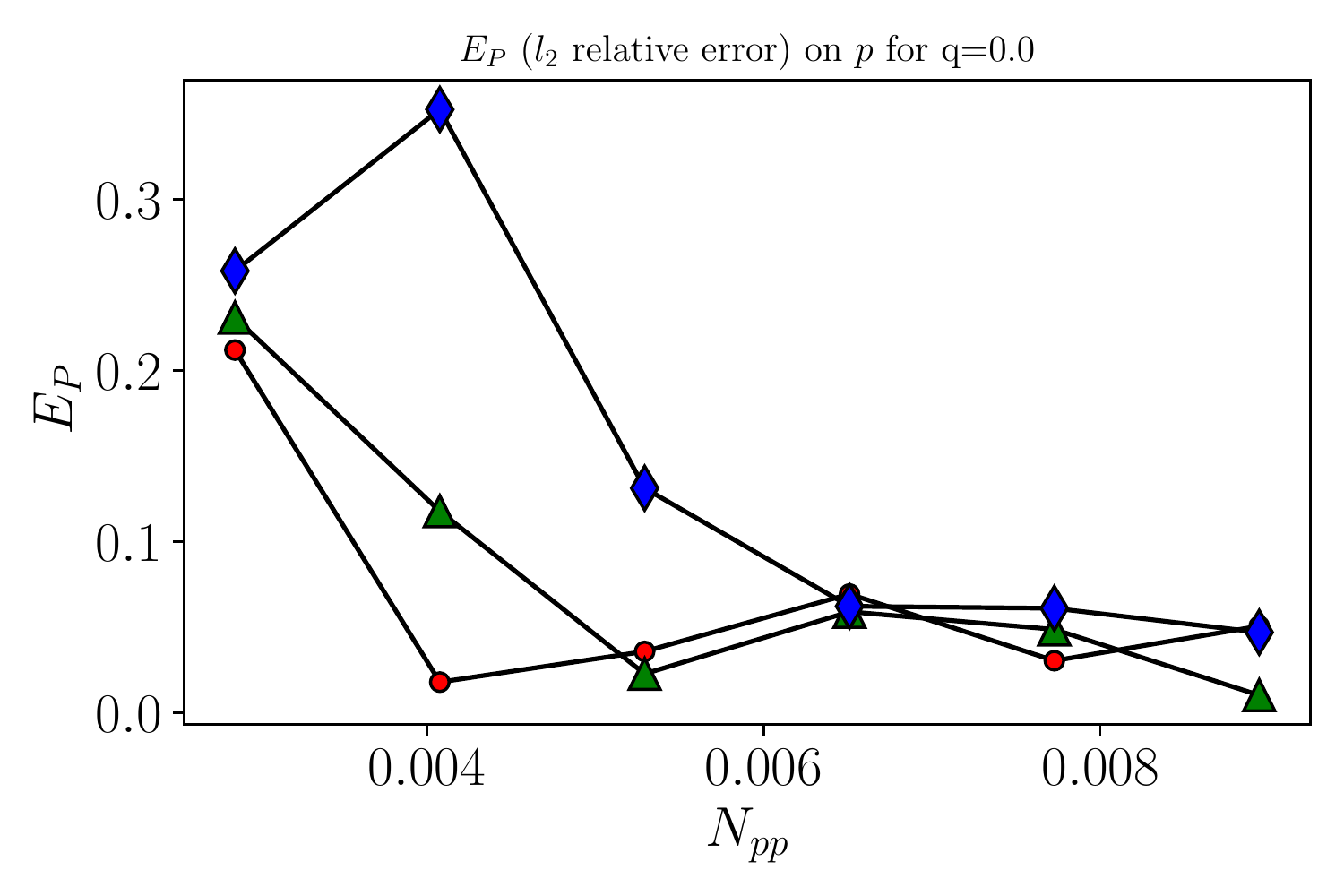}
				}
			\end{subfigure}
			\begin{subfigure}[\label{Vq005}]{
					\includegraphics[width=0.45\textwidth]{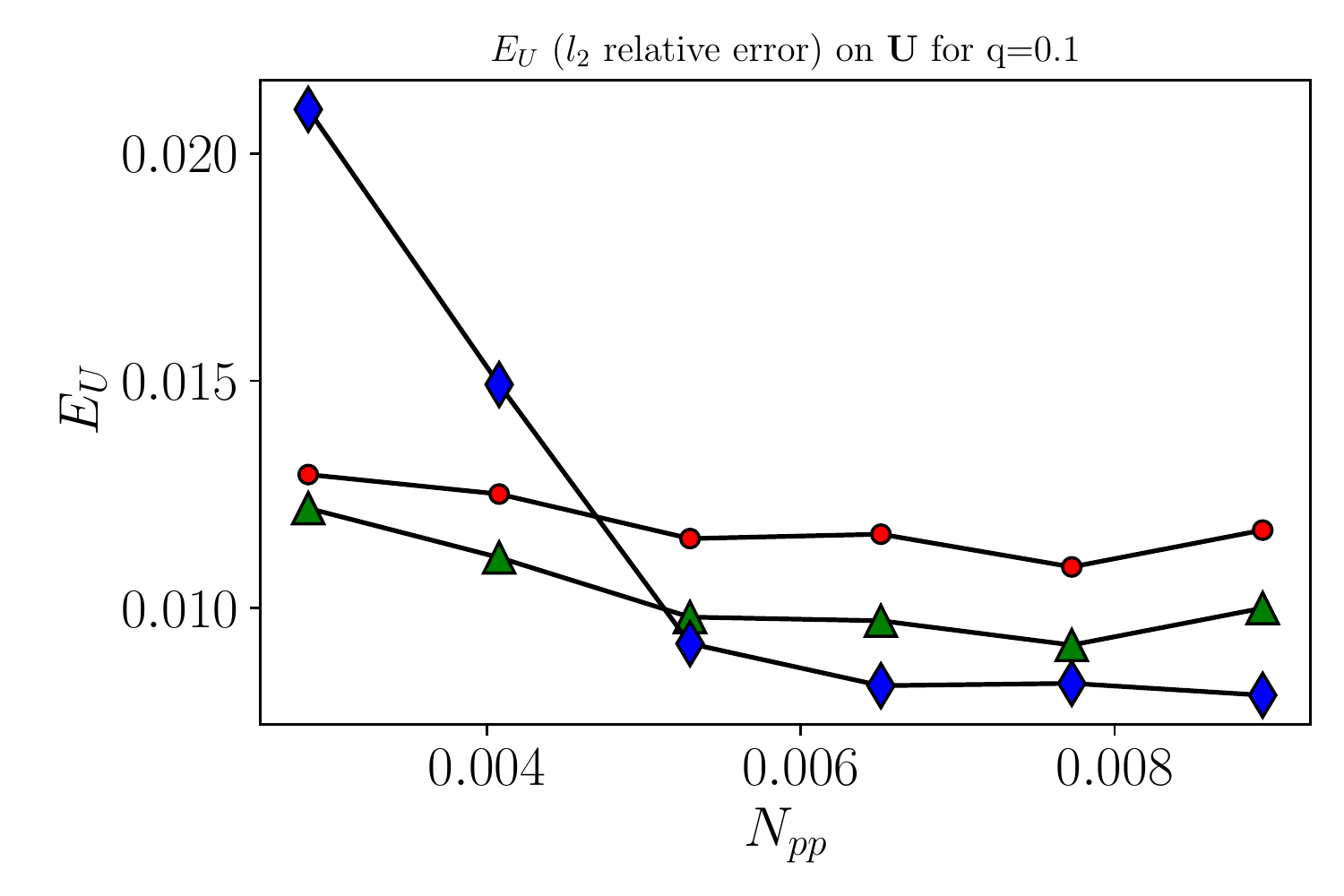}
				}
			\end{subfigure}
			\begin{subfigure}[\label{Pq005}]{
					\includegraphics[width=0.45\textwidth]{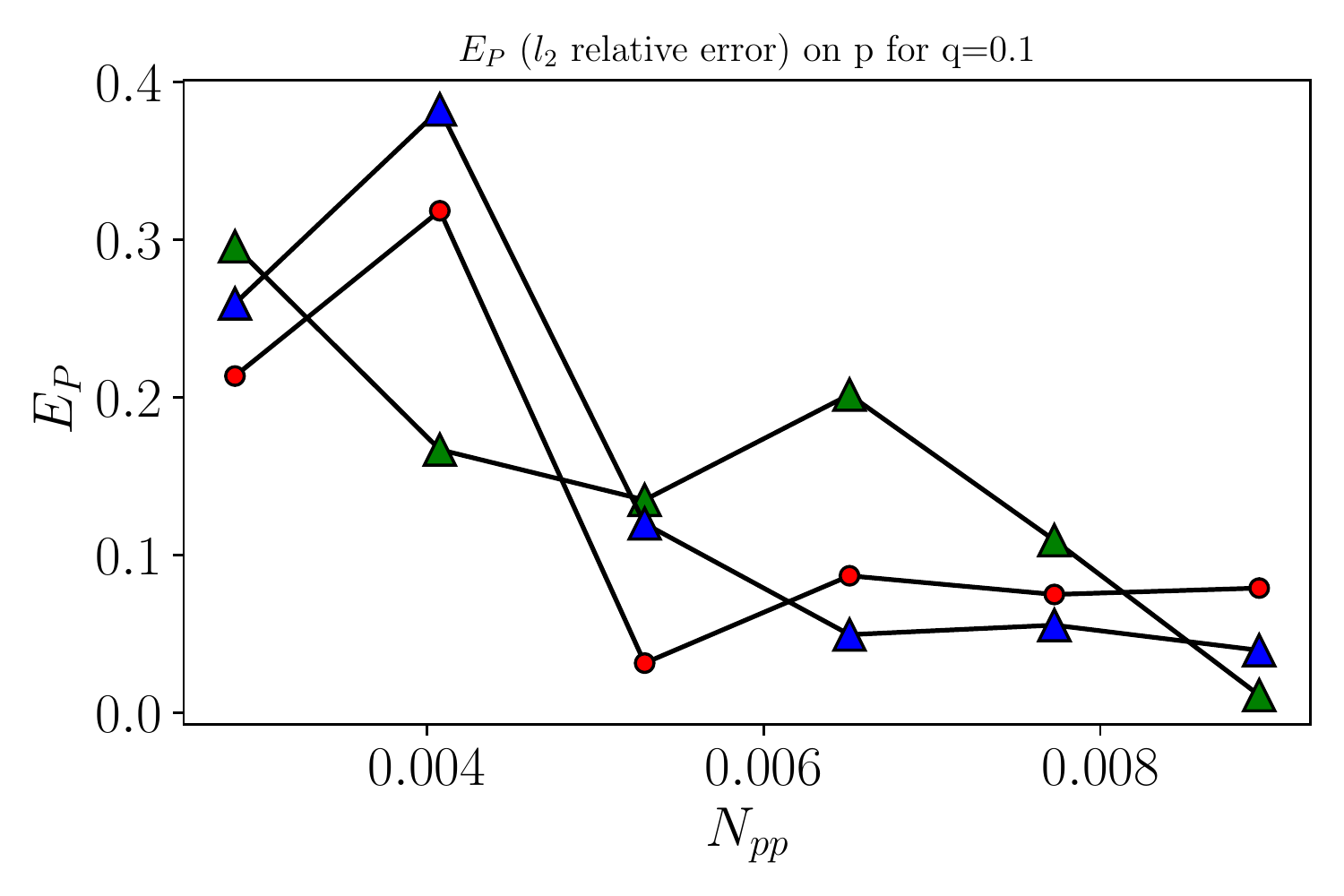}
				}
			\end{subfigure}
			\caption{Test case 2. Global error for the velocity regression (left) and for the pressure integration (right). The figures on the top collects results for the noise-free cases ($q=0$) while the figures on the bottom are related to the noisy cases ($q=0.1$).} 
			\label{Res_Cylinder_Analysis}
		\end{figure}

		Figure \ref{Res_Cylinder_Analysis} shows the behaviour of the global convergence error as a function of the image density for both the velocity regression and the pressure computation. Three clustering approaches are compared, namely $\mathbf{n}_K=[4,10,20]$, $\mathbf{n}_K=[6,10,20]$ and $\mathbf{n}_K=[10,10,20]$ (see Sec.\ref{sec:2p4}).
		everFigs. \ref{Vq0} and \ref{Pq0} show the results in absence of noise. Although the coarser clustering (blue triangles markers) suffers at the lowest seeding densities (because at low seeding produce overly large RBFs),it is worth noticing that the global error for the velocity in noise free conditions (Figure \ref{Vq0}) is never larger than 2\% and settles at 0.5\% if sufficient particles are available. \textcolor{black}{In the presence of noise (Figure \ref{Vq005}), the impact of the clustering method is more important and, for a sufficiently large number of particles, the error stabilizes in the range 0.5-1.2\% depending on the collocation points.} These results confirm the robustness of the proposed approach for physics-informed (constrained) regression.
		
		\begin{figure}[ht]
			\centering
			\includegraphics[width=\textwidth]{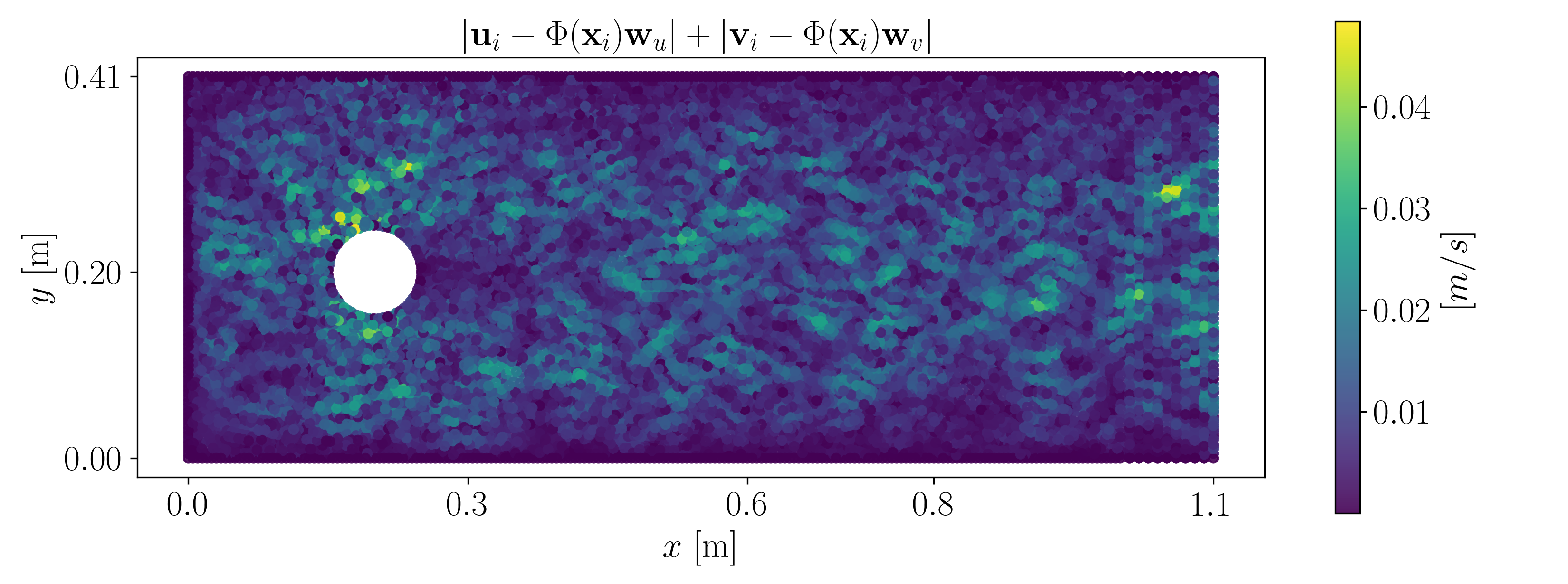}\\
			\includegraphics[width=\textwidth]{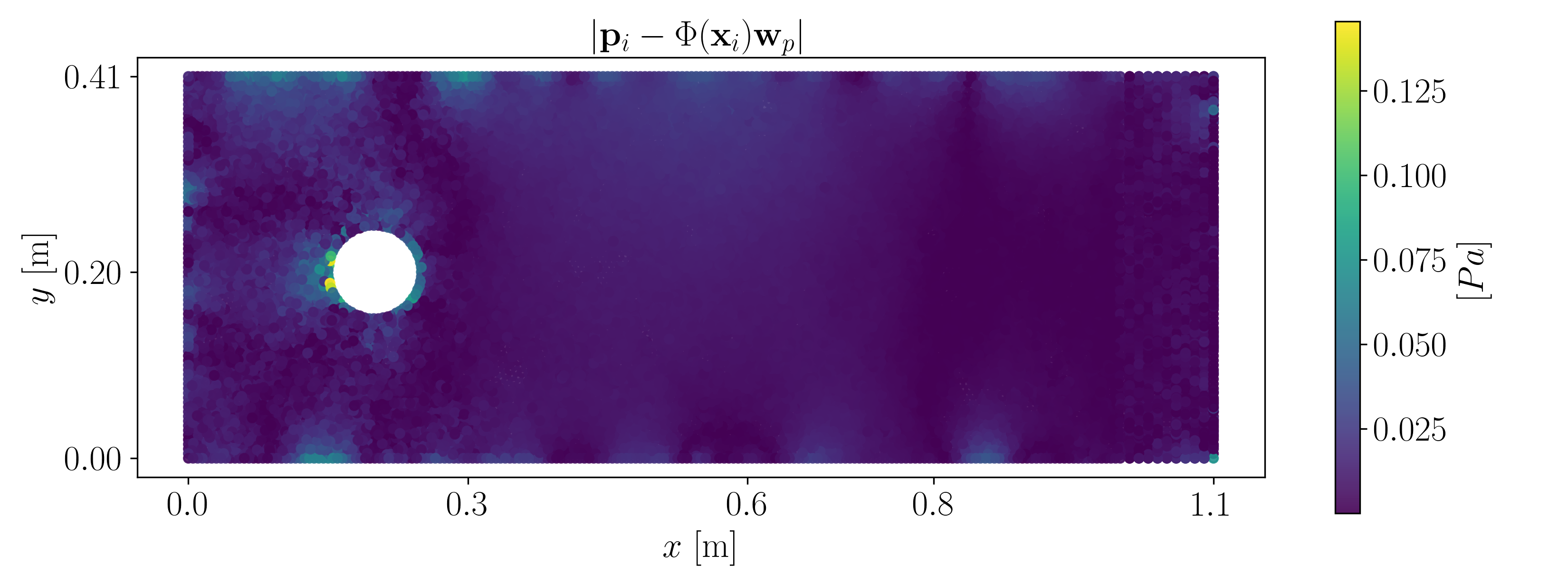}
			\caption{Test case 2. Local absolute error distribution for the velocity regression (top) and the pressure integration (bottom) for the 2D cylinder wake with a very noisy field $q=0.1$ and the largest seeding concentration ($N_p=0.089$, i.e $n_p=18755$).}
			\label{Res_1_Cylin}
		\end{figure}
		
		On the other hand, the pressure integration is more vulnerable and fails at the lowest seeding density for the coarser clustering (with global error up to $40\%$). Nevertheless, the global error drops to about $2\%$ with the intermediate clustering if a reasonable amount of particles is available. As expected, the pressure integration appears to be more sensitive to noise and the clustering strategy versus the number of particles available. This is evident as the finest clustering (red circular markers) leads to failure over a wide range of concentrations. Among the tested clustering approaches, the intermediate one with $\mathbf{n}_K=[6,10,20]$ (green diamond markers) appears to be the most robust and is the one kept for the remainder of this section.
		
\begin{figure}[ht]
			\centering
			\includegraphics[width=.35\textwidth]{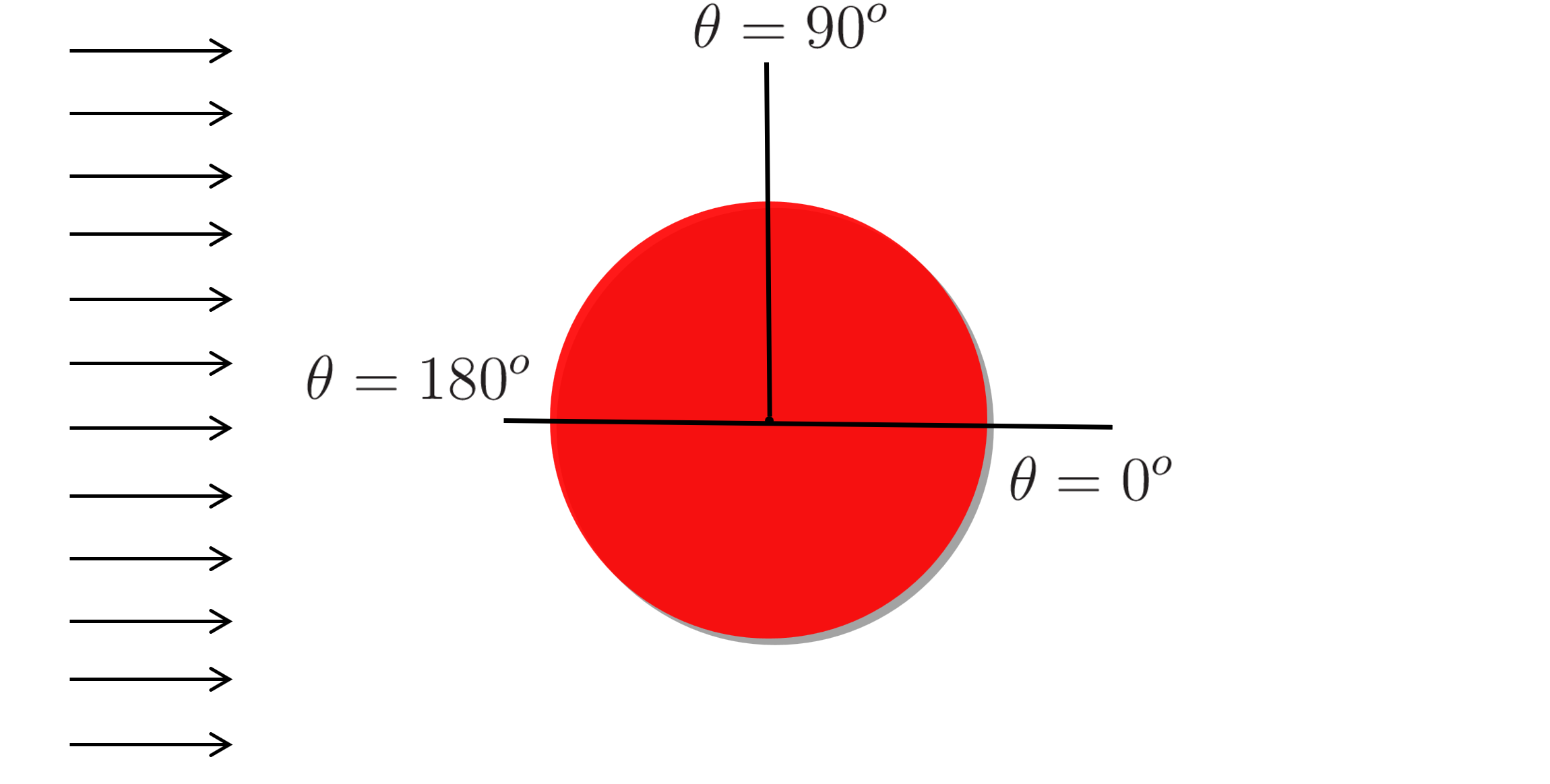}\\
			\vspace{1mm}
			\includegraphics[width=.31\textwidth ]{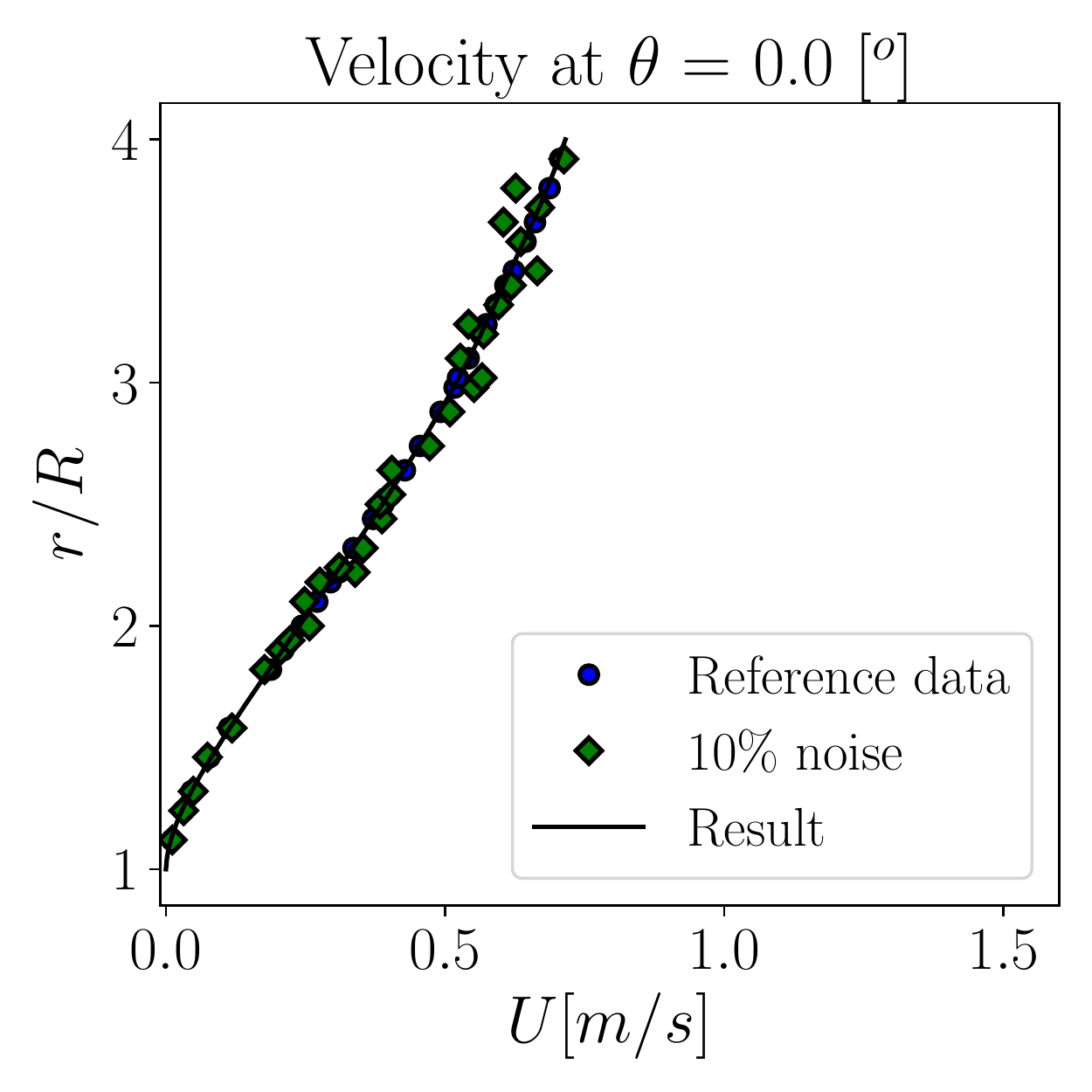}
			\hspace{1mm}
			\includegraphics[width=.31\textwidth ]{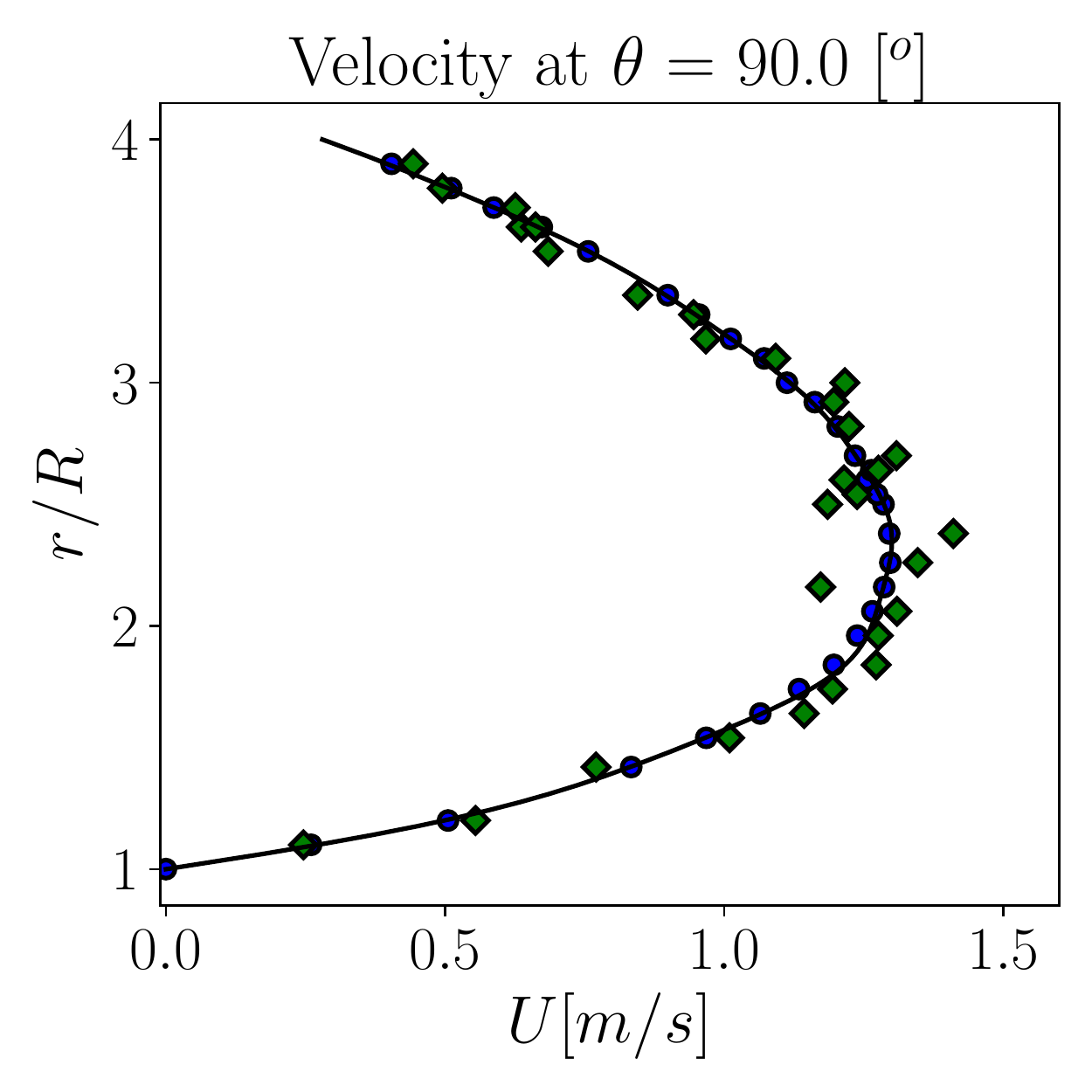}
			\hspace{1mm}
			\includegraphics[width=.31\textwidth ]{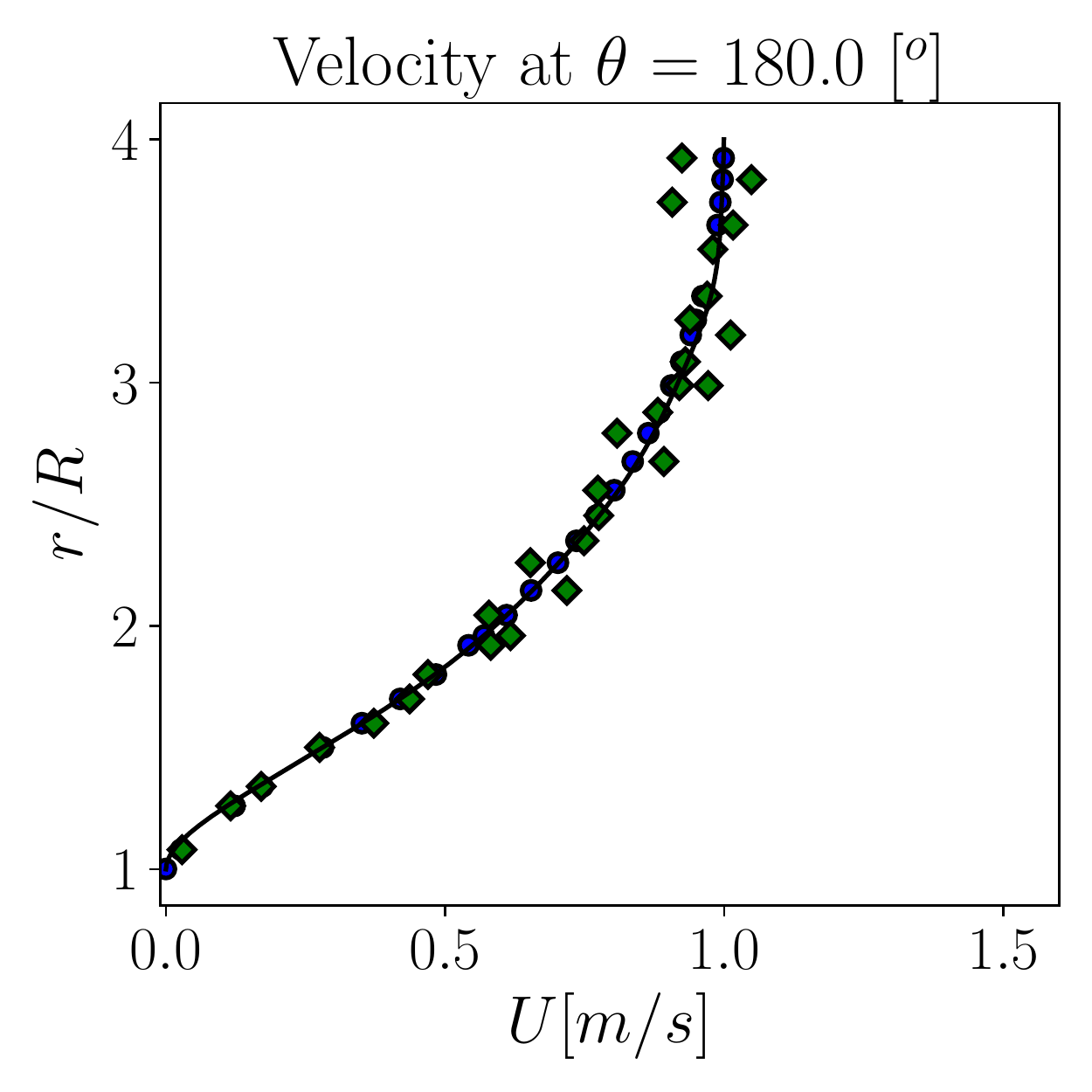}\\
			\includegraphics[width=.31\textwidth ]{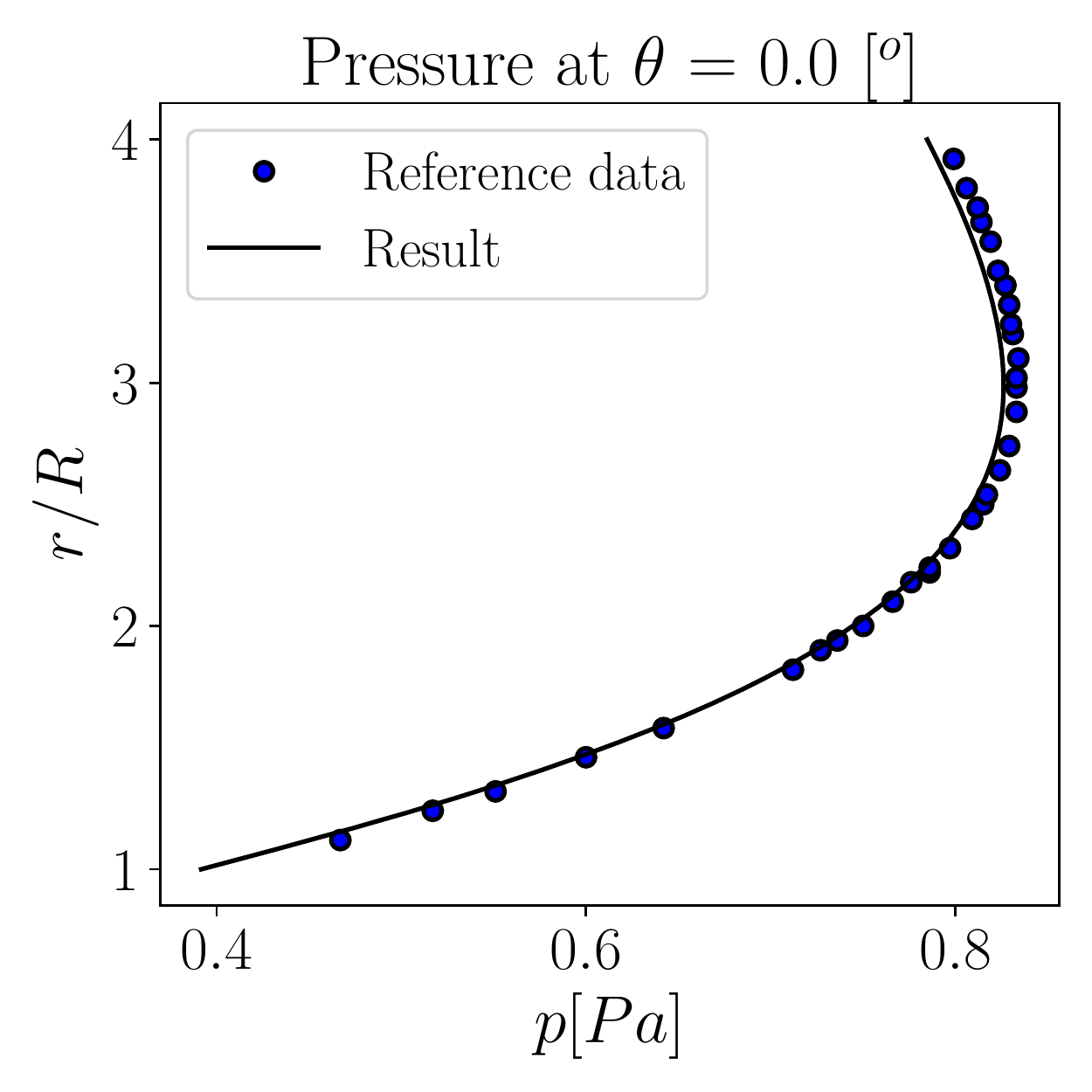}
			\hspace{1mm}
			\includegraphics[width=.31\textwidth ]{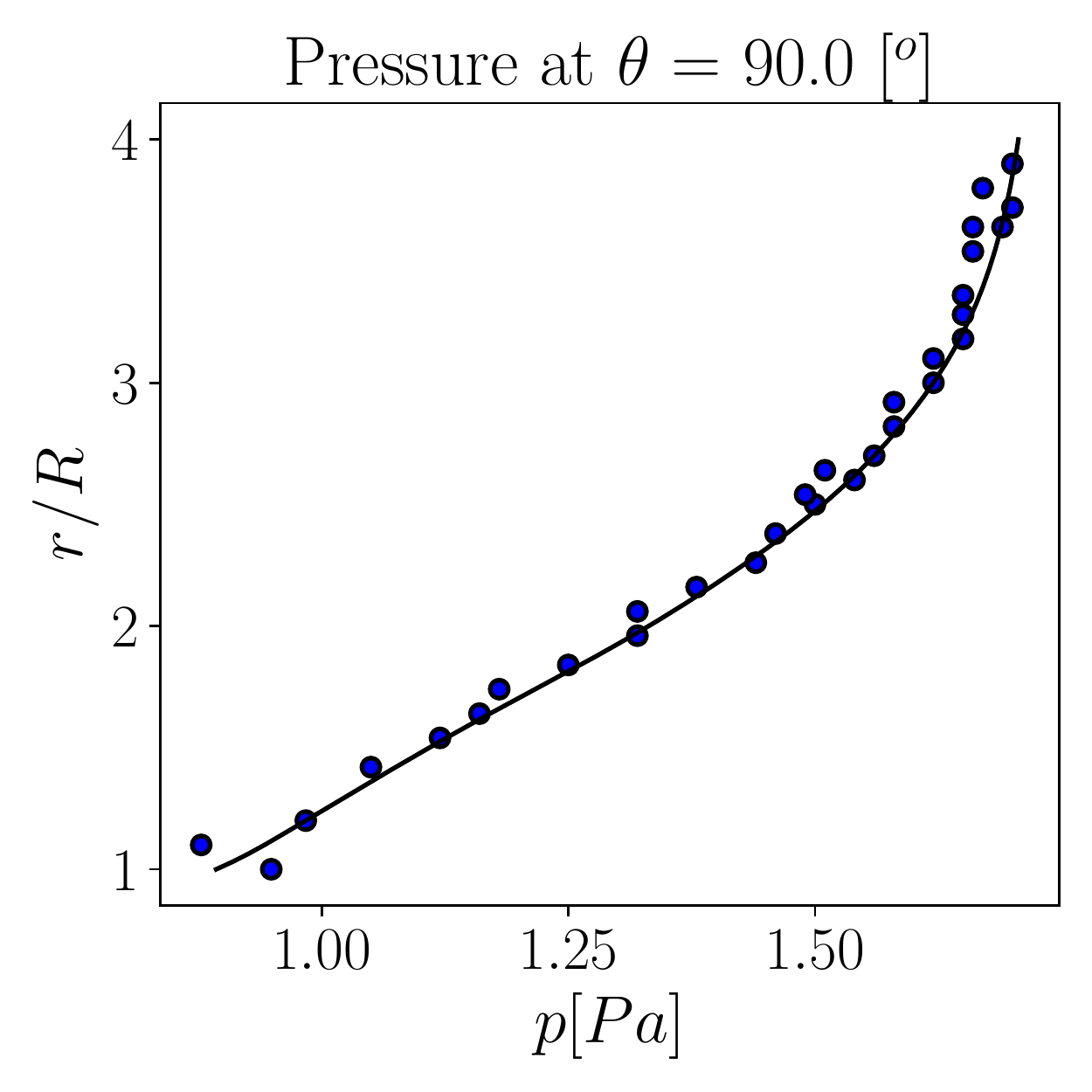}
			\hspace{1mm}
			\includegraphics[width=.31\textwidth ]{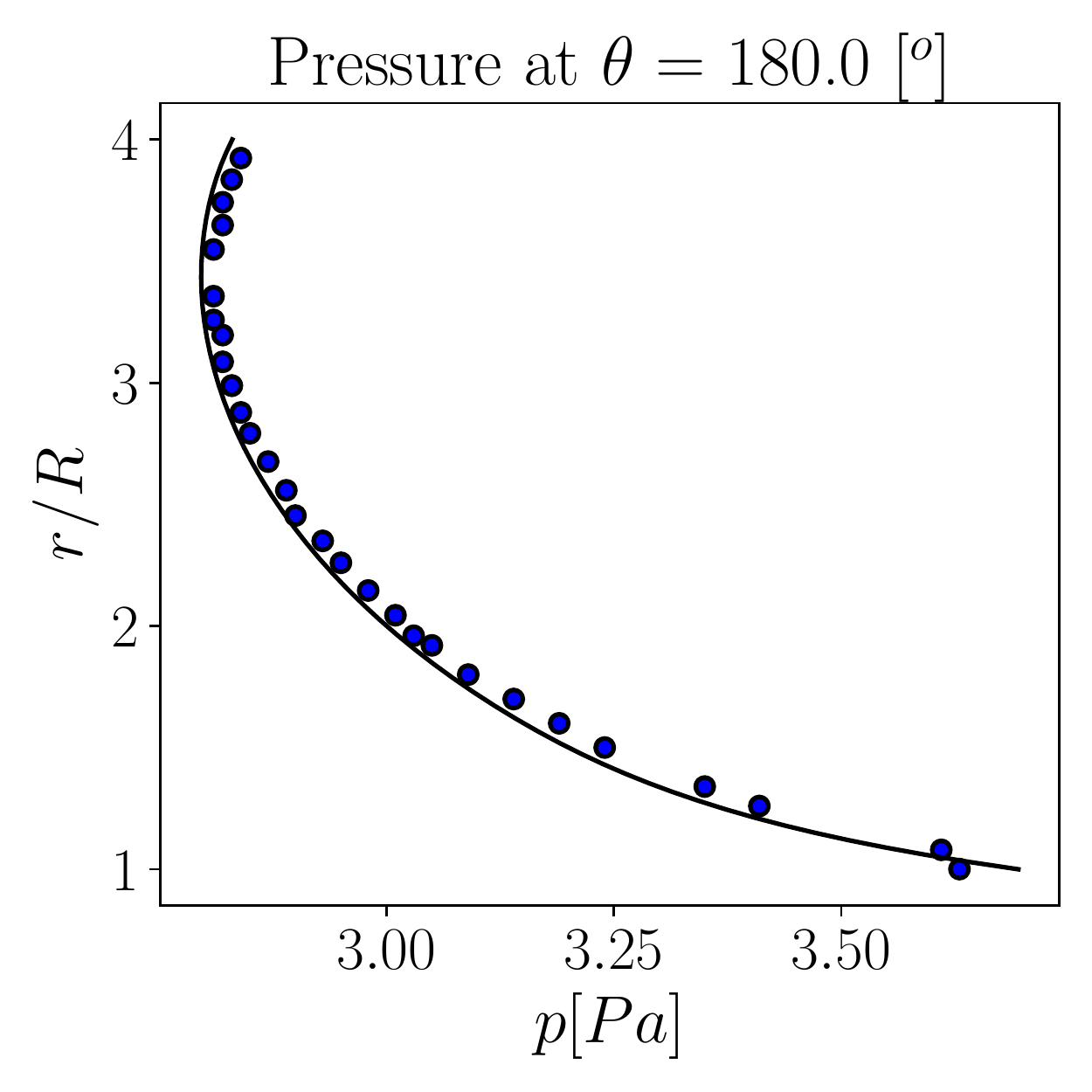}
			\caption{Test case 2. Velocity profiles (second row) and pressure profiles (third row)  extracted at three planes with $\theta=0^o,90^o,180^o$. The top panel recalls the flow orientation. Round blue markers are used for the CFD data, while green diamonds are used for the data employed in the (noisy) regression. The (analytic) result of the regression is shown with a continuous line. } 
			\label{Res_2_Cyl_vel_P}
		\end{figure}

		Considering a case with a significant amount of noise ($q=0.1$) and the largest seeding density ($N_p=0.089$, corresponding to $n_p=18755$), Figures \ref{Res_1_Cylin} shows the distribution of the error for the velocity regression (top) and the pressure integration (bottom). Both provide excellent results in the entire domain, with errors (defined as in eq. \ref{errors}) of the order of $1\%$. For the same seeding and noise conditions, Figure \ref{Res_2_Cyl_vel_P} offers a closer view of the velocity regression and the pressure computation in the most challenging region, namely close to the cylinder walls. Three profiles are extracted at different angles for the velocity magnitude and the pressure. For each of the lines identified by the polar angle $\theta$, velocity and pressure values are taken as the ones lying within the two planes, parallel to the radial direction and $\pm 2 mm$ apart from the original line. The velocity and pressure available from the RBFs expansion are analytical and computed on the associated planes. The comparison is thus primarily qualitative, as the reference data is not taken precisely at the sampled planes.
		
		For the velocity regression, the figures shows both the available CFD data (blue circular markers) and the data used for the regression (green diamonds markers), which was polluted by noise. The results shows that the regression removes the noise and provide an analytical solution (black continuous line) that satisfies the boundary conditions. This is in excellent agreement with the CFD data, where available, and shows a realistic fit in the remaining portions.

		\begin{figure}[ht]
			\centering
			\includegraphics[width=.27\textwidth ]{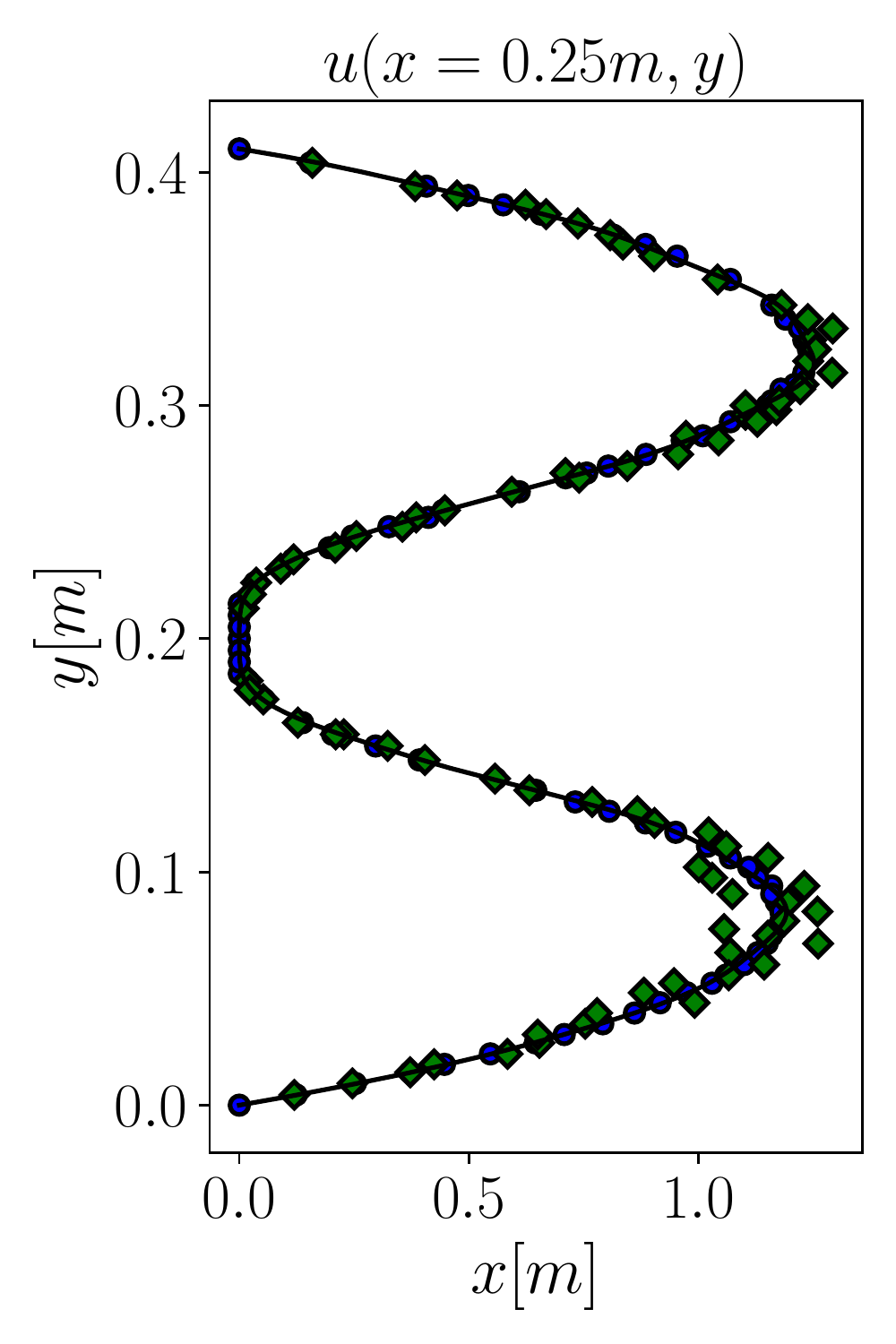}
			\hspace{1mm}
			\includegraphics[width=.27\textwidth ]{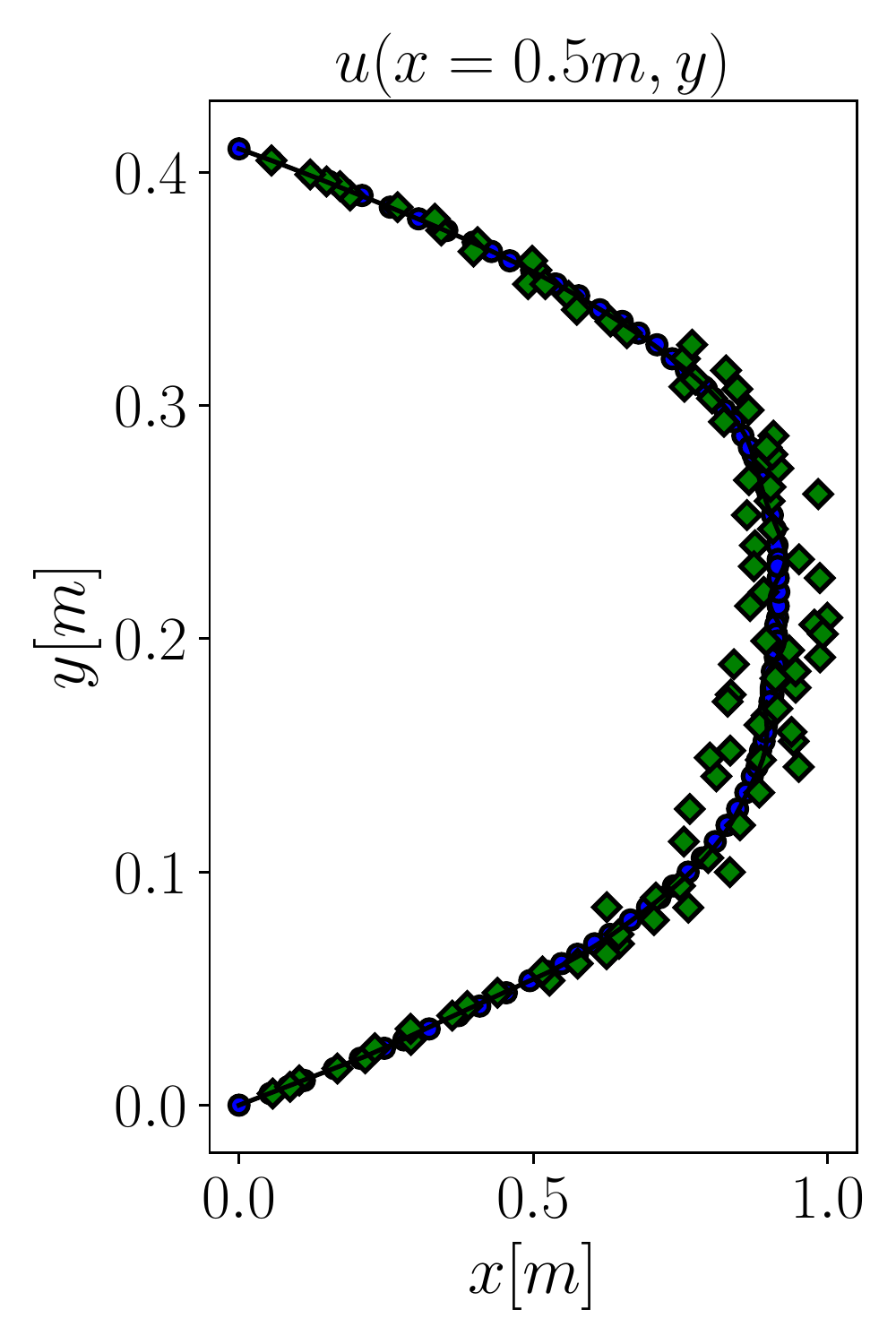}
			\hspace{1mm}
			\includegraphics[width=.27\textwidth ]{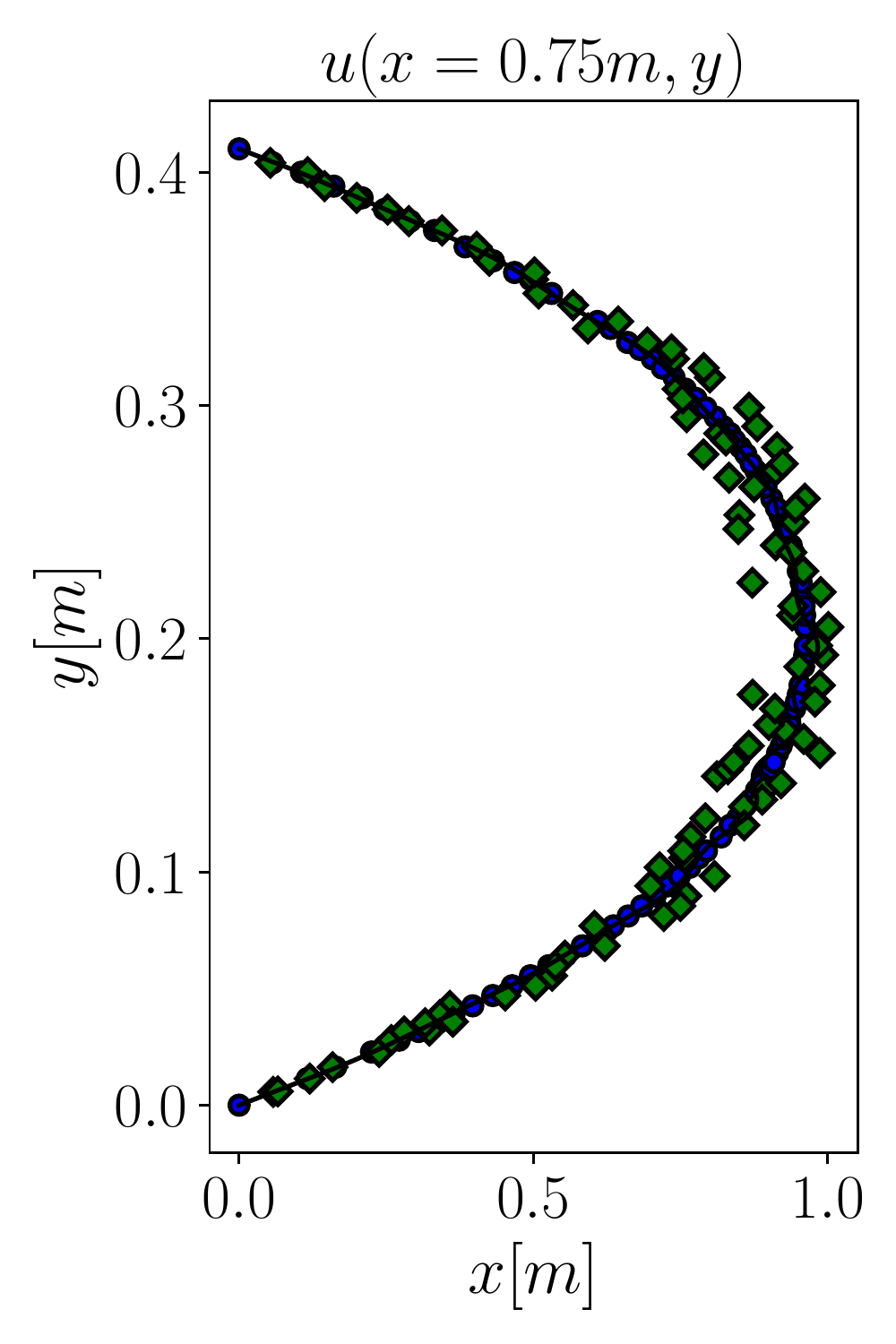}
			\caption{Test case 2. Vertical velocity profiles (u component only) taken at $x=0.25,0.5,0.75$[mm]. The same legend used in Figure \ref{Res_2_Cyl_vel_P} is used. } 
			\label{Vel_Profiles}
		\end{figure}
		
		
		The vertical velocity profiles far from the cylinder walls are shown in Figure \ref{Vel_Profiles} for three positions, namely $x=0.25,0.5,0.75$. The matching with the reference data is remarkable. To evaluate the accuracy of the proposed approach, it is worth comparing the errors obtained in this work with the errors obtained via Physics Informed Neural networks (PINNS) by \cite{Rao2020}. Considering the case in which all the dataset is used, the number of training data used by the proposed RBF regression is equal to $n_p=18755$ while \cite{Rao2020} uses $n_b=50000$ points were used in \cite{Rao2020} using a Latin hypercube sampling (LHS). Yet, the global error obtained by the RBF regression of the velocity reconstruction is $10$ times smaller ($E_U=0.014$ versus the $E_U=0.14$) than the error obtained via ANN-regression in \cite{Rao2020}). No information was found concerning the global error of the pressure field.
		
\begin{figure}[ht]
			\centering
			\includegraphics[width=0.55\columnwidth]{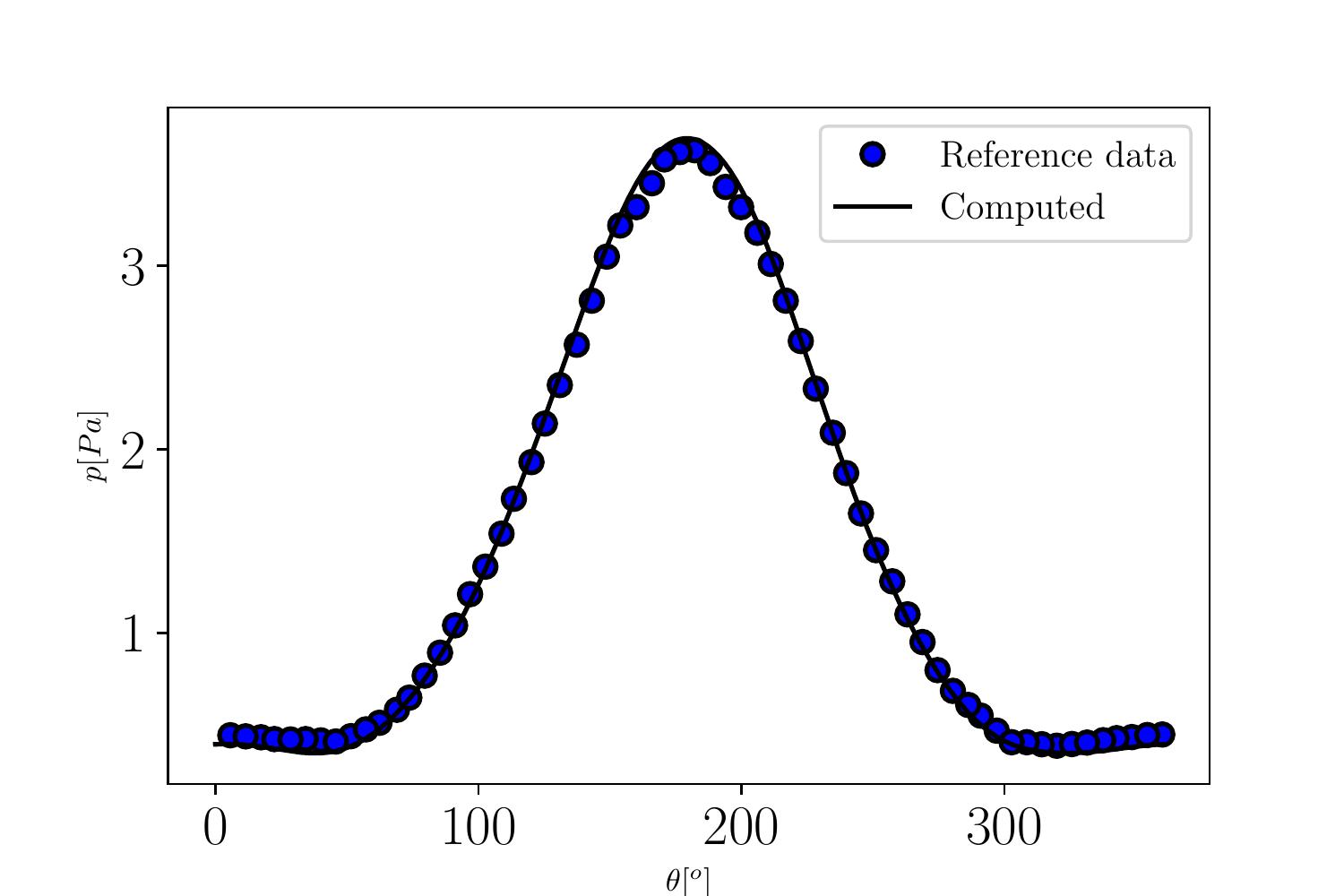}
			\caption{Test case 2. Pressure distribution around the cylinder wall, comparing the available CFD data (blue round markers) with the results of the mesh-less integration (black continuous line).}
			\label{Wall_P}
		\end{figure}

		Finally, Figure \ref{Wall_P} shows the pressure distribution around the cylinder walls in polar coordinates, comparing the results of the pressure integration (black continuous line) with the available CFD data. The results are in good agreement.

		\subsection{Test Case 3: The 3D Stokes Flow past a Sphere}\label{sec:4p3}
		
		We consider a case with $n_p=18300$ in the domain $\hat{r}\in[0.5,1]$, i.e. a thick spherical shell bounded on one side by the solid sphere. Using a clustering scheme with $\boldsymbol{n}_K=[6,10,20]$ results in $n_b=10381$ RBFs. A total of 2111 constraints are placed on the sphere's surface at $\hat{r}=0.5$ and 4879 on the outer surface at $\hat{r}=1$. These points are used to impose the divergence-free conditions and the points at the wall are also used to impose $\boldsymbol{U}=0$ at $\hat{r}=0.5$. This results in $n_D=2111$ and $n_{\nabla}=6990$, thus a total of  $n_{\lambda}=3 n_D+ n_{\nabla}=13323$ constraints for the velocity regression (cf. Table \ref{Table_I}). In addition to these, a large penalty of $\alpha_{\nabla}=25$ is used to promote the divergence-free condition over the entire domain. This results in a sub-optimal reconstruction of the velocity field but helps in the pressure integration.

		\begin{figure}[ht]
			\centering
			\begin{subfigure}[\label{Stokes_1a}]{
					\includegraphics[width=0.47\textwidth]{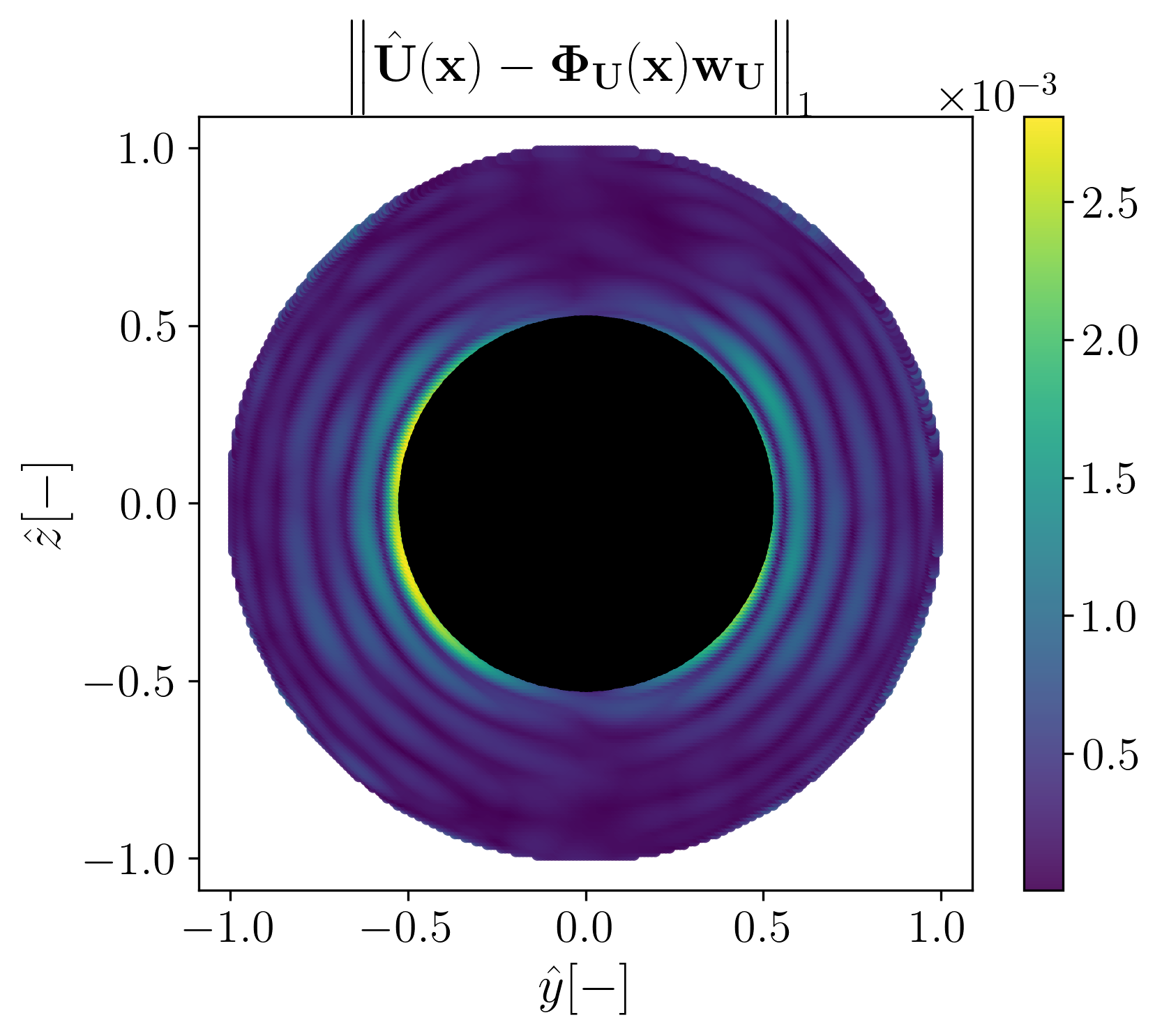}
				}
			\end{subfigure}
			\begin{subfigure}[\label{Stokes_1b}]{
					\includegraphics[width=0.47\textwidth]{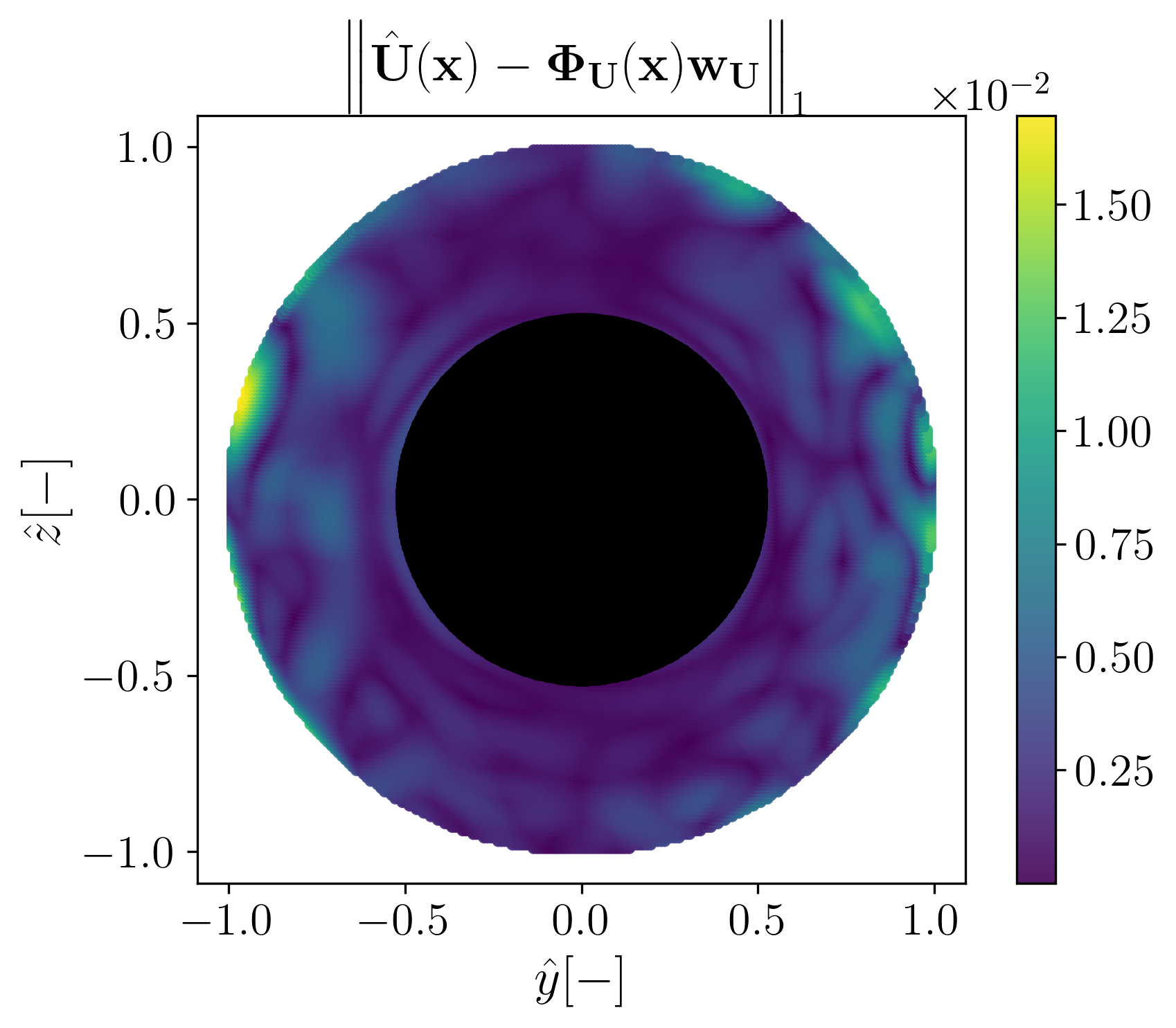}
				}
			\end{subfigure}
			\begin{subfigure}[\label{Stokes_3a}]{
					\includegraphics[width=0.47\textwidth]{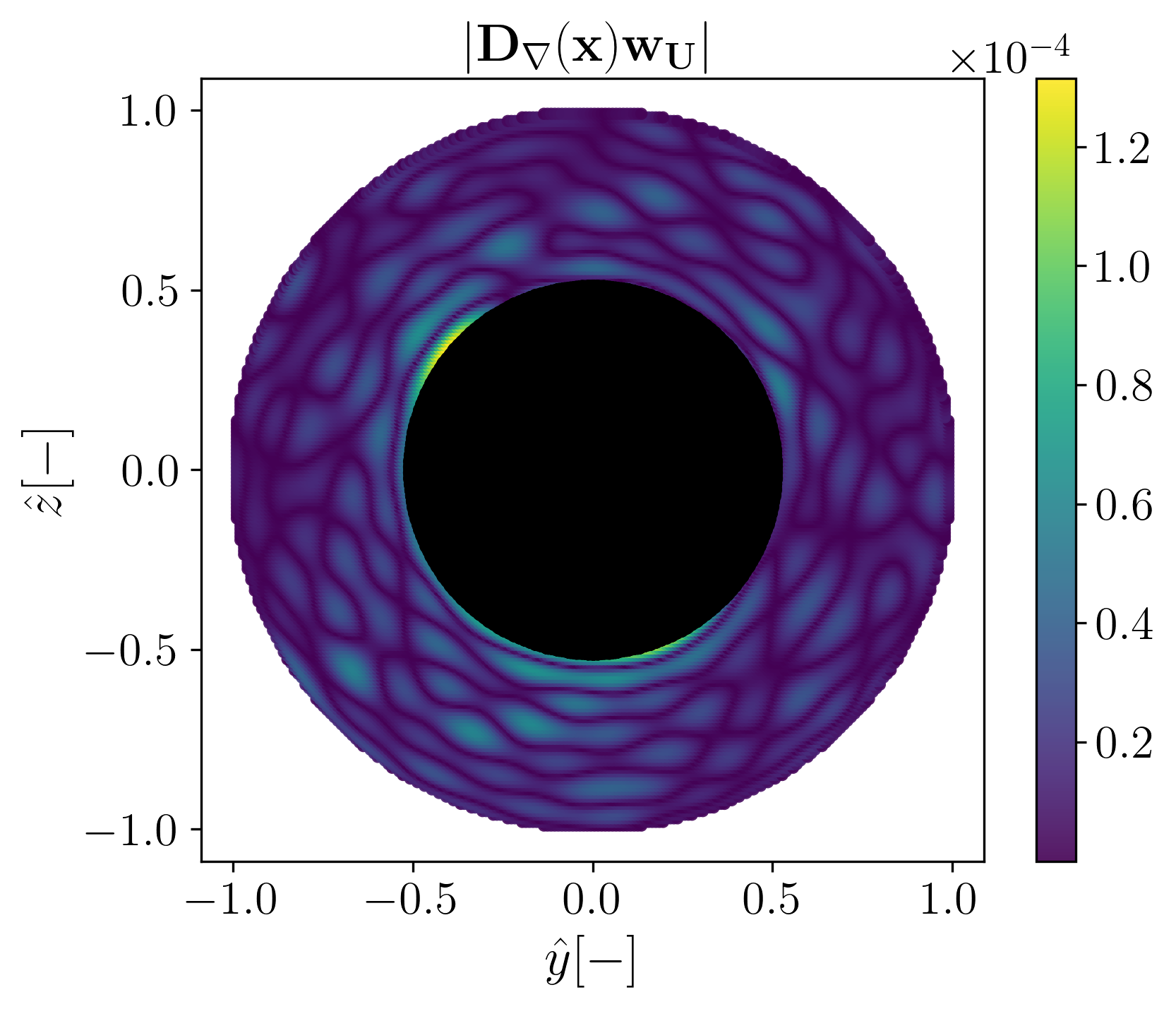}
				}
			\end{subfigure}
			\begin{subfigure}[\label{Stokes_3b}]{
					\includegraphics[width=0.47\textwidth]{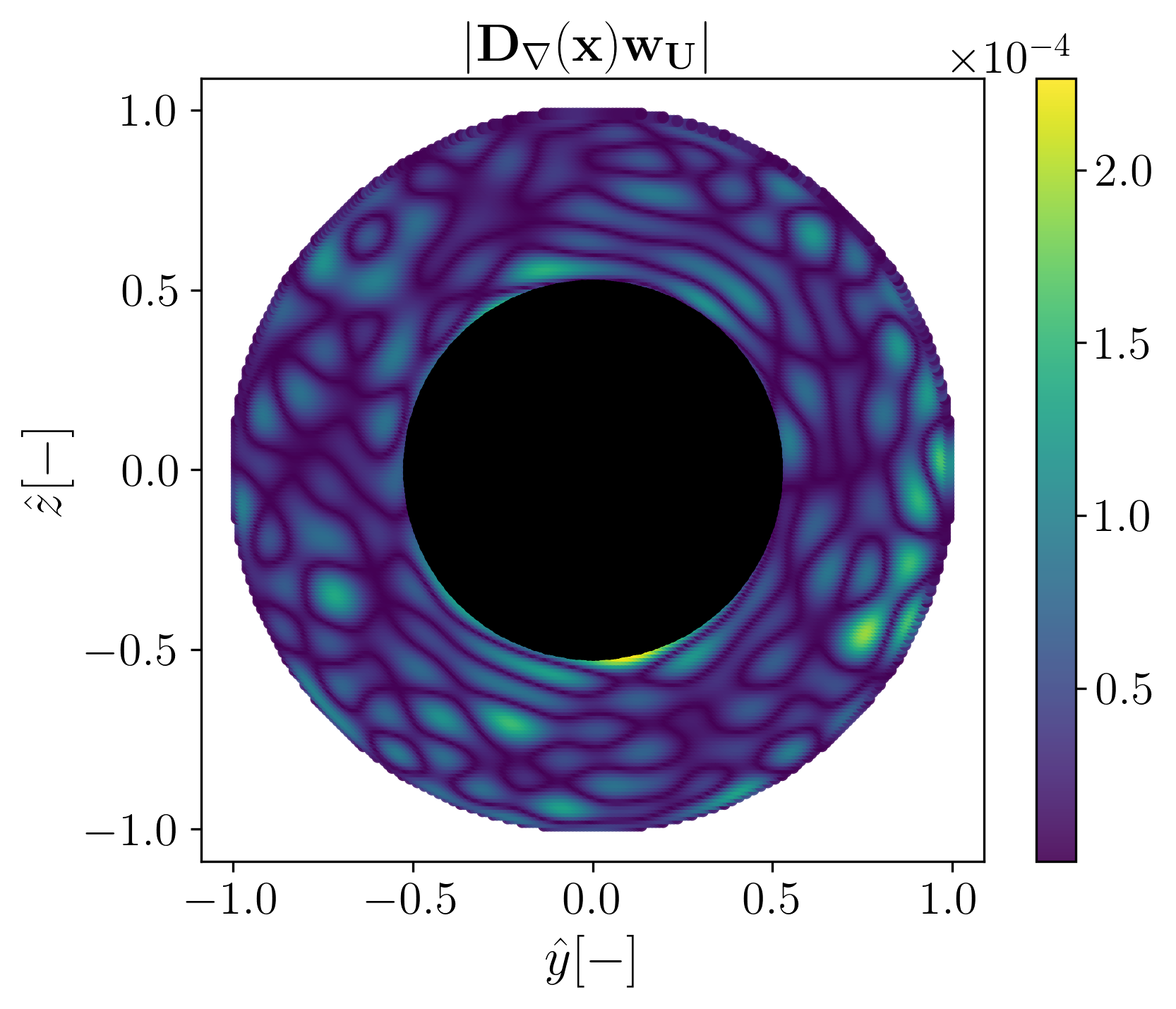}
				}
			\end{subfigure}
			\caption{Test case 3. Local absolute error distribution for the velocity regression. Fig. \ref{Stokes_1a} shows the results for the noiseless ($q=0$) test case, Fig. \ref{Stokes_1b} for the noisy ($q=0.05$) test case. Fig. \ref{Stokes_3a} and \ref{Stokes_3b} show the corresponding distribution of error for the flow divergence.} 
			\label{Stokes_1}
		\end{figure}

		The test case is analyzed with $q=0$ (no noise) and $q=0.05$. The local absolute error distribution is shown in Figure \ref{Stokes_1} for both ($q=0$ on the left and $q=0.05$ on the right). The global error for these test cases are $E_U=0.1\%$ and $E_U=0.6\%$ respectively. While in the first case the absolute errors are mostly produced next to the sphere's wall, in the second case these are mostly occurring in the outer surface. Nevertheless, the error is everywhere negligible in absence of noise and acceptable in the presence of noise. 
		Figures \ref{Stokes_3a} and \ref{Stokes_3b} show the corresponding absolute error for the divergence of the flow. These are everywhere negligible, confirming the quality of the regression over the full domain.
		
\begin{figure}[ht]
			\centering
			\begin{subfigure}[\label{Stokes_2a}]{
					\includegraphics[width=0.47\textwidth]{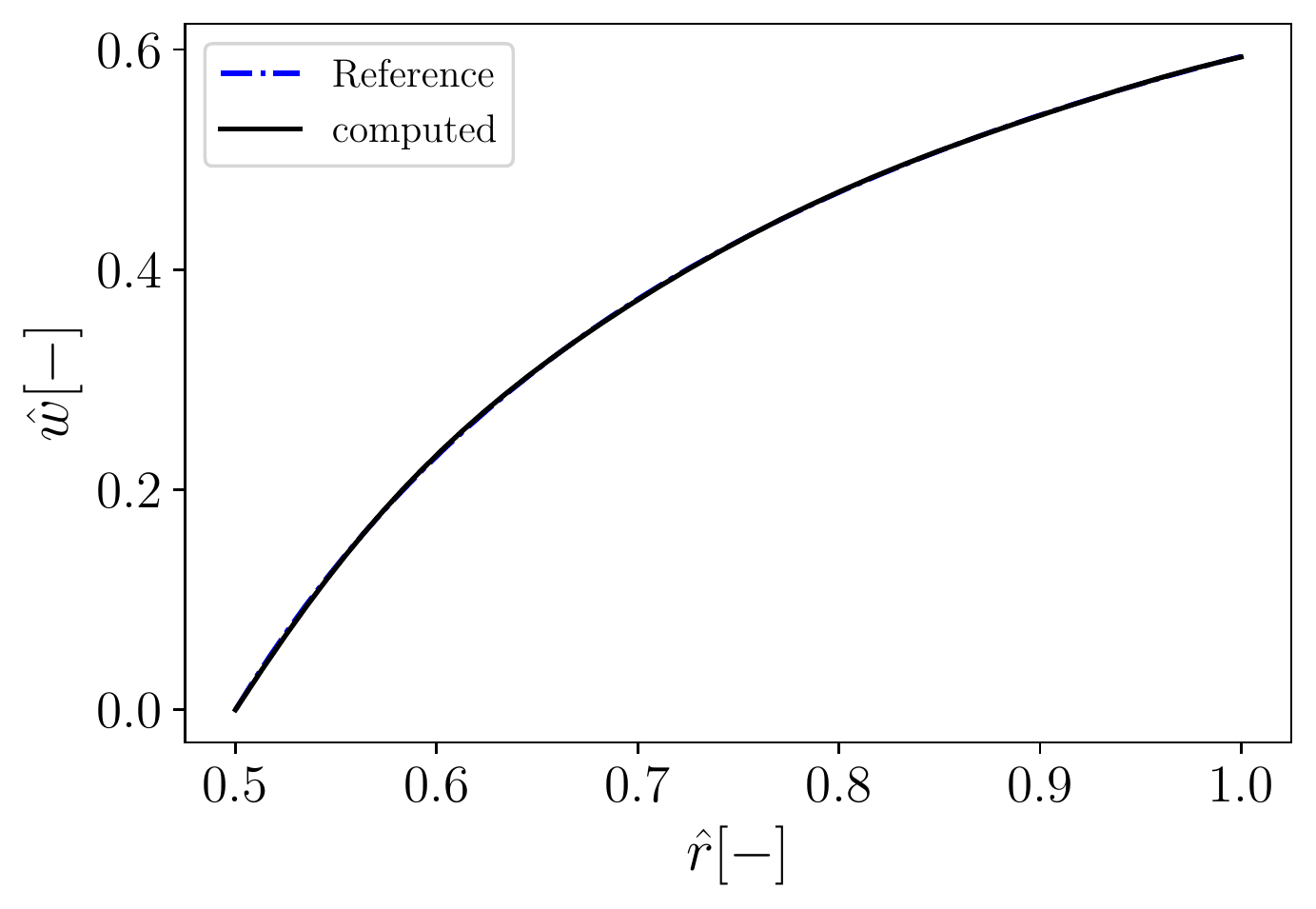}
				}
			\end{subfigure}
			\begin{subfigure}[\label{Stokes_2b}]{
					\includegraphics[width=0.47\textwidth]{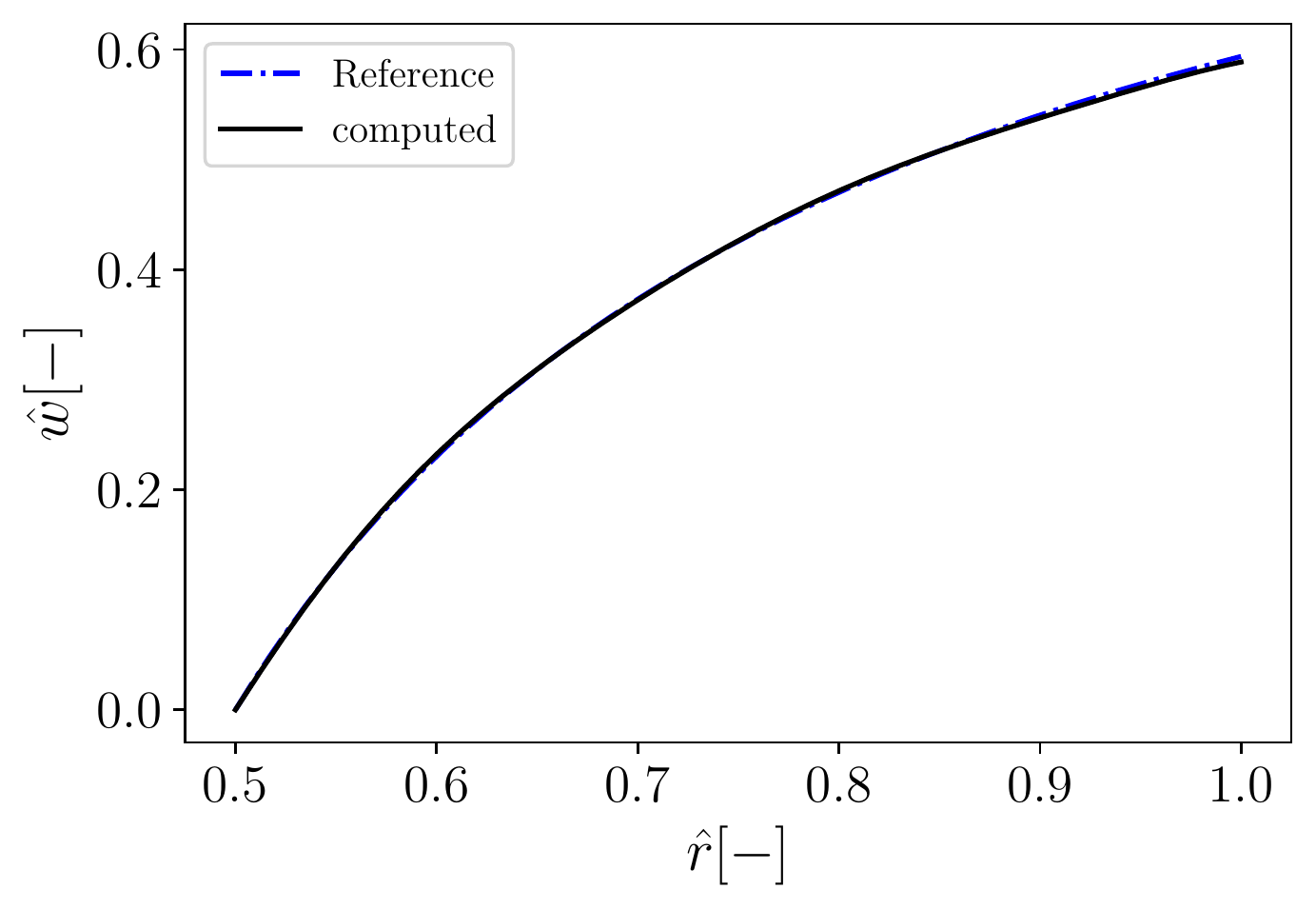}
				}
			\end{subfigure}
			
			\caption{Test case 3. Velocity profile at $\theta=0^o$ and $\varphi=0^o$, from $\hat{r}=0.5$, for the noiseless (Fig. \ref{Stokes_2a}) and the noisy (Fig. \ref{Stokes_2b}) test cases. The reference data (blue dashed-dotted line) and the RBF regression (black continuous line) are nearly indistinguishable. }
			\label{Stokes_2}
		\end{figure}

		To further illustrate the quality of the regression, Figure \ref{Stokes_2} shows the stream-wise velocity component at $\theta=0^o$ and $\varphi=0^o$, i.e. on an equatorial plane (taking the south pole on the stagnation point). The RBF regression (black continuous line) is indistinguishable from the analytical data (blue dashed-dotted line).

		The resulting fields were used to compute the pressure field, as usual, using the same RBF basis used for the velocity field. The same points used to constrain the velocity regression were used to impose Neumann conditions (cf. equation \eqref{Projection_P}) on the sphere's wall and on the outer surface. In addition to those, $6$ points were used to impose Dirichlet conditions on the sphere's surface. These points mimic pressure taps located on the sphere's wall and greatly enhance the stability of the pressure computation. \textcolor{black}{The importance of having Dirichlet conditions on some of the boundaries has been highlighted by \cite{Pan2016,Faiella2021,Pan2018}, who have thoroughly analyzed the error propagation in the pressure integration near boundaries with Neumann conditions. This explains why the natural approach for integrating the pressure in the flow past a blunt body consists in setting Dirichlet boundary conditions in the far-field (see for example \cite{McClure2017a} and the works cited in table 1 from \cite{Pan2016}), where the flow irrotationality allows for using the Bernoulli theorem; this approach is however not possible in this test case.}

		Reducing the number of pressure taps to $1$ has a detrimental effect on the pressure integration unless a much larger number of constraints is added on the velocity regression \textcolor{black}{(in line with the observations by \cite{Faiella2021})}. We leave a detailed analysis on the impact of number and location of Dirichlet boundary conditions (i.e. pressure taps) to future work and \textcolor{black}{we encourage the reader considering this test case in future studies on the pressure integration from image velocimetry.}
		
\begin{figure}[ht]
			\centering
			\begin{subfigure}[\label{Stokes_4a}]{
					\includegraphics[width=0.47\textwidth]{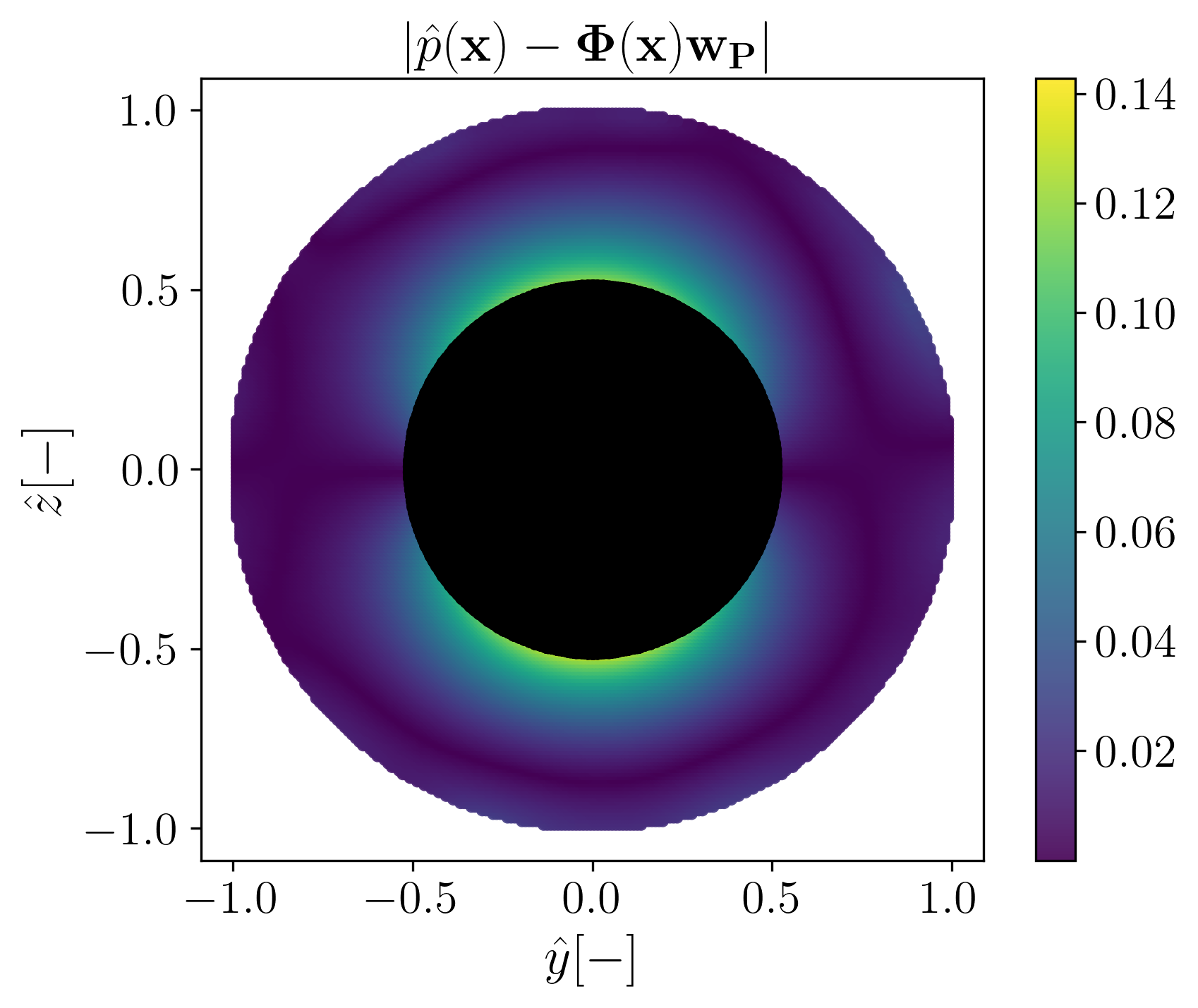}
				}
			\end{subfigure}
			\begin{subfigure}[\label{Stokes_4b}]{
					\includegraphics[width=0.47\textwidth]{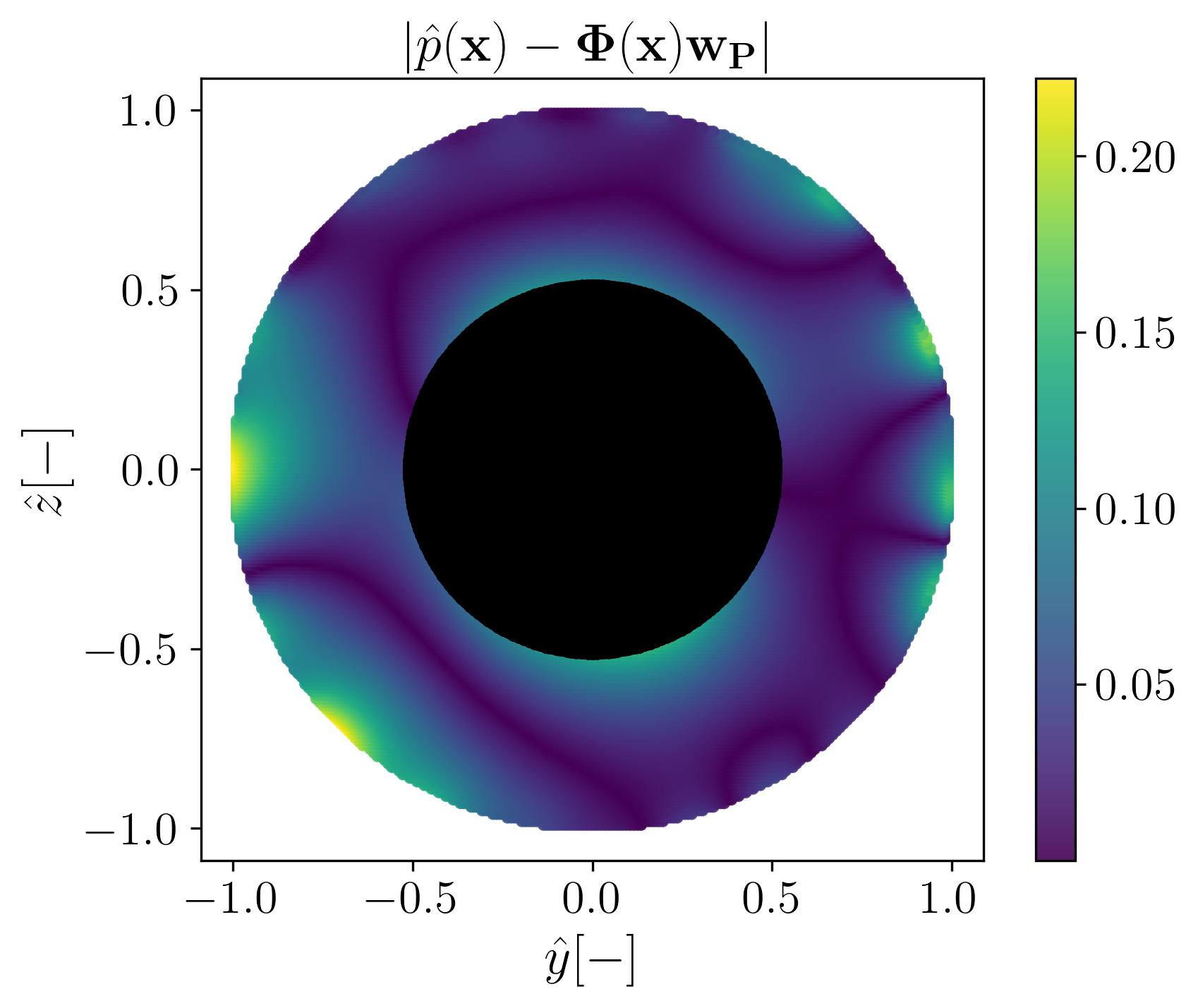}
				}
			\end{subfigure}
			\caption{Test case 3. Local absolute error distribution for the pressure field. Fig. \ref{Stokes_4a} shows the results for the noiseless ($q=0$) test case, Fig. \ref{Stokes_4b} for the noisy ($q=0.05$) test case.} 
			\label{Stokes_4}
		\end{figure}

		The local absolute error for the pressure field is shown in Figure \ref{Stokes_4}. These are characterized by a global error of $E_P=3.2\%$ and $E_P=8.2\%$ respectively. Although the differences in the velocity fields between the two cases are minor, the impact on the pressure integration is considerable. In both cases, the pressure error is located where the highest velocity error is produced (cf. Fig \ref{Stokes_1}), namely on the surface wall for the noiseless test case (Fig. \ref{Stokes_4a}) and on the outer surface for the noisy test case  (Fig. \ref{Stokes_4b}). While for the noiseless test case the impact is negligible, in the noisy test case the error is important far from the sphere's wall.

\begin{figure}[ht]
			\centering
			\begin{subfigure}[\label{Stokes_5a}]{
					\includegraphics[width=0.47\textwidth]{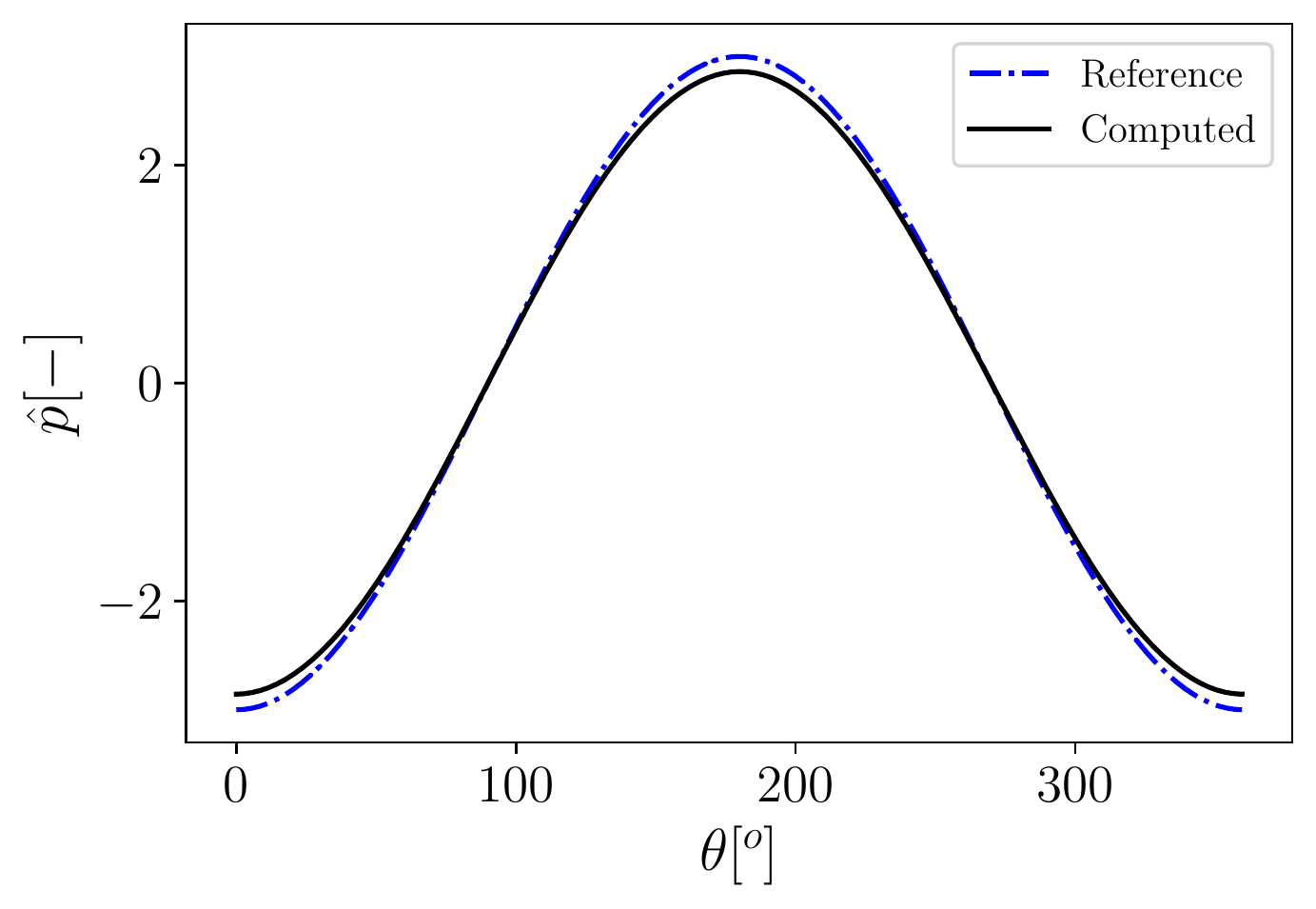}
				}
			\end{subfigure}
			\begin{subfigure}[\label{Stokes_5b}]{
					\includegraphics[width=0.47\textwidth]{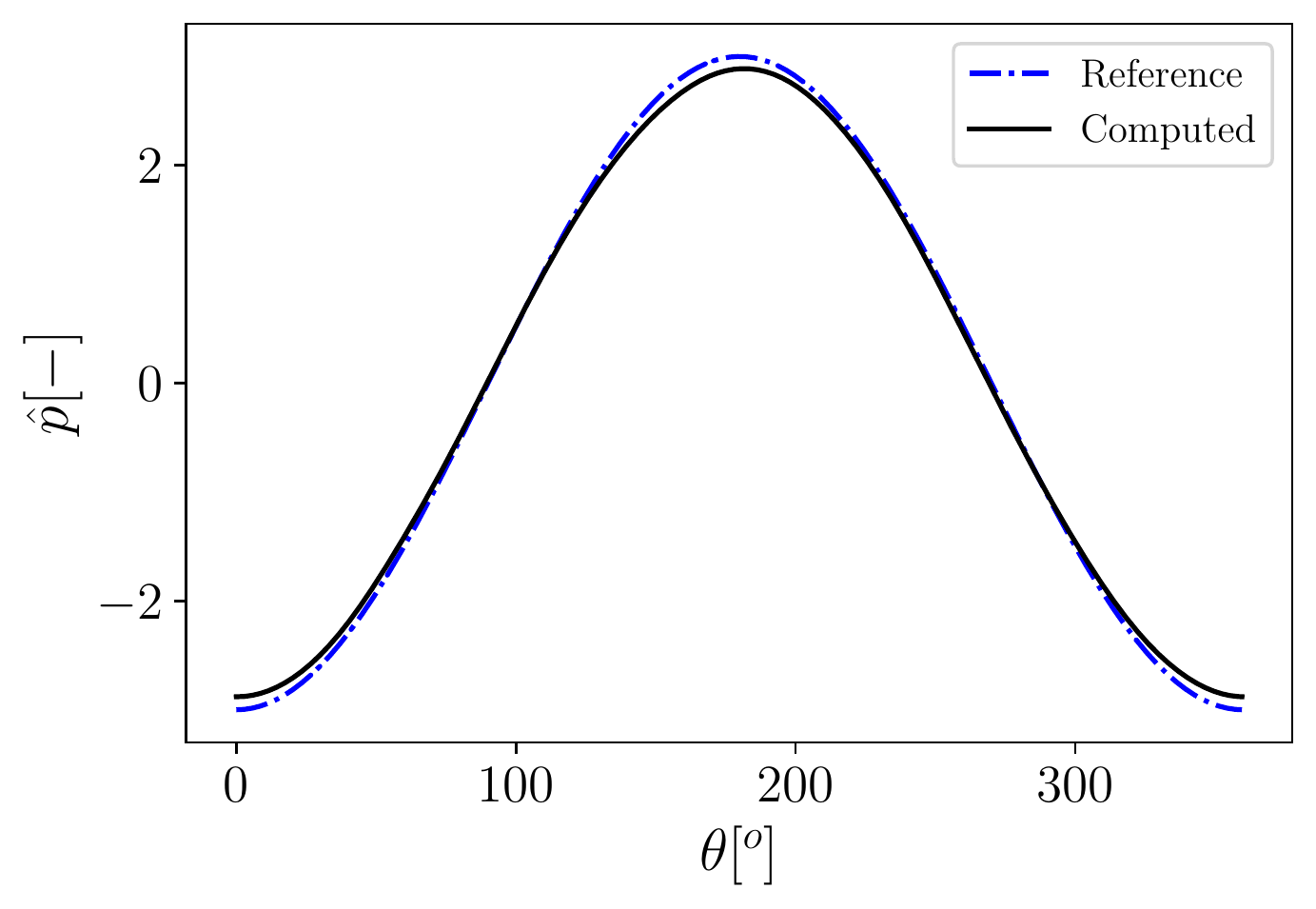}
				}
			\end{subfigure}\\
			\begin{subfigure}[\label{Stokes_6a}]{
					\includegraphics[width=0.47\textwidth]{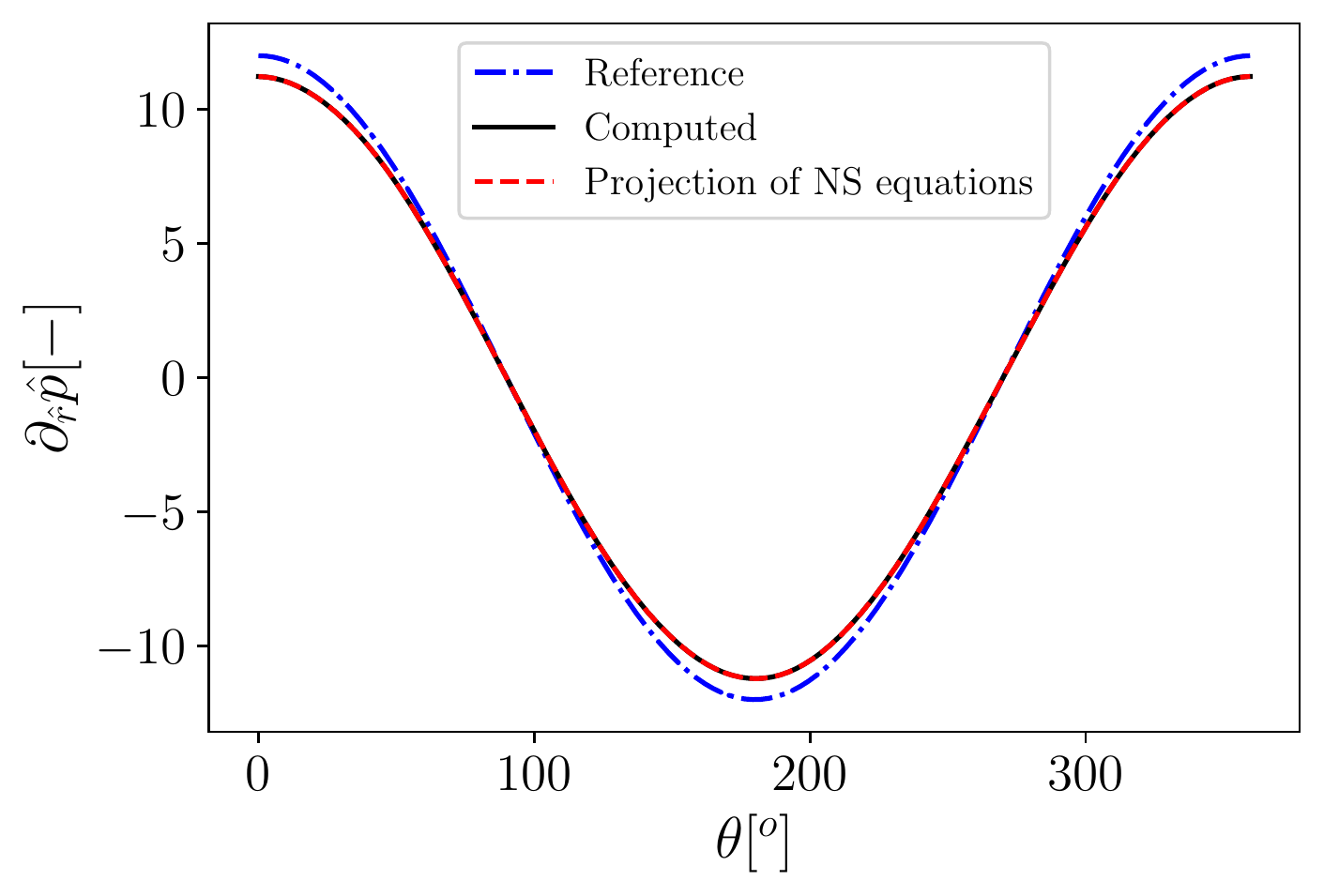}
				}
			\end{subfigure}
			\begin{subfigure}[\label{Stokes_6b}]{
					\includegraphics[width=0.47\textwidth]{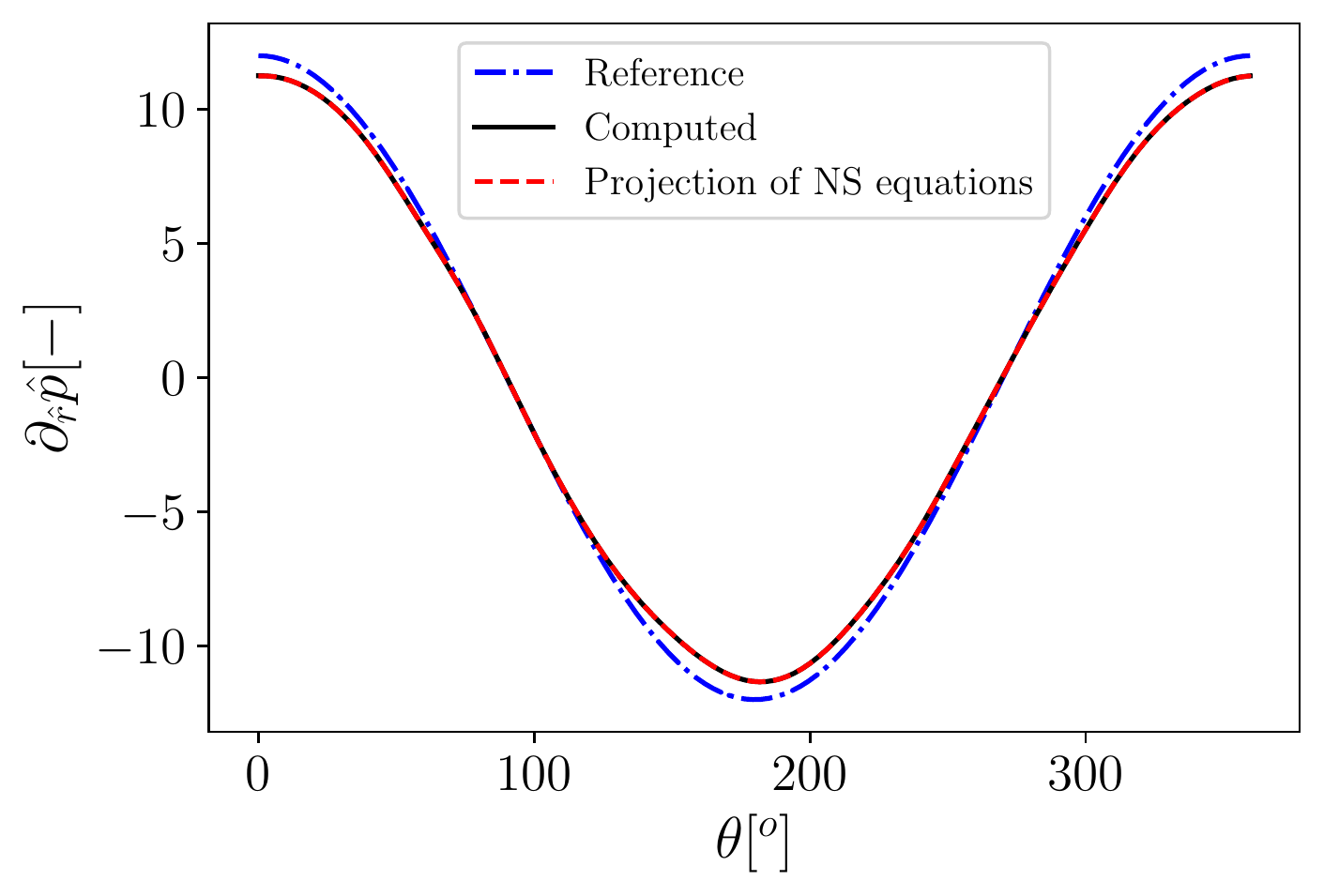}
				}
			\end{subfigure}
			\caption{Test case 3. Figs. \ref{Stokes_5a} and \ref{Stokes_5b} show the pressure distribution along the sphere's wall for the noiseless and the noisy case respectively. Similarly, \ref{Stokes_6a} and \ref{Stokes_6b} show the wall-normal pressure gradient.} 
			\label{Stokes_5}
		\end{figure}

		Nevertheless, it is worth highlighting that the pressure reconstruction is satisfactory in both cases in the proximity of the sphere, thanks to the contribution of the Dirichlet boundary conditions. Figure \ref{Stokes_5} shows the pressure distribution along the sphere for the noiseless and the noisy test cases. These are plotted as a function of $\theta$ because of the perfect axial symmetry of the regression. The impact of the noise appears negligible. 
		
		
		Finally, Figure \ref{Stokes_6a} and \ref{Stokes_6b} show the wall normal pressure gradient along the sphere's wall. In addition to the reference data (dashed-dotted blue line) and the RBF regression (black continuous line), the continuous red line shows the projection of the RBF velocity field according to equation \eqref{Projection_P}. The excellent agreement between these two shows that the (minor) discrepancy with respect to the reference data is due to the (minor) discrepancy in the reconstruction of the velocity field and not the pressure integration itself.

		\section{Conclusions and Perspectives}\label{sec:5}
		
		We presented a constrained Radial Basis Function (RBF) method for the regression of velocity fields and for the meshless computation of pressure fields in incompressible flows. While the derivation focuses on steady laminar flows, its extension to unsteady and turbulent flows is relatively straightforward and is currently under development. Moreover, the proposed methodology applies identically if the velocity data is available on Cartesian grids as in cross-correlation based velocimetry or on scattered points as in tracking-based velocimetry.
		
		We presented all the details of the mathematical framework and showed that both the velocity regression and the pressure integration can be cast as quadratic problems with linear constraint, resulting in two classic Karush-Kuhn-Tucker (KKT) systems. A simple direct solution method, based on Schur complements and Cholesky factorizations, was introduced together with a hierarchical clustering technique to automatically compute RBFs' collocation points and shape factors.
		
		The constraints are used to impose boundary conditions (e.g. no-slip) and physical priors (e.g. divergence-free or compliance with the Navier-Stokes equations). The result is an analytic expression for both velocity and pressure fields. Constraints allow a ``physics-informed'' regression, similar to what is done in ``physics-informed'' artificial neural networks (ANN). However, the RBF-based regression is considerably simpler than the ANN-based regression, and the conditions are introduced as hard constraints (with Lagrange multipliers) and not penalties.
		
		We illustrate the RBF approach on three test cases of growing complexity, from a small 2D problem to a fairly large 3D problem. All the selected test cases are numerical and therefore offer the possibility of comparing the result with reference data, validating the method and testing its robustness against noise and seeding concentration. The application to experimental data is currently under development and will be presented in a dedicated article.
		
		The results of the velocity regression proved to be remarkably robust in all test cases and all investigated conditions. On the other hand, the pressure integration proved to be strongly sensitive to small errors in the velocity regression if these occur in the proximity of boundaries at which Neumann boundary conditions are to be imposed from the Navier-Stokes equation. Nevertheless, even in the worst investigated test case (with $8\%$ of global error), the pressure reconstruction near solid boundaries (where some Dirichlet boundary conditions might be applied) is excellent. 
		
		Besides the aforementioned extension to unsteady and turbulent flows and the implementation of experimental data, ongoing work is focused on reducing the memory requirements following two research paths. On the one hand, by implementing iterative methods to solve the least square problems. On the other hand, developing techniques to promote sparsity of the basis matrix, combining compact supports RBFs with the Partition of Unity Method (PUM). 
		
		\appendix
		
		\section{Velocity Approximations in 3D}\label{AppendixA}
		In a 3D problem, the collected velocity field can be arranged in the column vector $\boldsymbol{U}=\left(\boldsymbol{u};\boldsymbol{v};\boldsymbol{w}\right) \in\mathbb{R}^{3n_p}$, where $\boldsymbol{u},\boldsymbol{v},\boldsymbol{w}\in\mathbb{R}^{n_p}$ collects the entries of the three velocity components $U_i=(u_i,v_i,w_i)$. In this section, the symbol $\boldsymbol{w}$ refers to the velocity in the $z$ direction and not the vector of weights as in \eqref{approx}. The definition of the collocation points are adapted from the 2D case to $\boldsymbol{x}_k^*=\left(x_k^*,y_k^*,z_k^*\right)$. Equation \eqref{approx_U} is extended to:
		\begin{equation}
			\boldsymbol{\Tilde{U}}=\left(  \begin{array}{c}
				\boldsymbol{\Tilde{u}}   \\
				\boldsymbol{\Tilde{v}}  \\
				\boldsymbol{\Tilde{w}}
			\end{array}\right) = \left(  \begin{array}{ccc}
				\boldsymbol{\Phi}(\boldsymbol{X}) & \boldsymbol{0} & \boldsymbol{0} \\
				\boldsymbol{0} & \boldsymbol{\Phi}(\boldsymbol{X})& \boldsymbol{0}\\
				\boldsymbol{0}& \boldsymbol{0}&\boldsymbol{\Phi}(\boldsymbol{X})
			\end{array}\right)\left(  \begin{array}{c}
				\boldsymbol{w_u}   \\
				\boldsymbol{w_v}   \\
				\boldsymbol{w_w}
			\end{array}\right)=\boldsymbol{\Phi}_U(\boldsymbol{X})\boldsymbol{w}_U\,,
		\end{equation}
		where $\boldsymbol{w_w}\in \mathbb{R}^{n_b}$ are the weights for the third velocity component. The involved matrices and arrays increase their size accordingly, and are thus $\boldsymbol{\Phi}_U(\boldsymbol{x}_i)\in \mathbb{R}^{3n_p\times3n_b}$ and $\boldsymbol{w}_U=\left(\boldsymbol{w_u};\boldsymbol{w_v};\boldsymbol{w_w}\right)\in \mathbb{R}^{3n_b}$.
		The RBF derivatives in the z-direction can be added to the two dimensional set  \eqref{eq10} :
		\begin{equation}
			\partial_{z} \varphi_{k}\left(\boldsymbol{X} \mid \boldsymbol{x}_{k}^{*}, c_{k}\right)=-2 c_{k}^{2}\left(z-z_{k}^{*}\right) \varphi_{k}\left(\boldsymbol{X} \mid \boldsymbol{x}_{k}^{*}, c_{k}\right).
		\end{equation}
		Analogously to the other velocities component \eqref{eq11}, it is possible to differentiate the third velocity:
		\begin{equation}
			\partial_{z} w\left(\boldsymbol{X}\right) \approx \sum_{k=0}^{n_{c}} w_{w k} \partial_{z} \varphi_{k}\left(\boldsymbol{X} \mid \boldsymbol{x}_{k}^{*}, c_{k}\right)=\boldsymbol{\Phi}_{z}\left(\boldsymbol{X}\right) \boldsymbol{w}_{w},
		\end{equation}
		where $\boldsymbol{\Phi}_{z}(\boldsymbol{X})\in \mathbb{R}^{n_p\times n_b}$ is defined as in \eqref{eq12}. In line with the notation in Section \ref{sec:2}, the divergence operator becomes:
		\begin{equation}
			\nabla \cdot\left(\begin{array}{c}
				\boldsymbol{u}\left(\boldsymbol{X}_{\nabla}\right) \\
				\boldsymbol{v}\left(\boldsymbol{X}_{\nabla}\right) \\
				\boldsymbol{w}\left(\boldsymbol{X}_{\nabla}\right)
			\end{array}\right) \approx\left(\begin{array}{lll}
				\boldsymbol{\Phi}_{x}\left(\boldsymbol{X}_{\nabla}\right) & \boldsymbol{\Phi}_{y}\left(\boldsymbol{X}_{\nabla}\right) &\boldsymbol{\Phi}_{z}\left(\boldsymbol{X}_{\nabla}\right) 
			\end{array}\right)\left(\begin{array}{c}
				\boldsymbol{w_{u}} \\
				\boldsymbol{w_{v}} \\
				\boldsymbol{w_w}
			\end{array}\right)=\boldsymbol{D}_{\nabla}\left(\boldsymbol{X}_{\nabla}\right) \boldsymbol{w_{U}}=\boldsymbol{0}
		\end{equation} where $\boldsymbol{X}_{\nabla}\in \mathbb{R}^{n_{\nabla}\times 3}$ and $\boldsymbol{D}_{\nabla}\in \mathbb{R}^{n_{\nabla}\times 3n_b}$.
		Denoting $\boldsymbol{n}=(n_x,n_y,n_z)$ as the normal vector to a surface and $\boldsymbol{n}_z\left(\boldsymbol{X}_N\right)\in \mathbb{R}^{n_N}$ with $\boldsymbol{X}_N\in \mathbb{R}^{n_N\times 3}$, the directional derivative is:
		\begin{equation}
			\boldsymbol{\Phi}_{\boldsymbol{n}}(\boldsymbol{X}_N)=\boldsymbol{\Phi}_x\left(\boldsymbol{X}_N\right)\odot \boldsymbol{n}_x \left(\boldsymbol{X}_N\right)+\boldsymbol{\Phi}_y\left(\boldsymbol{X}_N\right)\odot \boldsymbol{n}_y \left(\boldsymbol{X}_N\right)+\boldsymbol{\Phi}_z\left(\boldsymbol{X}_N\right)\odot \boldsymbol{n}_z \left(\boldsymbol{X}_N\right).
		\end{equation}
		Linear constraints on the normal derivatives on a surface with normal $\boldsymbol{n}$ (cf. equation \eqref{Neuvelocity} for the 2D) become:
		\begin{equation*}
			\left(  \begin{array}{ccc}
				\boldsymbol{\Phi}_{\boldsymbol{n}}(\boldsymbol{X}_N) & \boldsymbol{0}& \boldsymbol{0}  \\
				\boldsymbol{0} & \boldsymbol{\Phi}_{n}(\boldsymbol{X}_N)& \boldsymbol{0}\\
				\boldsymbol{0} & \boldsymbol{0}& \boldsymbol{\Phi}_{n}(\boldsymbol{X}_N)
			\end{array}\right)\left(  \begin{array}{c}
				\boldsymbol{w_u}   \\
				\boldsymbol{w_v} \\
				\boldsymbol{w_w}
			\end{array}\right)=\boldsymbol{N}_{\boldsymbol{n}}(\boldsymbol{X}_N)\, \boldsymbol{w}_U=\boldsymbol{c}_N,
		\end{equation*}
		While the linear constraints on the velocity field (cf. equation \eqref{Dirvelocity} for the 2D) become
		
		\begin{equation}
			\left(  \begin{array}{ccc}
				\boldsymbol{\Phi}(\boldsymbol{X}_D) & \boldsymbol{0} & \boldsymbol{0} \\
				\boldsymbol{0} & \boldsymbol{\Phi}(\boldsymbol{X}_D) & \boldsymbol{0} \\
				\boldsymbol{0} & \boldsymbol{0} & \boldsymbol{\Phi}(\boldsymbol{X}_D) \\
			\end{array}\right)\left(  \begin{array}{c}
				\boldsymbol{w_u}   \\
				\boldsymbol{w_v} \\
				\boldsymbol{w_w}
			\end{array}\right)=\boldsymbol{D}(\boldsymbol{X}_D)\, \boldsymbol{w}_U=\boldsymbol{c}_D\,,
		\end{equation} where $\boldsymbol{c}_D=(\boldsymbol{c}_{Du};\boldsymbol{c}_{Dv};\boldsymbol{c}_{Dw})\in\mathbb{R}^{3n_D}$, with $\boldsymbol{c}_{Dw}\in \mathbb{R}^{n_D}$ the values constraining the third velocity component and $\boldsymbol{X}_D\in \mathbb{R}^{n_D\times 3}$ collecting the points where the condition is set.
		Finally, the problem can be cast in the general form as in \eqref{Blocks_V}, here reported for completeness together with the related dimension:
		\begin{subequations}
			\begin{gather}
				\boldsymbol{A}_U= 2\boldsymbol{\Phi}_U^T\boldsymbol{\Phi}_U+2\alpha_\nabla \boldsymbol{D}_{\nabla}^T\boldsymbol{D}_{\nabla} \in\mathbb{R}^{3 n_b\times 3 n_b}\label{a_u_v_3}\\
				\boldsymbol{B}_U=\left (\boldsymbol{D}^T_\nabla,\boldsymbol{D}^T,\boldsymbol{N}^T_{\boldsymbol{n}}\right) \in\mathbb{R}^{3 n_b\times n_{\lambda}}\label{b_u_v_3}\\
				\boldsymbol{b}_{U1}=2\boldsymbol{\Phi}_U^T \boldsymbol{U}\in \mathbb{R}^{3n_b}\label{c_u_v_3}\\
				\boldsymbol{b}_{U2}=(\boldsymbol{0},\boldsymbol{c}_D\,,\boldsymbol{c}_N)^T\in\mathbb{R}^{n_\lambda },
			\end{gather}
		\end{subequations} where $n_\lambda=n_{\nabla}+3n_D+3n_N$.
		\section{Pressure Integration in 3D}\label{AppendixB}
		The problem of pressure integration is less affected by the increased dimension of the problem when moving from 2D to 3D. The main difference concerns the forcing term:
		\begin{equation}
			\nabla \cdot \left({U}\cdot\nabla{U}\right)=(\partial_x u)^2+(\partial_y v)^2+(\partial_z w)^2+2\partial_x v\partial_y u+2\partial_x w\partial_z u+2\partial_y w\partial_z v,
		\end{equation}
		which can then be written with the same formalism as \eqref{source}.
		Equation \eqref{Projection_P} requires particular attention as in 3D $\boldsymbol{n}=\left(n_{x}, n_{y}, n_{z}\right)$ and all three momentum equation must be projected.
		If the aforementioned changes are applied, the obtained matrices do not differ from the 2D and therefore equations \eqref{Blocks_P} and \eqref{Prob_P} hold true.
		
		\bibliography{Sperotto_Pieraccini_Mendez_MST_2022.bib}

	\end{document}